\pdfoutput=1
\documentclass[12pt,a4paper]{iopart} 

\usepackage{graphicx}
\usepackage{color}
\usepackage{amssymb}
\usepackage{iopams}
\usepackage{bm}
\usepackage{units}
\usepackage{fancyhdr}
\usepackage{multicol}
\usepackage{float}
\usepackage{url}

\usepackage[margin=1.5cm, bottom=1.5cm]{geometry}

\usepackage{harvard}

\definecolor{darkgreen}{rgb}{0.1,0.7,0.1}

\newcommand{\ontop}[2]{
  \renewcommand{\arraystretch}{0.2}
  \begin{array}{c}
  #1 \\ #2
  \end{array}
  \renewcommand{\arraystretch}{1.0}
}
\renewcommand{\lsim}{\ontop{<}{\sim}}
\newcommand{\gsim}{\ontop{>}{\sim}}

\newcommand{\listemph}[1]{\textbf{\emph{#1}}}
\newcommand{\mylistemph}[1]{\listemph{#1}}

\renewcommand{\vec}{\bm}

\renewcommand{\mat}[1]{\bm{\mathrm{#1}}}

\newcommand{\fourier}{\tilde}

\newcommand{\survey}[1]{{#1}}

\newcommand{\instrument}[1]{{#1}}

\newcommand{\citet}[2][]{#1\citeasnoun{#2}}

\newcommand{\citep}[2][]{(#1\citename{#2} \citeyear*{#2})}

\newcommand{\citeptwo}[3][]{(#1\citename{#2} \citeyear*{#2}, \citename{#3} \citeyear*{#3})}
\newcommand{\citeprefixed}[2]{(\citename{#1} \citeyear*{#1}#2)}

\newcommand{\begtwocol}{\doiftwocol{\begin{multicols}{2}}}
\newcommand{\stoptwocol}{\doiftwocol{\end{multicols}}}

\begin{document}

\citationmode{abbr} 

\def\aj{AJ}%
\def\araa{ARA\&A}%
\def\apj{ApJ}%
\def\apjl{ApJ}%
\def\apjs{ApJS}%
\def\ao{Appl.~Opt.}%
\def\apss{Ap\&SS}%
\def\aap{A\&A}%
\def\aapr{A\&A~Rev.}%
\def\aaps{A\&AS}%
\def\azh{AZh}%
\def\baas{BAAS}%
\def\jcap{JCAP}%
\def\jrasc{JRASC}%
\def\memras{MmRAS}%
\def\mnras{MNRAS}%
\def\na{New Astron.}%
\def\nar{New Astron.~Rev.}%
\def\pasa{PASA}
\def\pra{Phys.~Rev.~A}%
\def\prb{Phys.~Rev.~B}%
\def\prc{Phys.~Rev.~C}%
\def\prd{Phys.~Rev.~D}%
\def\pre{Phys.~Rev.~E}%
\def\prl{Phys.~Rev.~Lett.}%
\def\pasp{PASP}%
\def\pasj{PASJ}%
\def\qjras{QJRAS}%
\def\skytel{S\&T}%
\def\solphys{Sol.~Phys.}%
\def\sovast{Soviet~Ast.}%
\def\ssr{Space~Sci.~Rev.}%
\def\zap{ZAp}%
\def\nat{Nature}%
\def\iaucirc{IAU~Circ.}%
\def\aplett{Astrophys.~Lett.}%
\def\apspr{Astrophys.~Space~Phys.~Res.}%
\def\bain{Bull.~Astron.~Inst.~Netherlands}%
\def\fcp{Fund.~Cosmic~Phys.}%
\def\gca{Geochim.~Cosmochim.~Acta}%
\def\grl{Geophys.~Res.~Lett.}%
\def\jcp{J.~Chem.~Phys.}%
\def\jgr{J.~Geophys.~Res.}%
\def\jqsrt{J.~Quant.~Spec.~Radiat.~Transf.}%
\def\memsai{Mem.~Soc.~Astron.~Italiana}%
\def\nphysa{Nucl.~Phys.~A}%
\def\physrep{Phys.~Rep.}%
\def\physscr{Phys.~Scr}%
\def\planss{Planet.~Space~Sci.}%
\def\procspie{Proc.~SPIE}%
\let\astap=\aap
\let\apjlett=\apjl
\let\apjsupp=\apjs
\let\applopt=\ao

\pagestyle{fancy}
\lhead{\leftmark}
\rhead{\thepage}
\cfoot{}


\newcommand{\doifonecol}[1]{#1}
\newcommand{\doiftwocol}[1]{}


\review{Cosmology with cosmic shear observations: a review}

\author{Martin Kilbinger}

\address{Laboratoire AIM, CEA Saclay - CNRS - Paris 6, Irfu/SAp, F-91191 Gif-sur-Yvette, France}
\eads{\mailto{martin.kilbinger@cea.fr}}

\begin{abstract}


Cosmic shear is the distortion of images of distant galaxies due to weak
gravitational lensing by the large-scale structure in the Universe. Such images are
coherently deformed by the tidal field of matter inhomogeneities along the line
of sight. By measuring galaxy shape correlations, we can study the
properties and evolution of structure on large scales as well as the geometry
of the Universe. Thus, cosmic shear has become a powerful probe into the nature of dark matter
and the origin of the current accelerated expansion of the Universe. Over the
last years, cosmic shear has evolved into a reliable and robust cosmological
probe, providing measurements of the expansion history of the Universe and the
growth of its structure.

We review here the principles of weak gravitational lensing and show how cosmic
shear is interpreted in a cosmological context. Then we give an overview of
weak-lensing measurements, and present the main observational cosmic-shear results
since it was discovered 15 years ago,
as well as the implications for cosmology. We then conclude with
an outlook on the various future surveys and missions, for which cosmic shear
is one of the main science drivers, and discuss promising new weak cosmological
lensing techniques for future observations.

\end{abstract}

\noindent{\it Keywords: weak gravitational lensing, cosmology, large-scale structure of the universe}

\begtwocol

\tableofcontents



\vspace*{3em}

\section{Introduction}

On May, 29, 1919, during a solar eclipse, the deflection of light rays of stars
due to the Sun's gravitational field was measured \cite{1920RSPTA.220..291D},
marking the first successful test of the theory of general relativity
\citep[GR; ]{1916AnP...354..769E}. Then, the first discovery of extra-galactic
gravitational lensing was obtained in 1979, with the detection of a
doubly-imaged quasar lensed by a galaxy \cite{1979Natur.279..381W}. Lensing
distortions have been known since 1987 with the observation of giant arcs ---
strongly distorted galaxies behind massive galaxy clusters
\cite{1987A&A...172L..14S}. Three years later in 1990, weak gravitational
lensing was detected for the first time as statistical tangential alignments of
galaxies behind massive clusters \cite{1990ApJ...349L...1T}. It took another 10
years until, in 2000, coherent galaxy distortions were measured in blind
fields, showing the existence of weak gravitational lensing by the large-scale
structure, or cosmic shear
\cite{2000MNRAS.318..625B,kaiser00,2000A&A...358...30V,2000Natur.405..143W}.
And so, nearly 100 years after its first measurement, the technique of
gravitational lensing has evolved into a powerful tool for challenging GR on
cosmological scales.

All observed light from distant galaxies is subject to gravitational lensing.
This is because light rays propagate through a universe that is inhomogeneous
due to the ubiquitous density fluctuations at large scales. These fluctuations
create a tidal gravitational field that causes light bundles to be deflected
differentially. As a result, images of light-emitting galaxies that we observe
are distorted. The direction and amount of distortion is directly related to
the size and shape of the matter distribution projected along the line of
sight. The deformation of high-redshift galaxy images in random lines of sight
therefore provides a measure of the large-scale structure (LSS) properties,
which consists of a network of voids, filaments, and halos. The larger the
amplitude of the inhomogeneity of this cosmic web is, the larger the
deformations are. This technique of \emph{cosmic shear}, or \emph{weak
cosmological lensing} is the topic of this review.

The typical distortions of high-redshift galaxies by the cosmic web are on the
order of a few percent, much smaller than the width of the intrinsic shape and
size distribution. Thus, for an individual galaxy, the lensing effect is not
detectable, placing cosmic shear into the regime of \emph{weak gravitational
lensing}. The presence of a tidal field acting as a gravitational lens
results in a coherent alignment of galaxy image orientations. This alignment
can be measured statistically as a correlation between galaxy shapes.

Cosmic shear is a very versatile probe of the LSS. It measures the clustering
of the LSS from the highly non-linear, non-Gaussian sub-megaparsec
(Mpc) regime, out to very large, linear scales of more than a hundred Mpc. By
measuring galaxy shape correlations between different redshifts, the evolution
of the LSS can be traced, enabling us to detect the effect of dark energy on
the growth of structure. Together with the ability to measure the geometry of
the Universe, cosmic shear can potentially distinguish between dark energy and
modified gravity theories \cite{1999ApJ...522L..21H}. Since gravitational
lensing is not sensitive to the dynamical state of the intervening masses, it
yields a direct measure of the total matter, dark plus luminous. By adding
information about the distribution of galaxies, cosmic shear can shed light on
the complex relationship between galaxies and dark matter.

Since the first detection over a few square degrees of sky area a decade and a
half ago, cosmic shear has matured into an important tool for cosmology.
Current surveys span hundreds of square degrees, and thousands of square
degrees more to be observed in the near future. Cosmic shear is a major science
driver of large imaging surveys from both ground and space.

Various past reviews of weak gravitational lensing have covered the topic of
this review, \citet[e.g.~]{BS01}, \citet{SaasFee}, \citet{2008ARNPS..58...99H},
\citet{2008PhR...462...67M}, and \citet{2010CQGra..27w3001B}. Here, we will
present derivations of much of the basics of weak cosmological lensing, then
give an overview of the results of cosmic shear observations along with their
implications for cosmology.


\section{Cosmological background}
\label{sec:cosmo_back}

This section provides a very brief overview of the cosmological
concepts and equations relevant for cosmic shear. Detailed derivations
of the following equations can be found in standard cosmology
textbooks, \citet[e.g.~]{pee80}, \citet{CL:96}, \citet{2003moco.book.....D}. 

\subsection{Standard cosmological model}
\label{sec:standard_model}

In the standard cosmological model, the field equations of General Relativity
(GR) describe the relationship between space-time geometry and the
matter-energy content of the Universe governed by gravity. A solution to these
non-linear differential equations exists representing a homogeneous and
isotropic universe.

To quantify gravitational lensing, however, we need to consider light
propagation in an inhomogeneous universe. For a general metric that describes
an expanding universe including first-order perturbations, the line element
${\rm d} s$ is given as
\begin{equation}
\doiftwocol{\fl}
  {\rm d} s^2 = \left(1 + \frac {2 \Psi}{c^2} \right) c^2 {\rm d} t^2 -
  a^2(t) \left(1 - \frac {2 \Phi}{c^2} \right)
  {\rm d} l^2,
  \label{eq:metric_gen}
\end{equation}
where the scale factor $a$ is a function of cosmic time $t$ (we set
$a$ to unity at present time $t = t_0$), and $c$ is the speed of light.
The spatial part of the metric is given by the comoving coordinate $l$, which remains constant as
the Universe expands.
The two Bardeen gravitational potentials $\Psi$
and $\Phi$ are considered to describe weak fields,
$\Psi, \Phi \ll c^2$.
The potential of a lens with mass $M$ and radius $R$ can be approximated by
$G M / R = (c^2 / 2) (R_{\rm S} / R)$, where $G$ is Newton's gravitational constant
and $R_{\rm S}$ is the Schwarzschild radius. The weak-field condition is fulfilled for
most mass distributions, excluding only those very compact objects whose extent $R$
is comparable to their Schwarzschild radius.

In GR, and in the absence of anisotropic stress which is the case on
large scales, the two potentials are equal, $\Psi = \Phi$. If the
perturbations vanish, (\ref{eq:metric_gen}) reduces to the 
Friedmann-Lema\^itre-Robertson-Walker (FLRW) metric.

The spatial line element ${\rm d} l^2$ can be separated into a radial and angular
part, ${\rm d} l^2 = {\rm d} \chi^2 + f^2_K(\chi) {\rm d} \omega$. Here, $\chi$ is the comoving
coordinate and $f_K$ is the the comoving angular distance, the functional form of which
is given for the three distinct cases of three-dimensional space with curvature $K$ as
\doiftwocol{%
\end{multicols}
\par\noindent\rule{\dimexpr(0.5\textwidth-0.5\columnsep-0.4pt)}{0.4pt}%
\rule{0.4pt}{6pt}
}
\begin{equation}
  f_K(\chi)
  = \left\{ \begin{array}{llll} 
      K^{-1/2} \sin{\left( K^{1/2} \, \chi \right) } \;\;\;\;\;\;
      & \mbox{for} & K>0  & \mbox{(spherical)} \\
      \chi & \mbox{for} & K=0 & \mbox{(flat)} \\
      (-K)^{-1/2} \sinh{\left[ (-K)^{1/2} \, \chi \right]} & \mbox{for}
      & K<0 & \mbox{(hyperbolic)} \; .
    \end{array}
  \right.
\end{equation}
\doiftwocol{%
\vspace{\belowdisplayskip}\hfill\rule[-6pt]{0.4pt}{6.4pt}%
\rule{\dimexpr(0.5\textwidth-0.5\columnsep-1pt)}{0.4pt}
\begin{multicols}{2}
}
that are characterised by their corresponding equation-of-state relation
between pressure $p$ and density $\rho$, given by the parameter $w$ as
\begin{equation}
  p = w \, c^2 \rho.
  \label{eq:eos}
\end{equation}
The present-day density of each species is further scaled by
the present-day critical density of the Universe $\rho_{{\rm c}, 0} = 3 H_0^2 / (8
\pi G)$, for which the Universe has a flat geometry. The Hubble constant
$H_0 = H(a=1) = (\dot a / a)_{t=t_0} = 100 \, h \, \mbox{km} \, \mbox{s}^{-1} \mbox{Mpc}^{-1}$
denotes the present-day value of the Hubble parameter $H$, 
and the parameter $h \sim 0.7$ characterizes the uncertainty in our knowledge of $H_0$.
The density parameter
of non-relativistic matter is $\Omega_{\rm m} = \rho_{\rm m, 0} / \rho_{\rm crit,
0}$, which consists of cold dark matter (CDM), baryonic matter, and
possibly heavy neutrinos as $\Omega_{\rm m} = \Omega_{\rm c} + \Omega_{\rm
b} + \Omega_{\nu}$\footnote{Unless written as function of $a$, density parameters
are interpreted at present time; the subscript '0' is omitted.}.
Relativistic matter ($\Omega_{\rm r}$) consists of
photons, with the main contributors being the cosmic microwave background (CMB) radiation,
and light neutrinos. Finally, the
component driving the accelerated expansion (``dark energy'') is
denoted by $\Omega_{\rm de}$. Lacking a
well-motivated physical model, the dark-energy equation-of-state parameter $w$
is often parametrized by
the first or first few coefficients of a Taylor expansion, e.g.~$w(a) = w_0 +
w_1 (1 - a)$ \cite{2001IJMPD..10..213C,2003PhRvL..90i1301L}.
In the case of the cosmological constant,
$\Omega_{\rm de} \equiv \Omega_\Lambda$ and $w=-1$.

The sum of all density parameters defines the \emph{curvature density parameter}
$\Omega_K$, with $\Omega_{\rm m} + \Omega_{\rm de} + \Omega_{\rm r} = 1 - \Omega_K$,
where $\Omega_K = - (c/H_0)^2 K$ has opposite sign compared to the curvature $K$.

Alternative parametrizations of the density of a species $x$ are the \emph{physical density parameters}, which are defined as
$\omega_x = \Omega_x h^2$.

\subsection{Structure formation}
\label{sec:structure_formation}

In an expanding universe, density fluctuations evolve with time. Tiny quantum
fluctuations in the primordial inflationary cosmos generate small-amplitude
density fluctuations. Subsequently, these fluctuations grow into
the large structures we see today, in the form of clusters, filaments, and
galaxy halos.

At early enough times or on large enough scales, those density fluctuations
are small, and their evolution can be treated using linear
perturbation theory. Once those fluctuations grow to become non-linear, other
approaches to describe them are necessary --- for example higher-order
perturbation theory, renormalization group mechanisms, analytical models of
gravitational collapse, the so-called halo model, or $N$-body simulations
(see Sect.~\ref{sec:simuls}).

Fluctuations of the density $\rho$ around the mean density $\bar \rho$
are parametrized by the density contrast
\begin{equation}
  \delta = \frac{\rho - \bar \rho}{\bar \rho}.
  \label{eq:delta}
\end{equation}
For non-relativistic perturbations in the matter-dominated era on scales smaller than the horizon,
i.e.~the light travel distance since $t=0$,
Newtonian physics suffices to describe the evolution of $\delta$ \cite{pee80}.
The density contrast of an ideal fluid of zero pressure
is related to the gravitational potential via the Poisson
equation,
\begin{equation}
  \vec \nabla^2 \Phi = 4 \pi G a^2 \bar \rho \, \delta .
  \label{eq:Poisson}
\end{equation}
The differential equation describing the evolution of $\delta$ typically has to be solved
numerically, although in special cases analytical solutions exist. The
solution that increases with time is called \emph{growing mode}.
The time-dependent function is the \emph{linear growth
factor} $D_+$, which relates the density contrast at time $a$ to an earlier, initial epoch $a_{\rm i}$, with
$\delta(a) \propto D_+(a) \delta_(a_{\rm i})$.
In a matter-dominated Einstein-de-Sitter Universe,
$D_+$ is proportional to the scale factor $a$. The presence of dark
energy results in a suppressed growth of structures.

In the early Universe, when radiation is the dominant species, perturbations with
comoving scales smaller than the horizon do not grow.
Super-horizon perturbations grow as $\delta \propto a^2$ until
the time when, due to the expansion of the Universe, these perturbations ``enter
the horizon''. This leads to a difference of growth as a function of
perturbation scale, which is quantified in the transfer function $T$, expressed as
a function of the scale $k$ in Fourier space.
$T$ describes the evolution of the density contrast at scale $k$ compared to the super-horizon case
at an arbitrary large scale $k=0$ \cite{1998ApJ...496..605E},
\begin{equation}
T(k) = \left( \frac{\tilde \delta(k, a=1)}{\tilde \delta(k, a_{\rm i})} \right) \Bigg/ \left( \frac{\tilde \delta(k=0, a=1)}{\tilde \delta(k=0, a_{\rm i})} \right).
\end{equation}
Here, the tilde denotes Fourier transform.
On large scales, $T(k \rightarrow
0)$ approaches unity, since at early enough time all perturbations live
outside the horizon. At small scales, $T(k) \propto k^{-2}$. The details of
this function depend on the matter content and its equation of state (\ref{eq:eos})
\cite{bbks86,1995ApJS..100..281S,1998ApJ...496..605E}.

\subsection{Modified gravity models}
\label{sec:mod_grav}

A very general, phenomenological characterisation of deviations from
GR is to add parameters to the Poisson equation, and to
treat the two Bardeen potentials as two independent quantities. This
leads to two modified, distinct Poisson equations, which, expressed in
Fourier space, are \cite{2006astro.ph..5313U,2008JCAP...04..013A}
\begin{eqnarray}
k^2 \tilde \Psi(k, a) & = & 4 \pi G a^2 \left[ 1 + \mu(k, a) \right] \rho
\, \tilde \delta(k, a);
\label{eq:Poisson_mod_Psi} \\
k^2 \left[ \tilde \Phi(k, a) + \tilde \Psi(k, a) \right] & = & 8  \pi G a^2
\left[ 1 + \Sigma(k, a) \right] \rho \, \tilde \delta(k, a).
\label{eq:Poisson_mod_Phi_plus_Psi}
\end{eqnarray}
Non-zero values of the free functions $\mu$ and $\Sigma$ represent
deviations from GR. This flexible parametrization can account for a
variety of modified gravity models, for example a change in the
gravitational force from models with extra-dimensions as in DGP
\nocite{2000PhLB..484..112D}
(Dvali, Gabadadze \& Porrati 2000),
massive gravitons \cite{1994PhRvL..73.2950Z},
$f(R)$ extensions of the Einstein-Hilbert action
\cite{2010LRR....13....3D}, or Tensor-Vector-Scalar (TeVeS) theories
\cite{2009CQGra..26n3001S}.
Non-zero anisotropic stress is predicted from a variety of
higher-order gravity theories, but also expected from models of
clustered dark energy \cite{1998ApJ...506..485H,2011PhRvD..83b3011C}.
See \citet{2012PhR...513....1C} and \citet{2012IJMPD..2130002Y} for further models of modified gravity.

The above-introduced parametrization has the advantage of separating the effect
of the metric on non-relativistic particles (which are influenced by density
fluctuations through (\ref{eq:Poisson_mod_Psi})), and light deflection (which
is governed by both geometry and density fluctuations via
(\ref{eq:Poisson_mod_Phi_plus_Psi}), see e.g.~\citet{2001PhRvD..64h3004U},
\citet{2008PhRvD..78f3503J}). Thus, data from galaxy clustering, redshift-space
distortions, and velocity fields (testing the former relation on the one hand)
and weak-lensing observations (testing the latter equation on the other hand)
are complementary in their ability to constrain modified gravity models.

Alternative parametrizations for modifying GR have also been used, such as: the
ratio of potentials $\zeta = 1 - \Phi / \Psi$ (``gravitational slip''); the
growth index $\gamma$ defined by ${\rm d} \ln D_+ / {\rm d} \ln a =
{\Omega_{\rm m}}^\gamma(a)$; the parameter $E_{\rm G}$, which is a
galaxy-bias-independent ratio of the matter--galaxy correlation, the galaxy
auto-correlation, and the redshift distortion parameter $\beta$
\cite{2007PhRvL..99n1302Z}. All those parametrizations can be expressed in
terms of $\mu$ and $\Sigma$ \cite{CFHTLenS-mod-grav}.


\section{Weak cosmological lensing formalism}

This section introduces the basic concepts of cosmic shear, and
discusses the relevant observables and their relationships to theoretical models
of the large-scale structure. More details about those concepts can be found
in \citet[e.g. ]{BS01}.

\subsection{Light deflection and the lens equation}
\label{sec:light_deflection}

There are multiple ways to derive the equations describing the deflection of light rays
in the presence of massive bodies. An intuitive approach is the use of Fermat's principle
of minimal light travel time \cite{1992grle.book.....S,1985A&A...143..413S,1986ApJ...310..568B}.

Photons propagate on null geodesics, given by a vanishing line
element ${\rm d} s$. In the case of GR we get the light ray travel time from
the metric (\ref{eq:metric_gen}) as
\begin{equation}
  t = \frac 1 c \int \left(1 - \frac{2 \Phi}{c^2} \right) {\rm d} r,
\end{equation}
where the integral is along the light path in physical or proper coordinates ${\rm d} r$.
Analogous to geometrical optics, the potential acts as a medium with variable
refractive index $n = 1 - 2 \Phi / c^2$ (with $\Phi < 0$), changing the direction
of the light path. (This effect is what gives gravitational \emph{lensing} its name.)
We can apply Fermat's principle, $\delta t = 0$, to
get the Euler-Lagrange equations for the refractive index.
Integrating these equations along the light path results in the \emph{deflection angle}
$\vec{\hat \alpha}$ defined as the difference between the directions of emitted and received
light rays,
\begin{equation}
  \vec{\hat \alpha} = - \frac{2}{c^2} \int \vec \nabla_\perp^{\rm p} \Phi \, {\rm d} r.
  \label{eq:alpha_hat}
\end{equation}
The gradient of the potential is taken perpendicular to the light path,
with respect to physical coordinates.
The deflection angle is twice the classical prediction in Newtonian dynamics if
photons were massive particles \cite{Soldner1804}.

\subsection{Light propagation in the universe}
\label{sec:light_propagation}

In this section we quantify the relation between light deflection and
gravitational potential on cosmological scales. 
To describe differential propagation of rays within an
infinitesimally thin light bundle, we consider the difference
between two neighbouring geodesics, which is given by the
\emph{geodesic deviation equation}. In a homogeneous FLRW Universe,
the transverse comoving separation $\vec x_0$ between two light rays
as a function of comoving distance from the observer $\chi$ is proportional to the 
comoving angular distance
\begin{equation}
  \vec x_0(\chi) = f_K(\chi) \bm \theta,
  \label{eq:prop-hom}
\end{equation}
where the separation vector $\vec x_0$ is seen by the observer under the (small) angle $\btheta$
\cite{1992grle.book.....S,1994CQGra..11.2345S}.

This separation vector is modified 
by density perturbations in the Universe.
We have already seen (\ref{eq:alpha_hat}) that a light ray is deflected by an amount
${\rm d} \vec{\hat \alpha} = -2/c^2 \, \bm \nabla_\perp \Phi(\vec x, \chi^\prime) {\rm d} \chi^\prime$
in the presence of the potential $\Phi$ at distance $\chi^\prime$ from the observer.
Note that this equation is now expressed in a comoving frame, as well as the gradient.
From the vantage point of the deflector the induced change in separation vector at
source comoving distance $\chi$ is ${\rm d} \vec x = f_K(\chi - \chi^\prime) {\rm d} \vec
{\hat \alpha}$ (see Fig.~\ref{fig:propagation} for a sketch). The total separation is
obtained by integrating over the line of sight along $\chi^\prime$. Lensing
deflections modify the path of both light rays, and we denote with the superscript $^{(0)}$
the potential along the second, fiducial ray. The result is
\begin{equation}
\vec{x}(\chi) = f_K(\chi) \vec \theta - \frac{2}{c^2} \int_0^\chi {\rm d}\chi^\prime
   f_K(\chi-\chi^\prime) \left[ \vec \nabla_\perp\Phi(\vec{x},
     \chi^\prime), \chi^\prime) - \vec \nabla_\perp\Phi^{(0)}(\chi^\prime) \right].
\label{eq:prop-int}
\end{equation}
In the absence of lensing the separation vector $\vec x$ would be seen by the
observer under an angle
$\vec \beta = \vec x(\chi) / f_K(\chi)$. The difference between the apparent
angle $\vec \theta$ and $\vec \beta$
is the total, scaled deflection angle $\vec \alpha$, defining the \emph{lens equation}
\begin{equation}
  \vec \beta = \vec \theta - \vec \alpha ,
  \label{eq:lens}
\end{equation}
with
\begin{equation}
  \vec \alpha = \frac{2}{c^2} \int_0^\chi {\rm d}\chi^\prime
  \frac{f_K(\chi-\chi^\prime)}{f_K(\chi)} \left[ \vec \nabla_\perp\Phi(\vec{x},
    \chi^\prime), \chi^\prime) - \vec \nabla_\perp\Phi^{(0)}(\chi^\prime)
  \right].
  \label{eq:deflection_angle}
\end{equation}
\Eref{eq:lens} is analogous to the standard lens equation in the case
of a single, thin lens, in which case $\vec \beta$ is the source
position.

\stoptwocol
\begin{figure}
  \begin{center}
    \resizebox{0.7\hsize}{!}{
      \includegraphics{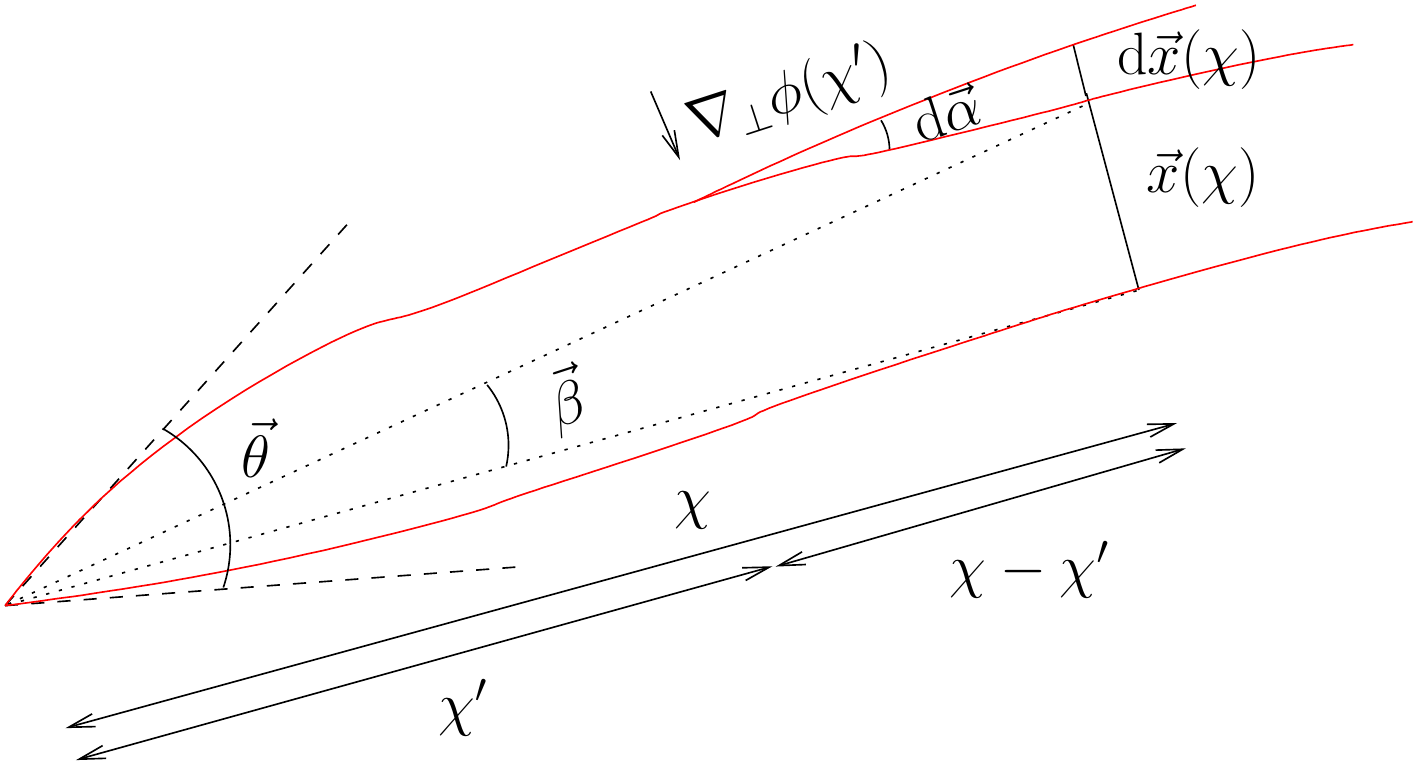}
    }
  \end{center}

  \caption{Propagation of two light rays (red solid lines), converging
    on the observer on the left. The light rays are separated by the transverse comoving distance 
    $\vec x$, which varies with distance $\chi$ from the observer. An exemplary deflector at distance $\chi^\prime$
    perturbes the geodescics proportional to the transverse gradient $\vec \nabla_\perp \phi$
    of the potential.
    The dashed lines indicate the apparent direction of the light rays,
    converging on the observer under the angle $\vec \theta$. The dotted lines show the unperturbed
    geodesics, defining the angle $\vec \beta$ under which the unperturbed transverse comoving separation $\vec x$
    is seen.}

  \label{fig:propagation}

\end{figure}
\begtwocol

\subsection{Linearized lensing quantities}
\label{sec:linear_lensing}

The integral equation (\ref{eq:prop-int}) can be approximated by substituting
the separation vector
$\vec x$ in the integral by the
$0^{\rm th}$-order solution $\vec x_0(\chi) = f_K(\chi) \vec \theta$ (\ref{eq:prop-hom}).
This corresponds to integrating the potential gradient along the
unperturbed ray, which is called the \emph{Born approximation} (see Sect.~\ref{sec:corrections}
for higher-order corrections). 
Further, we linearise the lens equation (\ref{eq:lens}) and define the (inverse) amplification matrix as the Jacobian $\mat
A = \partial \vec \beta / \partial \vec \theta$, which describes a
linear mapping from lensed (image) coordinates $\vec \theta$ to
unlensed (source) coordinates $\vec \beta$,
\begin{eqnarray}
  A_{ij} & = & \frac{\partial \beta_i}{\partial \theta_j} = \delta_{ij} -
  \frac{\partial \alpha_i}{\partial \theta_j}
  \nonumber \\
  & = & \delta_{ij} - \frac{2}{c^2} \int_0^\chi {\rm d}\chi^\prime
  \frac{f_K(\chi-\chi^\prime) f_K(\chi^\prime)}{f_K(\chi)} \frac{\partial^2}{\partial x_i
\partial x_j} \Phi(f_K(\chi^\prime) \vec \theta, \chi^\prime) .
  \label{eq:Jacobi}
\end{eqnarray}
The second term in (\ref{eq:deflection_angle}) drops out since it does not
depend on the
angle $\vec \theta$.

In this approximations the deflection angle can be written as the gradient of a
2D potential, the \emph{lensing potential} $\psi$,
\begin{equation}
  \psi(\vec \theta, \chi) = \frac 2 {c^2} \int_0^\chi {\rm d} \chi^\prime
    \frac{f_K(\chi - \chi^\prime)}{f_K(\chi) f_K(\chi^\prime)}\,
  \Phi(f_K(\chi^\prime) \vec \theta, \chi^\prime) .
  \label{eq:lensing_potential}
\end{equation}
With this definition, the Jacobi matrix can be expressed as
\begin{equation}
  A_{ij} = \delta_{ij} - \partial_i \partial_j \psi, 
  \label{eq:Jacobi_psi}
\end{equation}
where the partial derivatives are understood with respect to $\vec \theta$.
The symmetrical matrix ${\mat A}$ is parametrized in terms of the scalar
\emph{convergence}, $\kappa$, and the two-component spin-two
\emph{shear}, $\gamma = (\gamma_1, \gamma_2)$, as
\begin{equation}
  \mat A = \left( \begin{array}{cc} 1 - \kappa - \gamma_1 & - \gamma_2
    \\ - \gamma_2 & 1 - \kappa + \gamma_1 \end{array} \right).
\label{eq:jacobi}
\end{equation}
This defines the convergence and shear as second derivatives of the potential,
\begin{equation}
\kappa = \frac 1 2 \left( \partial_1 \partial_1 + \partial_2 \partial_2 \right) \psi
         = \frac 1 2 \nabla^2 \psi; \;\;;
  \gamma_1 = \frac 1 2 \left( \partial_1 \partial_1 - \partial_2 \partial_2 \right) \psi;
  \;\;
  \gamma_2 = \partial_1 \partial_2 \psi.
  \label{eq:kappa_gamma_psi}
\end{equation}
The inverse Jacobian $\mat A^{-1}$ describes the local mapping of the source
light distribution to image coordinates. The convergence, being the diagonal
part of the matrix, is an isotropic increase or decrease of the observed size
of a source image. Shear, the trace-free part, quantifies an anisotropic
stretching, turning a circular into an elliptical light distribution.

It is mathematically convenient to write the shear as complex number, $\gamma =
\gamma_1 + \rm i \gamma_2 = |\gamma| \exp(2 \rm i \varphi)$, with $\varphi$
being the polar angle between the two shear components. Shear transforms as a
spin-two quantity: a rotation about $\pi$ is the identity transformation of an
ellipse (see Fig.~\ref{fig:wheel} for an illustration).

\doiftwocol{\begin{figure}[H]
  \begin{center}
   \resizebox{0.8\hsize}{!}{
	  \includegraphics{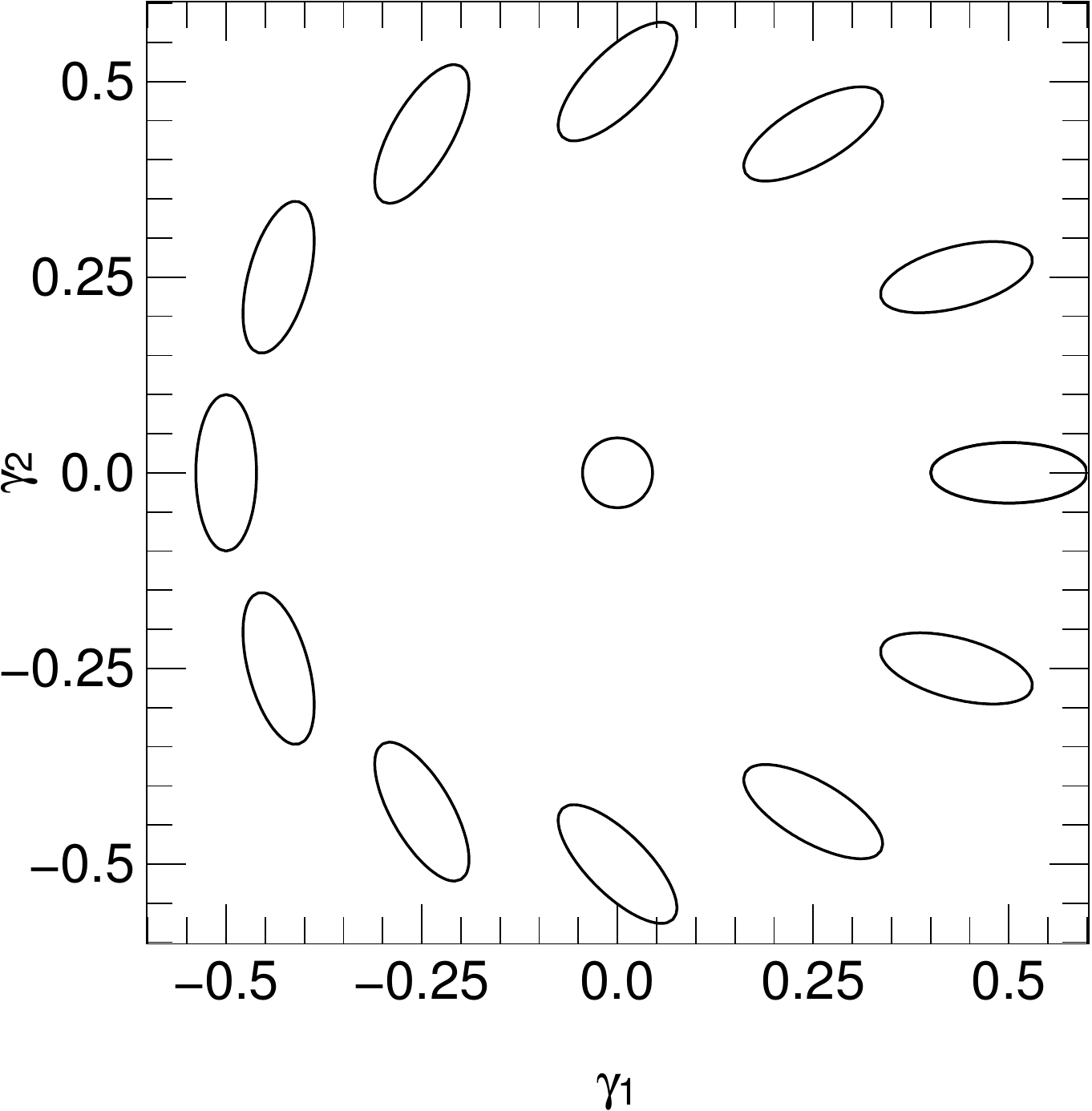}
   }
  \end{center}
}

\doifonecol{\begin{figure}
  \begin{center}
   \resizebox{0.45\hsize}{!}{
	  \includegraphics{figures/wheel.pdf}
   }
  \end{center}
}

   \caption{The orientation of the ellipses given by the Cartesian
      coordinates ${\gamma_1}$ and ${\gamma_2}$ of the
      shear. While the polar angle $\varphi$ passes through the range
      $[0;\, 2\pi]$, the shear ellipse rotates around
      $\pi$.}

   \label{fig:wheel}

\end{figure}

In the context of cosmological lensing by large-scale structures,
images are very weakly lensed, and the values of $\kappa$ and $\gamma$
are on the order of a few percent or less. Each source is mapped uniquely onto one image,
there are no multiple images, and the matrix $\mat A$ is indeed invertible.

We can factor out $(1-\kappa)$ from $\mat A$ (\ref{eq:jacobi}), since this multiplier
only affects the size but not the shape of the source. Cosmic
shear is based on the measurement of galaxy shapes (see Sect.~\ref{sec:shapes}), and
therefore the observable in question is not the shear $\gamma$ but the
\emph{reduced shear},
\begin{equation}
  g = \frac{\gamma}{1 - \kappa},
  \label{eq:reduced_shear}
\end{equation}
which has the same spin-two transformation properties as shear.
Weak lensing is the regime where the effect of gravitational lensing is very small,
with both the convergence and the shear much smaller than unity.
Therefore, shear is a good approximation of
reduced shear to linear order (see Sect.~\ref{sec:corrections} for
its validity). Magnification, which is an estimator of $\kappa$,
will be discussed in Sect.~\ref{sec:magnification}.

\subsection{Projected overdensity}
\label{sec:projected_overdensity}

Since the convergence $\kappa$ is related to the lensing potential $\psi$
(\ref{eq:lensing_potential}) via a
2D Poisson equation (\ref{eq:kappa_gamma_psi}), it can be interpreted
as a (projected) surface density. To introduce the 3D density contrast $\delta$,
we apply the 2D Laplacian
of the lensing potential (\ref{eq:lensing_potential}) to the 3D potential $\Phi$
and add the second-order
deriviate along the comoving coordinate, $\partial^2 / \partial \chi^2$.
This additional term vanishes, since positive and negative contributions cancel out
to a good approximation when integrating along the line of sight.
Next, we replace the
3D Laplacian of $\Phi$ with the over-density $\delta$ using the Poisson
equation (\ref{eq:Poisson}), and $\bar \rho \propto a^{-3}$. Writing the mean matter
density in terms of 
the critical density, we get
\begin{equation}
  \kappa(\vec \theta, \chi) = \frac{3 H_0^2 \Omega_{\rm m}}{2 c^2}
  \int\limits_0^\chi \frac{{\rm d} \chi^\prime}{a(\chi^\prime)} 
  \frac{f_K(\chi - \chi^\prime)}{f_K(\chi)} f_K(\chi^\prime) \,
\delta(f_K(\chi^\prime) \vec \theta, \chi^\prime).
  \label{eq:kappa_chi}
\end{equation}
This expression is a projection of the density along comoving coordinates,
weighted by geometrical factors involving
the distances between source, deflector, and observer. In the case of a flat universe,
the geometrical weight $(\chi - \chi^\prime) \chi^\prime$ is a parabola with maximum at
$\chi^\prime = \chi/2$.
Thus, structures at around half the distance to the source are most efficient to generate
lensing distortions.

The mean convergence from a population of source galaxies is obtained by weighting the above
expression with the galaxy probability distribution in comoving distance, $n(\chi) {\rm d} \chi$,
\begin{equation}
  \kappa(\vec \theta) = \int\limits_0^{\chi_{\rm lim}} {\rm d} \chi \, n(\chi) \,
  \kappa(\vec \theta, \chi).
  \label{eq:kappa}
\end{equation}
The integral extends out to the limiting comoving distance $\chi_{\rm lim}$ of
the galaxy sample. Inserting (\ref{eq:kappa_chi}) into (\ref{eq:kappa}) and
interchanging the integral order results
in the following expression, 
\begin{equation}
  \kappa(\vec \theta) = \frac{3 H_0^2 \Omega_{\rm m}}{2 c^2} \int\limits_0^{\chi_{\rm lim}}
  \frac{{\rm d} \chi}{a(\chi)}
  q(\chi) f_K(\chi) \, \delta(f_K(\chi) \vec \theta, \chi).
  \label{eq:kappa_final}
\end{equation}
The lens efficiency $q$ is defined as
\begin{equation}
  q(\chi) = \int\limits_\chi^{\chi_{\rm lim}} {\rm d} \chi^\prime \, n(\chi^\prime)
  \frac{f_K(\chi^\prime - \chi)}{f_K(\chi^\prime)},
  \label{eq:lens_efficiency}
\end{equation}
and indicates the lensing strength at a distance $\chi$ of the combined
background galaxy distribution. Thus, the convergence is a linear measure of the total matter density, projected
along the line of sight with dependences on the geometry of the universe via the
distance ratios, and the source galaxy distribution $n(\chi) {\rm d} \chi = n(z)
{\rm d} z$. 
The latter is usually obtained
using photometric redshifts (Sect.~\ref{sec:photo-z}). We will see in
Sects.~\ref{sec:tomography} and \ref{sec:mass_recon} how to recover
information in the redshift direction.

By construction, the expectation value of shear and convergence are zero, since
$\langle \delta \rangle = 0$. The first non-trivial statistical measure of the
distribution of $\kappa$ and $\gamma$ are second moments. Practical estimators
of weak-lensing second-order statistics in real and Fourier-space are discussed
in Sects.~\ref{sec:power_spectrum} and \ref{sec:real_space_2nd}.

\subsection{Estimating shear from galaxies}
\label{sec:estim_shear}

In the case of cosmic shear, not the convergence but the shear is measured from
the observed galaxy shapes, as discussed in this section. Theoretical
predictions of the convergence (\ref{eq:kappa_final}) can be related to the
observed shear using the relations (\ref{eq:kappa_gamma_psi}). Further, a
convergence field can be estimated by reconstruction from the observed galaxy
shapes, see Sect.~\ref{sec:mass_maps_formalism}. Alternatively, the convergence
can be estimated using \emph{magnification}, as discussed in
Sect.~\ref{sec:magnification}.

We can attribute an intrinsic, complex \emph{source ellipticity}
$\varepsilon^{\rm s}$
to a galaxy. Cosmic shear modifies this ellipticity as a function of the complex reduced shear,
which depends on the definition of $\varepsilon^{\rm s}$. If we define this quantity for an image
with elliptical isophotes, minor-to-major axis ratio $b/a$, and position angle $\phi$, as
$\varepsilon = (a - b)/(a + b) \times \exp(2 {\rm i} \phi)$,
the observed ellipticity $\varepsilon$ (for $|g| \le 1)$ is given as \cite{1997A&A...318..687S}
\begin{equation}
  \varepsilon = \frac{\varepsilon^{\rm s} + g}{1 + g^\ast\varepsilon^{\rm s}}.
  \label{eq:eps_g}
\end{equation}
The asterisk ``$^\ast$'' denotes complex conjugation.
In the weak-lensing regime, this relation is approximated to
\begin{equation}
 \varepsilon \approx \varepsilon^{\rm s} + \gamma .
  \label{eq:eps_eps_s_gamma}
\end{equation}

If the intrinsic ellipticity of galaxies has no preferred orientation, the
expectation value of $\varepsilon^{\rm s}$ vanishes, $\langle \varepsilon^{\rm
s} \rangle = 0$, and the observed ellipticity is an unbiased estimator of the
reduced shear,
\begin{equation}
\left\langle \varepsilon \right\rangle = g .
\label{eq:estimator_shear}
\end{equation}
This relation breaks down in the presence of intrinsic galaxy alignments (Sect.~\ref{sec:ia}).

Another commonly used ellipticity estimator has been proposed by
\cite{1995A&A...294..411S}. This estimator has a slightly simpler dependence on
second moments of galaxy images, which have been widely used for shape estimation,
see Sect.~\ref{sec:shapes}. However, it it does not provide an unbiased
estimator of $g$, but explicitly depends on the intrinsic ellipticity
distribution.

In the weak-lensing regime, the shear cannot be detected from an individual
galaxy. With distortions induced by the LSS of the order $\gamma \sim 0.03$, and the
typical intrinsic ellipticity rms of $\sigma_\varepsilon = \langle |
\varepsilon |^2 \rangle^{1/2} \sim 0.3$, one needs to average over a number of
galaxies $N$ of at least a few hundred to obtain a signal-to-noise ratio $S/N =
\gamma \times N^{1/2} / \sigma_\varepsilon$ of above unity. 

\subsection{E- and B-modes}
\label{sec:E_B_modes}

The Born approximation introduced in Sect.~\ref{sec:linear_lensing} results in the
definition of the convergence and shear to be functions of a single scalar
potential (\ref{eq:lensing_potential}). The two shear components
defined in that way (\ref{eq:kappa_gamma_psi}) are not independent, and the
shear field cannot have an arbitrary form. We can define a vector field $\vec u$
as the gradient of the ``potential'' $\kappa$, $\vec u = \vec \nabla \kappa$.
By definition, the curl of this gradient vanishes, $\vec \nabla \times \vec u =
\partial_1 u_2 - \partial_2 u_1 = 0$.
Inserting the relations between $\kappa, \gamma$ and $\psi$ (\ref{eq:kappa_gamma_psi})
into this equality results in second-derivative constraints for $\gamma$.
A shear field fulfilling those relations is called an \emph{E-mode} field,
analogous to the electric field. In real life however, $\vec u$ obtained from observed data
is in general not a pure gradient field but has a non-vanishing curl component.
The corresponding convergence field can be decomposed into its E-mode
component, $\kappa^{\rm E}$, and B-mode, $\kappa^{\rm B}$, given by
$\nabla^2 \kappa^{\rm E} = \vec \nabla u$ and $\nabla^2 \kappa^{\rm B} = \vec \nabla \times u$..
The \emph{B-mode} component can have various origins:
\begin{enumerate}
\item Higher-order terms in the light-propagation equation (\ref{eq:prop-int}),
      e.g.~lens-lens coupling and integration
      along the perturbed light path
      (\ref{eq:jacobi}) \cite{2010A&A...523A..28K}.
\item Other higher-terms beyond usual approximations of relations such as between shear and reduced shear,
      or between shear and certain ellipticity estimators (see Sect.~\ref{sec:shapes})
       \cite{2010A&A...523A..28K}.
\item Lens galaxy selection biases, such as size and magnitude bias \cite{2003ApJ...583...58W,2009ApJ...702..593S},
      or clustering of lensing galaxies \cite{1998A&A...338..375B,2002A&A...389..729S}.
\item Correlations of the intrinsic shapes of galaxies with each other, and with the structures
    that induce weak-lensing
    distortions (intrinsic alignment, Sect.~\ref{sec:ia}) \cite{2002ApJ...568...20C}.
\item Image and data analysis errors such as PSF correction residuals, systematics in the astrometry.
\end{enumerate}

The astrophysical effects (i) - (iv) cause a B-mode at the percent-level
compared to the E-mode. The intrinsic alignment B-mode amplitude is the least
well-known since the model uncertainty is large \cite{2013MNRAS.435..194C}. Up
to now, cosmic shear surveys do not have the statistical power to reliably
detect those B-modes. Until recently, the amplitude of a B-mode detection has
exclusively been used to assess the quality of the data analysis, assuming that
(v) is the only measurable B-mode contributor. While this is a valid approach,
it only captures those systematics that create a B-mode. A B-mode non-detection
might render an observer over-confident to believe that also the E-mode is
uncontaminated by systematics. Further, the ratio of B- to E-mode should not be
used to judge the data quality, since this ratio is not cosmology-independent
and can bias the cosmological inference of the data.

\subsection{The lensing power spectrum}
\label{sec:power_spectrum}

The basic second-order function of the convergence (\ref{eq:kappa_final}) is the
two-point correlation function (2PCF)
$\langle \kappa(\vec \vartheta) \kappa(\vec \vartheta + \vec \theta) \rangle$.
The brackets denote ensemble average, which can be replaced by a spatial average
over angular positions $\vec \vartheta$.
With the assumption that the density field $\delta$ on large scales is
statistically homogeneous and isotropic,
which follows from the cosmological principle, the same holds for the convergence.
The 2PCF is then invariant under translation and rotation, and therefore a
function of only the modulus of the separation vector between the two lines of
sight $\theta$. Expressed in Fourier space, the two-point correlation
function
defines the convergence power spectrum $P_\kappa$ with
\begin{equation}
  \left \langle \fourier \kappa(\vec \ell) \fourier \kappa^\ast(\vec \ell^\prime) \right \rangle
  = (2\pi)^2 \delta_{\rm D}(\vec \ell - \vec \ell^\prime) P_\kappa(\ell).
  \label{eq:p_kappa_def}
\end{equation}

Here, $\delta_{\rm D}$ is the Dirac delta function. The
complex Fourier transform $\fourier \kappa$ of the convergence is a function
of the 2D wave vector $\vec \ell$, the Fourier-conjugate of $\vec \theta$.
Again due to statistical homogeneity and isotropy, the power spectrum only
depends on the modulus $\ell$. For simplicity, we ignore the curvature of the
sky in this expression. For lensing on very large scales, and for 3D lensing
(Sect.~\ref{sec:tomography}), the curvature has to be accounted for by more
accurate expressions \cite{2008PhRvD..78l3506L}, or by applying spherical
harmonics instead of Fourier transforms.

If the convergence field is decomposed into an E-mode $\kappa^{\rm E}$ and B-mode component
$\kappa^{\rm B}$, two expressions analogous to (\ref{eq:p_kappa_def}) define the E- and B-mode power spectra,
$P_\kappa^{\rm E}$ and $P_\kappa^{\rm B}$.

Taking the square of (\ref{eq:kappa_final}) in Fourier space, we get the power spectrum
of the density contrast, $P_\delta$, on the right-hand side of the equation. Inserting the 
result into (\ref{eq:p_kappa_def}) we obtain
the convergence power spectrum in terms of the density power spectrum as
\begin{equation}
  P_\kappa(\ell) = \frac 9 4 \, \Omega_{\rm m}^2 \left( \frac{H_0}{c} \right)^4
  \int_0^{\chi_{\rm lim}} {\rm d} \chi \,
  \frac{g^2(\chi)}{a^2(\chi)} P_\delta\left(k = \frac{\ell}{f_K(\chi)}, \chi \right).
  \label{eq:p_kappa_limber}
\end{equation}
This simple result can be derived using a few approximations:
the Limber projection is applied, which only collects modes that lie
in the plane of the sky, thereby neglecting correlations along the line of
sight
\cite{1953ApJ...117..134L,1992ApJ...388..272K,2007A&A...473..711S,2012MNRAS.422.2854G}.
In addition, the small-angle approximation (expanding to first order
trigonometric functions of the angle) and the flat-sky limit (replacing
spherical harmonics by Fourier transforms) are used. A further assumption 
is the absence of galaxy clustering, therefore ignoring
source-source \cite{2002A&A...389..729S}, and source-lens
\cite{1998A&A...338..375B,H02} clustering. Theoretical predictions for the
power spectrum are shown in Fig.~\ref{fig:Pkappa_dPkappa_dp}, using linear
theory, and the non-linear fitting formulae of \citet{2012ApJ...761..152T}. See
Sect.~\ref{sec:tomography} for the definition of the tomographic redshift bins.

The projection (\ref{eq:p_kappa_limber}) mixes different 3D $k$-modes into 2D
$\ell$ wavemodes along the line-of-sight integration, thereby washing out many
features present in the 3D density power spectrum. For example, baryonic
acoustic oscillations are smeared out and are not seen in the lensing spectrum
\cite{2006ApJ...647L..91S,2009NewA...14..507Z}. This reduces the sensitivity of
$P_\kappa$ with respect to cosmological parameters, for example compared to the
CMB anisotropy power spectrum. Examples for some parameters are shown in
Fig.~\ref{fig:Pkappa_dPkappa_dp}. Then two main response modes of $P_\kappa$
for changing parameters are an amplitude change, caused by $\sigma_8$,
$\Omega_{\rm m}$, and $w_0$, and a tilt, generated by $n_{\rm s}$, and $h$
(and, consequently, shifts are seen when varying the physical density
parameters $\omega_{\rm m}$ and $\omega_{\rm b}$). The parameter combination
that $P_\kappa$ is most sensitive to is $\sigma_8 \Omega_{\rm m}^\alpha$, with
$\alpha \approx 0.75$ in the linear regime \cite{1997A&A...322....1B}.

\stoptwocol
\begin{figure}

  \begin{center}
  \hspace*{1em}
    \resizebox{\hsize}{!}{
      \includegraphics[bb=40 10 500 232]{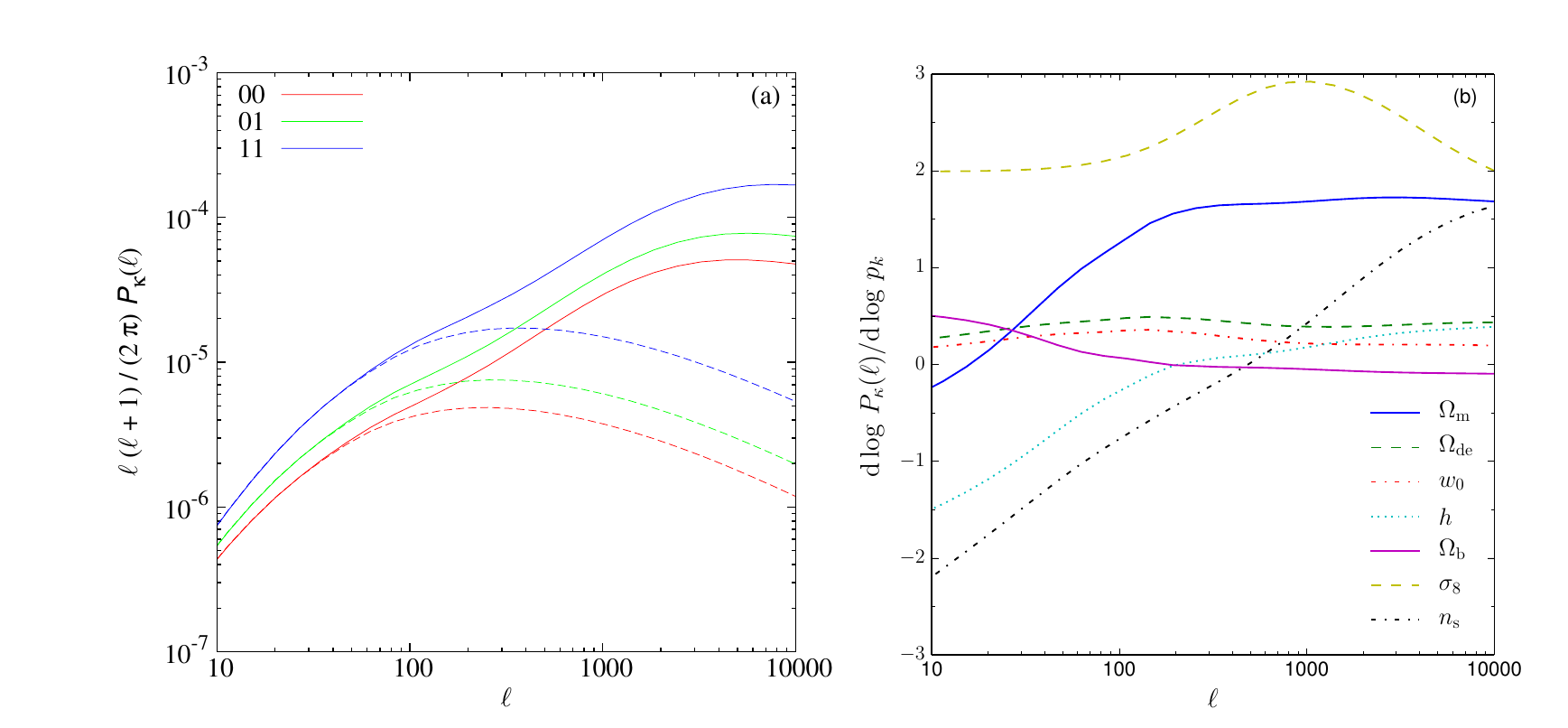}
    }%
  \end{center}

  \caption{(a) The scaled tomographic convergence auto- and cross-power
spectrum $\ell (\ell+1)/(2\pi) P_{\kappa, ij}(\ell)$ for two redshift bins $i, j$
  with redshift ranges $z = [0.5; 0.7]$, and $[0.9; 1.1]$, respectively. Solid (dashed)
  lines correspond to the non-linear (linear) model.
  (b) Derivatives ${\rm d} \log P_\kappa / {\rm d} \log p_k$ of the convergence
power spectrum with respect to various cosmological parameters $p_k$, as
indicated in the figure. The corresponding redshift
 bin is $[0.9; 1.1]$.
          }

  \label{fig:Pkappa_dPkappa_dp}

\end{figure}
\begtwocol

Writing the relations between $\kappa$, $\gamma$ and the lensing potential $\psi$
(\ref{eq:kappa_gamma_psi}) in Fourier space, and using complex notation for the shear,
one finds for $\ell \ne 0$
\begin{equation}
  \fourier \gamma(\vec \ell) = \frac{ \left(\ell_1 + {\rm i} \, \ell_2\right)^2}{\ell^2}
  \fourier \kappa(\vec \ell)
  = {\rm e}^{2{\rm i}\beta} \fourier \kappa(\vec \ell),
  \label{eq:gamma_kappa_Fourier}
\end{equation}
with $\beta$ being the polar angle of the wave-vector $\vec \ell = (\ell_1,
\ell_2)$, written as complex quantity. 
Therefore, we get the very useful fact that the power spectrum of the
shear equals the one of the convergence, $P_\gamma = P_\kappa$.

The shear power spectrum can in principle be obtained directly from observed
ellipticities \citeaffixed{2001ApJ...554...67H}{e.g.}, or via pixellised convergence
maps in Fourier space that have been reconstructed (see
Sect.~\ref{sec:mass_maps_formalism}) from the observed ellipticities, e.g.
\citet{1998ApJ...506...64S}. However, the simplest and most robust way to
estimate second-order shear correlations are in real space, which we will
discuss in the following section.

\subsection{The shear correlation function}
\label{sec:real_space_2nd}

The most basic, non-trivial cosmic shear observable is the real-space shear
two-point correlation function (2PCF), since it can be estimated by simply
multiplying the ellipticities of galaxy pairs and averaging.

The two shear components of each galaxy are conveniently decomposed into
\emph{tangential component}, $\gamma_{\rm t}$, and cross-component,
$\gamma_\times$. With respect to a given direction vector $\vec \theta$
whose polar angle is $\phi$,
they are defined as
\begin{equation}
  \gamma_{\rm t} = - \Re \left( \gamma \, {\rm e}^{-2{\rm i}\phi} \right); \quad
  \gamma_\times = - \Im \left( \gamma \, {\rm e}^{-2{\rm i}\phi} \right).
  \label{eq:gamma_tx}
\end{equation}
The minus sign, by convention, results in a positive value of $\gamma_{\rm t}$ for the tangential
alignment around a mass overdensity. Radial alignment around underdensities
have a negative $\gamma_{\rm t}$. A positive cross-component shear is rotated by
$+\pi/4$ with respect to the tangential component.

Three two-point correlators can be formed from the two shear components,
$\langle \gamma_{\rm t} \gamma_{\rm t} \rangle$, $\langle \gamma_\times
\gamma_\times \rangle$ and $\langle \gamma_{\rm t} \gamma_\times \rangle$.
The
latter vanishes in a parity-symmetric universe, where the shear field is
statistically invariant under a mirror transformation. Such a transformation
leaves $\gamma_{\rm t}$ invariant but
changes the sign of $\gamma_\times$.
The two non-zero two-point correlators are
combined into the two components of the shear 2PCF \cite{1991ApJ...380....1M},
\begin{eqnarray}
\eqalign
  \xi_+(\theta) 
  & = \langle \gamma \gamma^\ast \rangle(\theta)  &
  = \langle \gamma_{\rm t} \gamma_{\rm t} \rangle(\theta) + \langle \gamma_\times \gamma_\times \rangle(\theta); \quad
  \nonumber \\
  \xi_-(\theta)
  & = \Re \left[ \langle \gamma \gamma \rangle(\theta) {\rm e}^{-4{\rm i} \phi} \right] &
  = \langle \gamma_{\rm t} \gamma_{\rm t} \rangle(\theta) - \langle \gamma_\times \gamma_\times \rangle(\theta) . 
  \label{eq:xi_pm}
\end{eqnarray}
The two components are plotted in Fig.~\ref{fig:2pcf}. 
We note here that from the equality of the shear and convergence power spectrum and Parseval's theorem, it follows that
$\xi_+$ is identical to the two-point correlation function of $\kappa$.

An estimator of the 2PCF \cite{SvWKM02} is
\begin{equation}
  \hat \xi_\pm(\theta) = \frac{ \sum_{ij} w_i w_j \left( \varepsilon_{{\rm t}, i} \varepsilon_{{\rm t}, j} \pm
                                \varepsilon_{\times, i} \varepsilon_{\times, j} \right)}{ \sum_{ij} w_i w_j} .
  \label{eq:estim_xi_pm}
\end{equation}
The sum extends over pairs of galaxies ($i, j$) at positions on the sky $\vec
\vartheta_i$ and $\vec \vartheta_j$, respectively,
whose separation $|\vec \vartheta_i - \vec \vartheta_j|$ lies in an angular
distance bin around $\theta$. Each galaxy has a measured ellipticity $\varepsilon_i$, and an
attributed weight $w_i$, which may reflect the measurement uncertainty.
Using the weak-lensing relation (\ref{eq:eps_eps_s_gamma})
and taking the expectation value of (\ref{eq:estim_xi_pm}), we get terms of the following
type, exemplarily stated for $\xi_+$:
\begin{equation}
\langle \varepsilon^{(\rm s)}_{i} {\varepsilon_j^{(\rm s)}}^\ast \rangle;
\langle \varepsilon^{(\rm s)}_i \gamma_j^\ast \rangle;
\langle \gamma_i {\varepsilon_j^{(\rm s)}}^\ast \rangle;
 \quad \mbox{and} \quad \langle \gamma_i \gamma_j^\ast \rangle.
  \label{eq:eps_eps_four_terms}
\end{equation}
We discuss the first three terms in Sect.~\ref{sec:ia}, in the context of intrinsic alignment (IA).
In the absence of IA, those three terms vanish and the last term is equal to
$\xi_+(|\vec \vartheta_i - \vec \vartheta_j|))$. The analogous case holds for $\xi_-$.

The main advantage of the simple estimator (\ref{eq:estim_xi_pm}) is that it does not require
the knowledge of the mask geometry, but only whether a given galaxy is within
the masked area or not.
For that reason, many other second-order estimators that we discuss in the
following are based in this one.

Using (\ref{eq:p_kappa_def}) and (\ref{eq:gamma_kappa_Fourier}), we write the 2PCF as Hankel transforms of the
convergence power spectrum,
\begin{eqnarray}
  \xi_+(\theta) 
  &
  = \frac 1 {2\pi} \int {\rm d} \ell \, \ell {\rm J}_0(\ell
   \theta)
  [ P_\kappa^{\rm E}(\ell) + P_\kappa^{\rm B}(\ell)];
  \quad
  \nonumber \\
   \xi_-(\theta)
  &
  = \frac 1 {2\pi} \int
   {\rm d} \ell \, \ell {\rm J}_4(\ell \theta)
  [ P_\kappa^{\rm E}(\ell) - P_\kappa^{\rm B}(\ell) ].
   \label{eq:xi_pm_pkappa}
\end{eqnarray}
These expressions can be easily and quickly integrated numerically using fast
Hankel transforms \cite{2000MNRAS.312..257H}.

\doifonecol{%
\begin{figure}

  \begin{center}
    \resizebox{0.6\hsize}{!}{
     \includegraphics{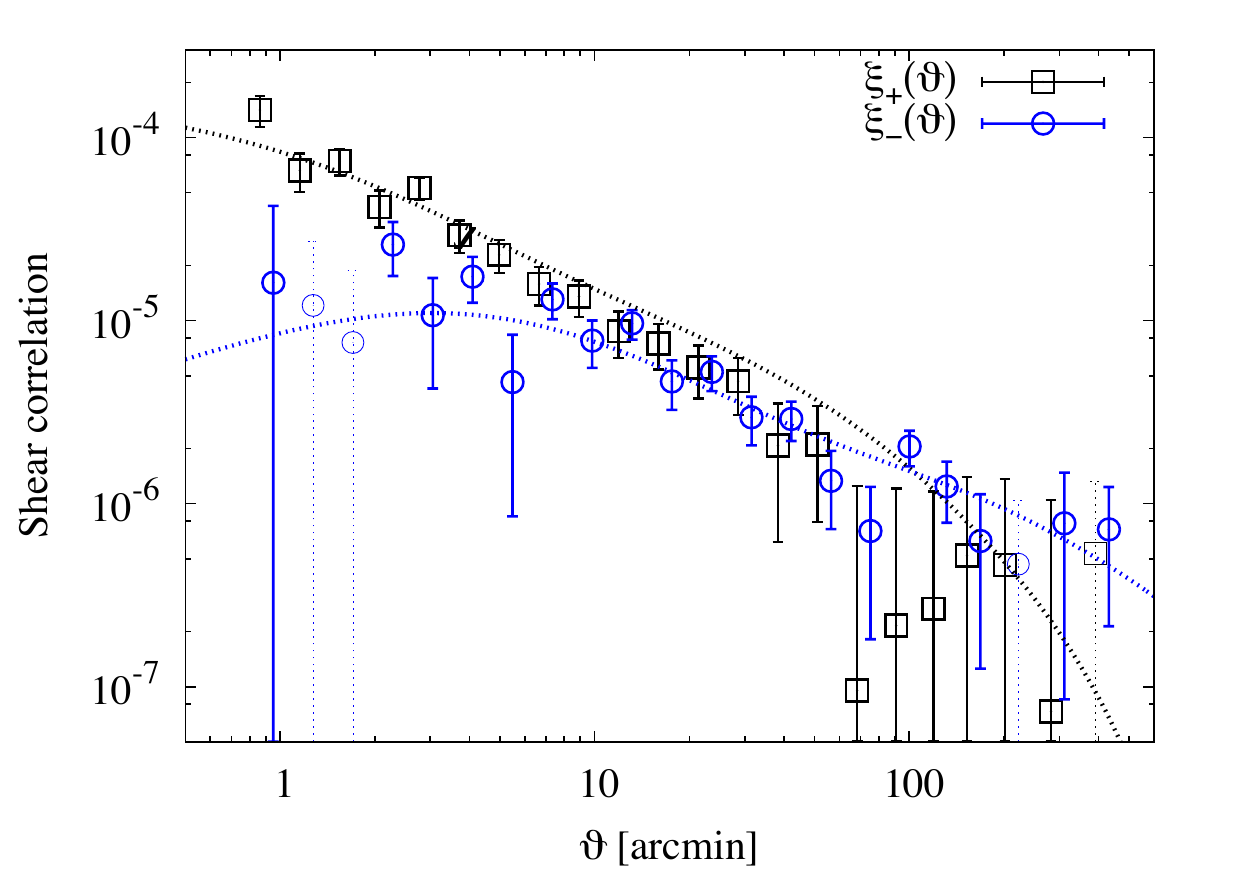}
    }
  \end{center}
}

\doiftwocol{%
\begin{figure}[H]

  \begin{center}
    \resizebox{\hsize}{!}{
      \includegraphics[bb=5 0 330 252]{figures/fig4pdf}
    }
  \end{center}
}

  \caption{2PCF components $\xi_+$ and $\xi_-$ (\ref{eq:xi_pm}) measured in
CFHTLenS. The dotted lines show the WMAP7 model prediction
\cite{2010arXiv1001.4538K}. From \citet{CFHTLenS-2pt-notomo}. Copyright 2013 Oxford University Press.}

  \label{fig:2pcf}

\end{figure}

The two 2PCF components mix E- and B-mode power spectra in two different ways.
To separate the two modes, a further filtering of the 2PCF is necessary, which
will be discussed in the following section.

\subsection{Derived second-order functions}
\label{sec:other_2nd_order}

Apart from the 2PCF (\ref{eq:xi_pm}), other, derived second-order functions
have been widely used to measure lensing correlations in past and present
cosmic shear surveys. The motivation for derived statistics are to construct
observables that (1) have high signal-to-noise for a given angular scale, (2)
show small correlations between different scales, and (3) separate into E- and
B-modes. In particular the latter property is of interest, since the B-mode can
be used to assess the level of (certain) systematics in the data as we have
seen in Sect.~\ref{sec:E_B_modes}.

All second-order functions can be written as filtered integrals over the
convergence power spectrum, and the corresponding filter functions define their
properties. In this section, we introduce the most widely used shear
second-order functions, and briefly discuss their properties, see also
\citet{CFHTLenS-2pt-notomo} for an overview.

The \emph{top-hat shear dispersion} is defined as the mean shear rms in an
aperture of radius $\theta$, $\langle | \gamma |^2 \rangle(\theta)$. The
signal-to-noise of this measure for a given $\theta$ is high, since it is a
broad low-pass band of $P_\kappa$, at the expense of exhibiting very strong
correlations over the whole range of angular scales. This function has been
used mainly in early cosmic-shear results, where the overall signal-to-noise
was low.

Another popular statistic is the \emph{aperture-mass dispersion}, denoted as
$\left\langle M_{\rm ap}^2 \right\rangle(\theta)$ (Fig.~\ref{fig:map2}). First,
one defines the \emph{aperture mass} as mean tangential shear with respect to
the centre $\vec \vartheta$ of a circular region, weighted by a filter function
$Q_\theta$ with characteristic scale $\theta$,
\begin{equation}
  M_{\rm ap}(\theta, \vec \vartheta)
  = \int {\rm d}^2 \vartheta^\prime \,
  Q_\theta(|\vec \vartheta - \vec \vartheta^\prime|) \,
  \gamma_{\rm t}(\vec \vartheta^\prime)
  = \int {\rm d}^2 \vartheta^\prime \,
  U_\theta(|\vec \vartheta - \vec \vartheta^\prime|) \,
  \kappa(\vec \vartheta^\prime).
  \label{eq:map}
\end{equation}
The second equality can be derived from the relations between shear and convergence,
which defines the filter function $U_\theta$ in terms of $Q_\theta$ \cite{KSFW94,S96}.  The
aperture mass is therefore closely related to the local projected over-density,
and owes its name to this fact.  The function $U_\theta$ is compensated
(i.e.~the integral over its support vanishes, $\int {\rm d}^2 \vartheta \, U_\theta(\vec \vartheta) = 0$),
and filters out a constant mass sheet $\kappa_0 =
\mbox{const}$, since the monopole mode ($\ell = 0$) is not recoverable from the
shear (\ref{eq:gamma_kappa_Fourier}).  Two choices for the functions
$U_\theta$, and consequently $Q_\theta$, have been widely used for cosmic
shear, a fourth-order polynomial \cite{1998MNRAS.296..873S}, and a Gaussian
function \cite{2002ApJ...568...20C}.

By projecting out the tangential component of the shear, $M_{\rm ap}$ is
sensitive to the E-mode only. One defines $M_\times$ by replacing $\gamma_{\rm
t}$ with $\gamma_\times$ in (\ref{eq:map}) as a probe of the B-mode only.  The
variance of (\ref{eq:map}) between different aperture centres defines the
dispersion $\left\langle M_{\rm ap}^2 \right\rangle(\theta)$, which
can be interpreted as fluctuations of lensing strength between lines of sight,
and therefore have an intuitive connection to fluctuations in the projected
density contrast.

\doifonecol{%
\begin{figure}

  \centerline{\resizebox{0.6\hsize}{!}{
    \includegraphics{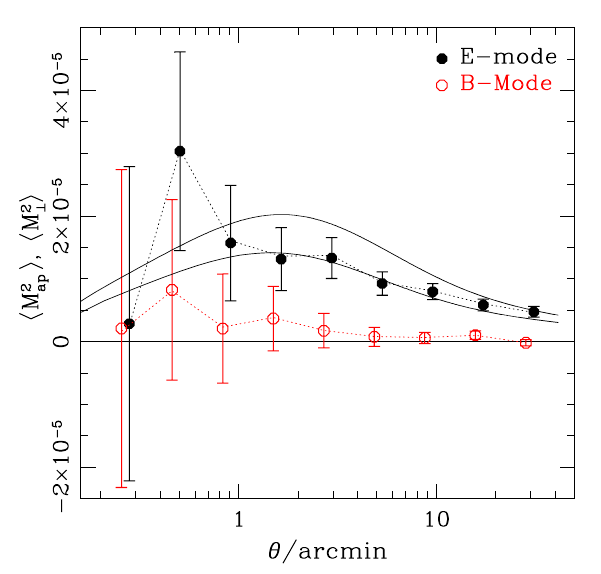}
  }}
}
\doiftwocol{%
\begin{figure}[H]

  \centerline{\resizebox{\hsize}{!}{
    \includegraphics[bb=10 0 162 160]{figures/SHJKS10-Map2.pdf}
  }}
}

  \caption{Aperture-mass dispersion measured in COSMOS. The two solid lines
correspond to predictions with $\sigma_8 = 0.7$ and $0.8$, respectively.
From \citet{SHJKS09}. Figure reproduced with permission from \citet{SHJKS09},
\emph{\aap}, \textbf{516}, A63. \copyright\ ESO.} 

  \label{fig:map2}

\end{figure}

The prospect of E-/B-mode separation motivated another derived shear
second-order functions, the \emph{E- and B-mode correlation functions} $\xi_+^{\rm E,
B}$ and $\xi_-^{\rm E, B}$ \cite{2002A&A...389..729S,2002ApJ...568...20C}.
Despite their name, they are also represented as integrals over the 2PCF.

Both top-hat shear rms and aperture-mass dispersion can in principle be estimated by
averaging over many aperture centres $\vec \vartheta$.
This is however not practical: The sky coverage of a galaxy survey is not
contiguous, but has gaps and holes due to masking. Apertures with overlap with
masked areas biases the result, and avoiding overlap results in a substantial
area loss. This is particularly problematic for filter functions whose support extend
beyond the scale $\theta$.
One possibility is to fill in the missing data, e.g.~with inpainting techniques
\cite{2009MNRAS.395.1265P}, resulting in a pixelised, contiguous convergence
map on which the convolution (\ref{eq:map}) can be calculated very efficiently
\cite{2012MNRAS.423.3405L}.
Alternatively, the dispersion measures can be expressed in
terms of the 2PCF, and are therefore based on the estimator
(\ref{eq:estim_xi_pm}) for which the mask geometry does not play a role.

In fact, every second-order statistic can be expressed as integrals over the
2PCF because, as mentioned above, all are functions of $P_\kappa$, and the
relation (\ref{eq:xi_pm_pkappa}) can be inverted. In general, they do not contain the
full information about the convergence power spectrum \cite{EKS08}, but
separate E- and B-modes.

The general expression for an E-/B-mode separating function $X_{\rm E, B}$ is
\begin{equation}
X_{\rm E, B} = \frac 1 {2\pi} \int_0^\infty {\rm d} \ell \, \ell \, P_\kappa^{\rm E, B}(\ell) \fourier U^2(\ell) .
\label{eq:X_EB_Fourier}
\end{equation}
A practical estimator using (\ref{eq:estim_xi_pm}) is 
\begin{equation} \hat X_{\rm E, B}
= \frac 1 2 \sum_i \vartheta_i \, \Delta \vartheta_i \left[ T_+\left(
{\vartheta_i} \right) \hat \xi_+(\vartheta_i) \pm T_-\left( {\vartheta_i}
\right) \hat \xi_-(\vartheta_i) \right] .
\label{eq:X_EB}
\end{equation}
Here, $\Delta \vartheta_i$ is the bin width, which can vary with $i$, for
example in the case of logarithmic bins.
The filter functions $T_\pm$ and $\fourier U^2$ are
Hankel-transform pairs, given by the integral relation
\cite{2002ApJ...568...20C,2002A&A...389..729S}
\begin{equation}
T_\pm(x) =
\int_0^\infty {\rm d} t \, t \, {\rm J}_{0,4}(x t) \fourier U^2(t).
\label{eq:T_pm}
\end{equation}
This implicit relation between $T_+$ and $T_-$ guarantees the separation into E- and B-modes
of the estimator (\ref{eq:X_EB}).

In some cases of $X_{\rm E, B}$, for example for the aperture mass and top-hat
shear dispersion, the power-spectrum filter $\fourier U$ is explicitely given as
the Fourier transform of a real-space filter function $U$, see
e.g.~(\ref{eq:map}) for the aperture mass.  In other cases
the functions $T_{\pm}$ are constructed first, and $\fourier U$ is
calculated by inverting the relation (\ref{eq:T_pm}). Model predictions of
$X_{\rm E}$ can be obtained from either (\ref{eq:X_EB_Fourier}),
or (\ref{eq:X_EB}). For the latter, one inserts a theoretical model for $\xi_{\pm}$,
and does not need to calculate $\fourier U$.

\subsubsection{E-/B-mode mixing}

None of the derived second-order functions introduced so far provide a pure
E-/B-mode separation. They suffer from a leakage between the modes, on small
scales, or large scales, or both. This mode mixing comes from the incomplete
information on the measured shear correlation: On very small scales, up to 10
arc seconds or so, galaxy images are blended, preventing accurate shape
measurements, and thus the shape correlation on those small scales is not
sampled. Large scales, at the order of degrees, are obviously only sampled up
to the survey size. This leakage can be mitigated by (i) extrapolating the
shear correlation to unobserved scales using a theoretical prediction (thereby
potentially biasing the result), or (ii) cutting off small and/or large scales
of the derived functions (thereby loosing information) \cite{KSE06}.

E-/B-mode mixing can be avoided altogether by defining derived second-order
statistics via suitable filter functions. For a pure E-/B-mode separation,
those filter functions need to vanish on scales where the shear correlation is
missing. Corresponding derived second-order quantities are the \emph{ring
statistics} \cite{SK07}, variations thereof \cite{2010A&A...510A...7E,FK10},
and the so-called COSEBIs \citeaffixed{COSEBIs}{Complete Orthogonal Sets of E-/B-mode integrals;},
see Fig.~\ref{fig:cosebis} for an example.
The latter quantities do not depend on a continuous angular
scale parameter $\theta$, but are a discrete set of modes
$E_n, B_n, n=1, 2 \ldots$ Typically, fewer than 10 COSEBI modes are sufficient
to capture all second-order E-mode information \cite{2012A&A...542A.122A}.

\doifonecol{%
\begin{figure}

  \centerline{\resizebox{0.6\hsize}{!}{
    \includegraphics{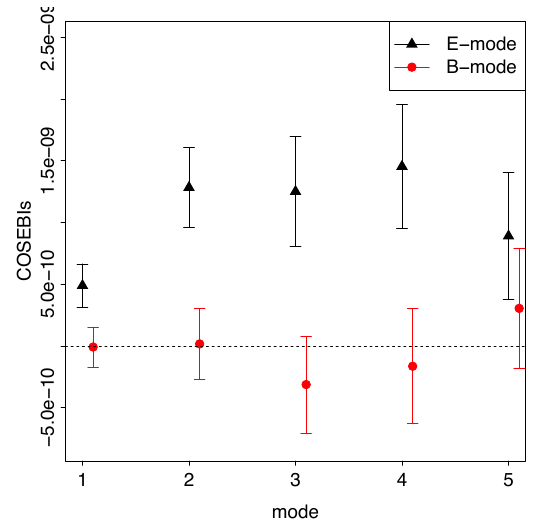}
  }}
}
\doiftwocol{%
\begin{figure}[H]

  \centerline{\resizebox{\hsize}{!}{
    \includegraphics[bb=10 0 148 160]{figures/Huff11-Cosebis.pdf}
  }}
}

  \caption{First five COSEBIs modes, measured in SDSS. From
\citet{2011arXiv1112.3143H}. Figure reproduced with permission from \citet{2011arXiv1112.3143H},
\emph{\mnras}, \textbf{440}, 1322. Copyright 2014 Oxford University Press.}

  \label{fig:cosebis}

\end{figure}

\subsection{Shear tomography and 3D lensing}
\label{sec:tomography}

The redshift distribution of source galaxies determines the redshift range over
which the density contrast is projected onto the 2D convergence and shear. By
separating source galaxies according to their redshift, we obtain lensing
fields with different redshift weighting via the lens efficiency (\ref{eq:lens_efficiency}),
thus probing different epochs in the history of the Universe with different weights.
Despite the two-dimensional aspect of
gravitational lensing, this allows us to recover a 3D
\emph{tomographic} view of the large-scale structure
In particular, it helps us to measure subtle
effects that are projected out in 2D lensing, such as the growth of structures,
or a time-varying dark-energy state parameter $w(z)$.

\subsubsection{Tomographic redshift binning}
\label{sec:tomo-binning}

If we denote the redshift distribution in each of $N_z$ bin with $p_i, i = 1
\ldots N_z$, we obtain a new lensing efficiency $q_i$
(\ref{eq:lens_efficiency}) for each case, and a resulting projected overdensity
$\kappa_i$. This leads to $N_z (N_z - 1) / 2$ convergence power spectra
$P_{\kappa, ij}, 1 \le i \le j \le N_z$, including not only the auto-spectra ($i=j$) but
also the cross-spectra ($i \ne j$). In (\ref{eq:p_kappa_limber}), $q^2$ is replaced by the product
$q_i q_j$ \cite{1998ApJ...506...64S,1999ApJ...522L..21H}.

The lensing efficiency is a very broad function in comoving distance, or
redshift, and therefore the tomographic power spectra are not independent:
Parts of the large-scale structure, despite it being weighted differently by
galaxies in different bins, contribute to all tomographic power spectra.
Moreover, photometric redshift errors cause galaxy distributions in different
redshift bins to overlap, increasing their correlation. Uncorrelated bins can
be obtained by linear re-weighting of the galaxy redshift distribution
\cite{HW05,2012MNRAS.423.3445S,2013arXiv1312.0430B}. These \emph{nulling}
techniques can also produce near-cancellations of power in certain regions of
$k$-space, for example to exclude highly non-linear, baryon-effect dominated
scales.

A relatively small number of redshift bins has been predicted to be sufficient
for a large improvement on standard cosmological parameters
\cite{1999ApJ...522L..21H,SKS03}. When more parameters are included that have
small effects on the growth rate, e.g.~dark energy and its time variation, the
number of necessary bins is higher. The photometric redshift uncertainty limits
the number of useful sub-divisions of the galaxies into bins: A large
dispersion washes out the binned redshift distribution resulting in information
loss \cite{2006ApJ...636...21M}. Catastrophic redshift outliers, galaxies whose
photometric redshift is many standard deviations away from its true redshift,
set even higher requirements on photometric redshifts, and on their calibration
from spectroscopic training samples
\cite{2006MNRAS.366..101H,2010MNRAS.401.1399B,2009ApJ...699..958S,2010ApJ...720.1351H}.
The presence of intrinsic alignment (see Sect.~\ref{sec:ia}) significantly adds to
the demands on photometric redshifts \cite{2007NJPh....9..444B}.

\subsubsection{3D lensing}
\label{sec:3d_lensing}

The technique of 3D lensing makes full use the redshift information of each
galaxy individually, instead of binning into redshift slices. 3D lensing
\cite{2003MNRAS.343.1327H,2005PhRvD..72b3516C,2011MNRAS.413.2923K,2011MNRAS.416.1629M}
samples the three-dimensional shear field by performing a spherical harmonics
transform of the field into $\ell$-modes on the sky, and $k$-modes along the
line of sight. In particular, it does not mix different $k$-modes by projecting
onto the 2D sky, and allows for a clear cut directly in $k$-space, e.g.~to
exclude small, non-linear scales from the analysis.

3D lensing uses the shear (in harmonics space) as signal, the expectation of
which is zero. The cosmological information is contained in the covariance,
which depends on the shear power spectrum. This is analogous to CMB, where the
temperature difference $\Delta T / T$ is the signal with zero mean, and the
information is extracted from the signal covariance, the angular power spectrum
$C_\ell$. The advantages of this approach are that only the Gaussian part of
the shear field is measured, involving statistics up to second order in
$\gamma$ (the covariance of $\gamma$), whereas for a traditional lensing
analysis, with second-order correlations as the observed signal, terms up to
order four in $\gamma$ have to be included for signal covariance. This allows
for a simple analytical calculation of the 3D lensing covariance without having
to resort to realisations of $N$-body simulations, the number of which needs to
be very large for an accurate estimation of the covariance
(Sect.~\ref{sec:cov_estim}).

To practically estimate the covariance from data, the mask geometry has to be
accounted for. Masked areas cause power to mix from scales affected by the mask
to other scales, contaminating the cosmological analysis. This is done by
either (i) applying the effect of the mask, expressed by a mixing matrix, to
the theoretical prediction, for example with the pseudo-$C_{\ell}$ method
\cite{2002ApJ...567....2H}, or (ii) by correcting the data for the mask,
requiring matrix inversion under a regularisation scheme. To successfully
account for the mask, the survey area needs to be relatively large; a 3D
lensing analysis of a survey of a few square degrees is very challenging. For
this reason, the first comprehensive cosmological analysis was performed
only recently, using about 120 square degrees of CFHTLenS \cite{CFHTLenS-3d}.

Unlike the case of the shear correlation function or power spectrum, the 3D
lensing observables in general include a non-zero B-mode, in addition to the
E-mode, due to mixing in presence of a mask and shear biases
\cite{CFHTLenS-3d}. In the absense of systematics (instrumental or
astrophysical), the B-mode is consistent with shot noise. This signal can be
predicted using a reference cosmology.

\subsection{Intrinsic alignment}
\label{sec:ia}

Shapes of galaxies can be correlated in the absence of gravitational lensing,
due to gravitational interactions between galaxies and the surrounding tidal
fields. The \emph{intrinsic alignment} (IA) of galaxy shapes adds an excess
correlation to the cosmic shear signal that, if not taken into account
properly, can bias cosmological inferences by tens of per cent. IA is difficult
to account for, since it cannot simply be removed by a sophisticated galaxy
selection, nor can it be easily predicted theoretically since it depends on
details of galaxy formation. First predictions of IA came about around the time
of the first detection of cosmic shear. These works used a wide range of
methods, including analytical calculations
\cite{2001ApJ...559..552C,2001MNRAS.320L...7C,2002MNRAS.332..788M} and $N$-body
simulations of dark matter, measuring alignments from dark matter halos only
\cite{2000ApJ...545..561C,2000MNRAS.319..649H}, or by populating halos with
galaxies using simple, semi-analytic prescriptions for their shape correlations
\cite{2000MNRAS.319..649H}. Unfortunately, different predictions do not agree
with each other, and the main difficulty remains to understand the alignment of
galaxies within their surrounding dark-matter structures. Galaxy formation can
strongly modify angular momentum and tidal stretching alignments, and erase
correlations that have been present in the dark matter. Only recently have
hydro-dynamical simulations attained a high enough resolution to form galaxies
with realistic morphologies, allowing for a direct measurement of galaxy
alignments
\cite{2010MNRAS.405..274H,2014arXiv1402.1165D,2014arXiv1406.4668C,2014MNRAS.441..470T}.

Two main mechanisms are believed to produce intrinsic shape correlations of galaxies that
are formed in the same tidal gravitational field: \mylistemph{First}, spin axes of (spiral)
galaxies become aligned during galaxy formation due to the exertion of a torquing
moment by the tidal field (tidal torquing). \mylistemph{Second}, halo shapes of (elliptical)
galaxies are coherently stretched by the tidal field. This can also include
anisotropic accretion e.g.~along filaments, \citet[see ]{2009IJMPD..18..173S} for a
comprehensive review of the theory of galactic angular momenta and their
correlations.

Due to IA, the intrinsic ellipticity of galaxies $\varepsilon^{\rm s}$ no
longer has a random orientation, or phase. This directly contributes to the
measured two-point shear correlation function (\ref{eq:estim_xi_pm}), as
follows. The first term in (\ref{eq:eps_eps_four_terms}) describes the
correlation of intrinsic ellipticities of two galaxies $i$ and $j$. This term
($II$, or shape-shape correlation) is non-zero only for physically close
galaxies. Its contribution to cosmic shear ($GG$, or shear-shear correlation),
the last term in (\ref{eq:eps_eps_four_terms}), can be suppressed by
down-weighting or omitting entirely galaxy pairs at the same redshift
\cite{HH03,KS02,KS03}.

The second and third term in (\ref{eq:eps_eps_four_terms}) correspond to the
correlation between the intrinsic ellipcitiy of one galaxy with the shear of
another galaxy. For either of these terms ($GI$, or shape-shear correlation) to be
non-zero, the foreground galaxy ellipticity has to be correlated via IA to
structures that shear a background galaxy. A lensing mass distribution causes
background galaxies to be aligned tangentially. Foreground galaxies at the same
redshift as the mass distribution are strechted radially towards the mass by
tidal forces. Therefore the ellipticities of background and foreground galaxies
tend to be orthogonal, corresponding to a negative $GI$ correlation. For
typical cosmic shear surveys with not too small redshift bins, $GI$ dominates
over $II$. It is therefore rather surprising that this term was discussed for
the first time only in 2004 \cite{2004PhRvD..70f3526H}.

Contrary to $II$, the $GI$ contribution is very hard to remove since galaxy
pairs at all line-of-sight separations are affected. To mitigate $GI$ in a
model-independent way, one can exploit the known dependence with source redshift
(which only involves angular distances). Weighted combinations of the measured shear
signal at different redshifts can be formed that provide a \emph{nulling} of
the $GI$ contribution. The disadvantage of this technique is a loss in
cosmological information, and a very strong requirement on photometric redshift
accuracy \cite{2008AA...488..829J,2009A&A...507..105J}. Alternatively, cosmic
shear data can be analysed by jointly modelling IA and cosmology
\cite{2005A&A...441...47K,2008arXiv0808.3400B,2010A&A...523A...1J}. This
however might introduce biases on cosmological parameters if a wrong or
restrictive IA model is chosen \cite{2012MNRAS.424.1647K}.
\citet[See ]{2014arXiv1407.6990T} for a review of IA in a weak-lensing context.

Additional observations of galaxy properties could help to infer their
intrinsic orientation, alleviating the contribution of intrinsic alignment, for
example from spectroscopic data \cite{2002ApJ...570L..51B,2013arXiv1311.1489H},
or polarization observations in the radio (see Sect.~\ref{sec:radio}).

\subsection{Galaxy-galaxy lensing}
\label{sec:ggl}

In contrast to cosmic shear, galaxy-galaxy lensing (GGL) correlates shapes of
high-redshift galaxies with positions of galaxies at lower redshift. The
resulting weak-lensing correlation singles out the mass associated with the
foreground galaxy sample. GGL has many applications in astrophysics and
cosmology, in particular when it is combined with other observations of 
properties of the foreground galaxy sample. GGL probes galaxy halos from
several kpc out to Mpc distances, providing insights about halo masses and
density profiles, e.g.~as function of stellar mass, luminosity, type, or
environment \cite{2006MNRAS.368..715M}.

GGL becomes of particular cosmological interest when complemented with
statistical measurements of other matter tracers, for example spatial galaxy
correlations (projected or in redshift space), cosmic shear, or galaxy velocity
correlations \cite{PhysRevD.81.023503,2012A&A...543A...2S}. The combination of those observables
allows for detailed and quasi-model-independent analyses of the relation
between luminous and dark matter, including the measurement of galaxy bias,
in particular its linearity, scale-dependence, and stochasticity. Since these
different correlations also probe different combinations of the Bardeen
potentials (\ref{eq:metric_gen}), they can test General Relativity.

GGL is usually quantified as the mean tangential shear $\langle \gamma_{\rm t} \rangle$
of background galaxies around foreground galaxies. It is a measure of the excess projected mass
within an aperture \cite{1991ApJ...370....1M,1996ApJ...473...65S}. The mean convergence in a
circular aperture of angular radius $\theta$ is
\begin{equation}
  \bar \kappa(\le \theta)
    = \frac{1}{\pi\theta^2} \int_{|\vec\theta^\prime| < \theta} {\rm d}^2 \theta^\prime \kappa(\vec \theta^\prime)
    = \frac{2}{\theta^2} \int_0^\theta {\rm d} \theta^\prime \, \theta^\prime \langle \kappa \rangle(\theta^\prime),
  \label{eq:kappa_bar}
\end{equation}
where the second equality defines the circularly averaged quantity
$\langle \kappa \rangle(\theta^\prime) = (2\pi)^{-1} \! \int_0^{2\pi} {\rm d} \varphi \,
\kappa(\theta^\prime, \varphi)$.
Since $2\kappa$ is equal to the divergence of the gradient of the lensing
potential $\psi$ (\ref{eq:kappa_gamma_psi}), one can apply Gauss' law and
eliminate the divergence. The resulting integral along the circle then depends
on the derivative of
$\psi$ in radial direction $\hat{\vec e}_\theta$, normal to the circle; the tangential part
along the aperture's circumference $\hat{\vec e}_\varphi$ is projected out.
Accounting for the circle length of $2\pi\theta$, one gets
\begin{equation}
  \bar \kappa(\le \theta) = \frac{1}{2\pi\theta} \int_0^{2\pi} {\rm d} \varphi \, \partial_\theta \psi(\theta, \varphi) .
\end{equation}
Calculating the derivative with respect to the radius $\theta$
introduces a second radial derivative, resulting in
\begin{equation}
  \frac{\partial [\theta \bar \kappa(\le \theta)]}{\partial \theta} =
    \int_0^{2\pi} \frac{{\rm d}\varphi}{2\pi} \,
  \partial_\theta \partial_\theta \psi(\theta, \varphi)
  = \left\langle \kappa \right\rangle (\theta) - \left\langle \gamma_{\rm t} \right\rangle (\theta) .
  \label{eq:dtk_dk_1}
\end{equation}
The second equality follows from (\ref{eq:kappa_gamma_psi}) for a local Cartesian coordinate system
$(\hat{\vec e}_\theta, \hat{\vec e}_\varphi)$.
As above, angle brackets denote circular averages.

On the other hand, calculating the derivative of (\ref{eq:kappa_bar}) yields
\begin{equation}
  \frac{\partial [\theta \bar \kappa(\le \theta)]}{\partial \theta} = - \bar \kappa(\theta) + 2 \langle \kappa \rangle(\theta) .
  \label{eq:dtk_dk_2}
\end{equation}
Equating the two expressions, we get
\begin{equation}
  \left\langle \gamma_{\rm t} \right\rangle (\theta)
  =  \bar \kappa (\le \theta) - \left\langle \kappa \right \rangle (\theta) .
  \label{eq:gamma_t_kappa}
\end{equation}
The mean tangential shear at radius $\theta$ is therefore a direct measure of
the total projected mass inside this radius, minus a boundary term. This result is very general
as it makes no assumption about the actual mass distribution.

We can express $\langle \gamma_{\rm t} \rangle$ in terms of a \emph{surface
mass excess} as follows. In the case of a single lens localised at angular
diameter distance $D_{\rm l}$, we can approximate (\ref{eq:kappa_chi}) with
\begin{equation}
  \kappa(\vec \theta) = \frac{4 \pi G}{c^2} \frac{D_{\rm l} D_{\rm ls}}{D_{\rm s}} \int\limits_{D_{\rm l} - \Delta D/2}^{D_{\rm l} + \Delta D/2}
  {\rm d} D \, \Delta \rho(D \vec \theta, D),
  \label{eq:kappa_single_lens}
\end{equation}
where $D_{\rm s}$ is the distance to the source, and $D_{\rm ls}$ the distance
between lens and source. All distances in this and subsequent equations are proper distance.
The integral over the lens mass density contrast
$\Delta \rho = \rho - \bar \rho = \bar \rho \delta$ is carried out along the physical
extent of the mass
concentration $\Delta D$. This integral is defined as \emph{surface mass density} $\Sigma$.
Introducing the
critical surface mass density $\Sigma_{\rm cr}^{-1} = (4\pi G/c^2) (D_{\rm l}D_{\rm ls}/D_{\rm s})$,
the convergence is simply
\begin{equation}
  \kappa(\vec \theta) = \frac {\Sigma(\vec \theta)} {\Sigma_{\rm cr}}. 
  \label{eq:kappa_Sigma}
\end{equation}
Then, (\ref{eq:gamma_t_kappa}) can be written in terms of the surface mass as
\begin{equation}
  \left \langle \gamma_{\rm t} \right\rangle (\theta) \, \Sigma_{\rm cr} = \Delta \Sigma(\le \theta) = \bar \Sigma(\theta) -
  \left \langle \Sigma \right \rangle (\theta) .
\end{equation}

If the convergence in (\ref{eq:kappa_single_lens}) is measured statistically with GGL, it is actually
a two-point correlation function, between background lensing and foreground galaxy over-density $\delta_{\rm g}$.
Thus,
\begin{eqnarray}
  \left\langle \kappa \right\rangle(\theta)
  &=& \left \langle \kappa(\vec \vartheta) \delta_{\rm g}(\vec \vartheta + \vec \theta) \right\rangle_{\vec \vartheta} \nonumber \\
  &=& \Sigma_{\rm cr}^{-1} \bar \rho \int {\rm D} \left\langle \delta(D \vec \theta, D) \delta_{\rm g}(D_{\rm l} \vec \theta, D_{\rm l})
    \right\rangle \nonumber\\
  &=& \Sigma_{\rm cr}^{-1} \bar \rho \int {\rm d} D \, \xi_{\delta \rm g} (\sqrt{(D \theta)^2 + (D - D_{\rm l})^2}) .
\end{eqnarray}
In the last step, the cross-correlation function between galaxy and matter
overdensity, $\xi_{\delta {\rm g}}$ was introduced. Alternatively, the
convergence can be expressed as a function of the Fourier transform of
$\xi_{\delta {\rm g}}$, the galaxy-matter power spectrum $P_{\rm \delta g}$.
The relation between $\xi_{\delta {\rm g}}$ and the matter correlation function
$\xi_{\delta \delta}$ depends on the model of the bias between galaxies and
matter. GGL thus offers a unique way to measure this relation.

The measurement of circularly averaged tangential shear
(\ref{eq:gamma_t_kappa}) makes it robust against anisotropic systematics,
e.g.~induced by PSF residuals, see Sect.~\ref{sec:PSF}. Further, diagnostic
null tests can easily be performed: First, the mean cross-component
$\gamma_\times$ around foreground galaxies violates parity and is therefore
expected to vanish. Second, the tangential shear should be zero around
random points, and around special points that are not associated with the foreground
sample, such as stars, field centres, or chip corners.

\subsection{The lensing bispectrum}
\label{sec:bispectrum}

The convergence power spectrum $P_\kappa$ (\ref{eq:p_kappa_limber}) only
captures the Gaussian component of the LSS. There is however substantial
complementary non-Gaussian information in the matter distribution, in particular on
small scales, where the non-linear evolution of structures
creates non-Gaussian weak-lensing correlations. On small and intermediate
scales, these non-linear structures are the dominant contributor to
non-Gaussian lensing signatures, compared to (quasi)-linear perturbations, or
potential primordial non-Gaussianity. Constraints on the latter from cosmic
shear alone can not compete with constraints from other probes such as CMB or
galaxy clustering \cite{2004MNRAS.348..897T,2011MNRAS.411..595P,2012MNRAS.426.2870H}.

To measure these non-Gaussian characteristics, one has to go beyond the second-order
convergence power spectrum. The next-leading order statistic is the bispectrum $B_\kappa$,
which is defined by the
following equation:
\begin{equation}
   \left\langle \fourier \kappa(\vec \ell_1) \fourier \kappa(\vec \ell_3) \fourier \kappa(\vec
      \ell_3) \right\rangle
      = (2 \pi)^2
  \delta_{\rm D}(\vec \ell_1 + \vec \ell_2 + \vec \ell_3)
   \left[ B_\kappa(\vec \ell_1, \vec \ell_2) +
    B_\kappa(\vec \ell_2, \vec \ell_3) + B_\kappa(\vec \ell_3, \vec \ell_1)
  \right].
  \label{eq:b_kappa}
\end{equation}
The bispectrum measures three-point correlations of the convergence defined on
a closed triangle in Fourier space. $B_\kappa$ can be related to the density
bispectrum $B_\delta$ via Limber's equation \cite{2001ApJ...548....7C}. Other
measures of non-Gaussianity are presented in Sect.~\ref{sec:peaks}.

The corresponding real-space weak-lensing observable is the shear three-point
correlation function (3PCF)
\cite{2003MNRAS.340..580T,tpcf1,2003ApJ...584..559Z,2006A&A...456..421B}.
Correlating the two-component shear of three galaxies sitting on the vertices
of a triangle, the 3PCF has $2^3 = 8$ components, and depends on three angular
scales. Those eight components can be combined into four complex \emph{natural}
components \cite{tpcf1,SKL05}.

A simple estimator of the 3PCF can be constructed analogous to
(\ref{eq:estim_xi_pm}), by summing up triplets of galaxy ellipticities at
binned triangles. The relations between the 3PCF and the bispectrum are
complex, and it is not straightforward to efficiently evaluate those
numerically. Most measurements and cosmological analyses of higher-order cosmic
shear have been obtained using the aperture-mass skewness $\langle M_{\rm ap}^3
\rangle$ \cite{2003ApJ...592..664P,JBJ04,SKL05}. $\langle M_{\rm ap}^3 \rangle$
is the skewness of (\ref{eq:map}), and can be written as pass-band filter over
the convergence bispectrum. Analogous to the second-order case, relations exist
to represent $\langle M_{\rm ap}^3 \rangle$ as integrals over the 3PCF,
facilitating the estimation from galaxy data without the need to know the mask
geometry. Corresponding filter functions have been found in case of the
Gaussian filter, but not the polynomial one \cite{JBJ04}. An extension
to filters with three different aperture scales, permitting to probe the
bispectrum for different $\ell_1, \ell_2, \ell_3$ has been obtained in
\citet{SKL05}. Pure E-/B-mode separating third-order functions have been
calculated \cite{2011A&A...533A..48S,2012MNRAS.423.3011K}, but a rigorous
treatment analogous to the second-order COSEBIs
(Sect.~\ref{sec:other_2nd_order}) are still lacking. This being said, the
leakage on small scales is less severe compared to the second-order case
\cite{2013arXiv1311.7035S}.

Convergence power- and bispectra show different dependencies on the geometry of
the universe, and on the growth of structures. This is true even for simple
models, where, inspired by perturbation theory, the bispectrum is given in
terms of products of the power spectrum. The combination of second- and
third-order statistics helps lifting parameter degeneracies, in particular the
one between $\Omega_{\rm m}$ and $\sigma_8$
\cite{1997A&A...322....1B,2004MNRAS.348..897T,KS05}. Combining weak-lensing
power- and bispectrum measures can also be used for self-calibration techniques
\cite{2006MNRAS.366..101H,2013MNRAS.434..148S}.

\subsection{Higher-order corrections}
\label{sec:corrections}

The approximations made in Sects.~\ref{sec:linear_lensing} and
\ref{sec:power_spectrum}, resulting in the convergence power spectrum, have to
be tested for their validity. Corrections to the linearised propagation
equation (\ref{eq:jacobi}) include couplings between lens structures at
different redshift (lens-lens coupling), and integration along the perturbed
ray (additional terms to the Born approximation). Further, higher-order
correlations of the convergence take account of the reduced shear as
observable. Similar terms arise from the fact that the observed size and
magnitudes of lensing galaxies are correlated with the foreground convergence
field \citeaffixed{2001MNRAS.326..326H,2009PhRvL.103e1301S}{magnification and
size bias; }. Over the relevant scale range ($\ell \le 10^4$) most of those
effects are at least two orders of magnitude smaller than the first-order
E-mode convergence power spectrum, and create a B-mode spectrum of similar low
amplitude. The largest contribution is the reduced-shear correction, which
attains nearly $10\%$ of the shear power spectrum on arc minute scales
\cite{1997A&A...322....1B,1998MNRAS.296..873S,2006PhRvD..73b3009D,2010A&A...523A..28K}. 

Thanks to the broad lensing kernel, the Limber approximation is very precise
and deviates from the full integration only on very large scales, for $\ell <
20$ \cite{2012MNRAS.422.2854G}. The full GR treatment of fluctuations together
with dropping the small-angle approximation was also found to make a difference
only on very large scales \cite{2010PhRvD..81h3002B}.

Additional corrections come from the clustering of galaxies, causing local
variations in the redshift distribution that are correlated with the density
and therefore with the lensing convergence. Both the self-clustering of source
galaxies \cite{2002A&A...389..729S} as well as associations between source
galaxies and lens structures \citeaffixed{2013arXiv1306.6151V}{source-lens
clustering;} cause sub-percent corrections for $\ell \le 10^4$.

Many of the above mentioned corrections are more important for higher-order
lensing statistics \cite{H02,2005PhRvD..72h3001D,2013arXiv1306.6151V}. This can
be seen by developing the density contrast into a perturbative series $\delta =
\delta^{(1)} + \delta^{(2)} + \ldots$, with $\delta^{(\nu)} \propto
[\delta^{(1)}]^\nu$. The first-order lensing power spectrum goes with the
square of the first-order density contrast, $P_\kappa \propto \langle
[\delta^{(1)}]^2 \rangle$. Lensing corrections typically add one order in
density, so corrections to $P_\kappa$ are proportional to $\langle
[\delta^{(1)}]^3 \rangle$, and are thus suppressed by one power of the density
contrast. However, the lowest third-order term $\langle [\delta^{(1)}]^3
\rangle$ vanishes since $\delta^{(1)}$ is assumed to be a Gaussian field. The
lensing bispectrum $B_\kappa$ in this perturbation theory approach is
proportional to $\langle \delta^{(2)} \delta^{(1)} \delta^{(1)} \rangle \propto
\langle [\delta^{(1)}]^4 \rangle$, which is of the same order as the
corrections $\propto \langle [\delta^{(1)}]^4 \rangle$ to $B_\kappa$.

The intrinsic alignment of galaxy orientations contribute to the lensing power
spectrum at up to 10\%. This is discussed in more detail in Sect.~\ref{sec:ia}.

\subsection{Weak-lensing mass maps}
\label{sec:mass_maps_formalism}

All information that can be extracted from weak-lensing distortions is contained in
the observed ellipticity of galaxies. In Sects.~\ref{sec:real_space_2nd},
\ref{sec:other_2nd_order} and \ref{sec:bispectrum}, we have seen how to extract
statistical information from the observed ellipcitites. In some cases however,
one wishes to estimate the convergence to create a local measure of the
projected overdensity.

The convergence can be obtained either indirectly by reconstruction from the
observed galaxy ellipticities, or directly from a measurement of the
magnification, see Sect.~\ref{sec:magnification}. The former method was first
proposed by \citet{1993ApJ...404..441K}: The relation between $\kappa$ and
$\gamma$ via the lensing potential $\psi$ (\ref{eq:kappa_gamma_psi}) can be
written as the integral
\begin{equation}
  \kappa(\vec \theta) = \frac 1 \pi \int {\rm d}^2 \theta^\prime \, {\cal D}^\ast(\vec \theta - \vec \theta^\prime)
  \, \gamma(\vec \theta^\prime) + \kappa_0.
  \label{eq:kappa_gamma_kernel}
\end{equation}
The kernel $\cal D$ is $\pi$ times the Fourier transform of the prefactor in
(\ref{eq:gamma_kappa_Fourier}). The convergence is therefore given as a 
convolution of the shear field, i.e.~there exist a linear relation between the two.
However, a few caveats complicate a simple
application of the above relation to obtain $\kappa$. \mylistemph{First}, replacing the
integral by a discrete sum over galaxies at measured positions results in
infinite noise, since the sampled uncorrelated intrinsic ellipticities are a white-noise
component, contributing a $1/\theta^2$-divergence to (\ref{eq:kappa_gamma_kernel}) \cite{1993ApJ...404..441K}.
Smoothing is therefore required, which results in a
decreased resolution and correlated noise, and requires accounting for
masks. \mylistemph{Second}, the integral (\ref{eq:kappa_gamma_kernel}) extends
over $\mathbb{R}^2$, and any reconstruction algorithm has to be modified to account
for finite observed fields, to avoid boundary artefacts in the reconstruction
\cite{1996A&A...305..383S}. \mylistemph{Third}, the convergence is obtained up to an
additive constant $\kappa_0$, corresponding to the $\ell = 0$ mode that is
undefined in (\ref{eq:gamma_kappa_Fourier}): A constant convergence does not
induce a shear, and is therefore unobservable without additional information such
as magnification to
lift this \emph{mass-sheet degeneracy}. Lastly, (\ref{eq:gamma_kappa_Fourier})
is only a first-order approximation, and becomes a non-linear relation if the
reduced shear $g = \gamma / (1 - \kappa)$ is used instead of the unobservable shear.

Despite of these difficulties, mass maps have several advantages over the shear
field. The convergence is a scalar quantity, and in some sense simpler than the
spinor shear, and also more directly related to (projected) mass. This
facilitates the cross-correlation with other maps of mass tracers, for example
galaxy overdensities, or the Sunyaev-Zel'dovich (SZ) effect. These measurements
help to understand properties of those tracers and their relation to the
underlying dark-matter environment. Further, higher-order correlations are
simpler using convergence maps; in fact, many reconstruction methods provide
the $\kappa$ field in Fourier space, allowing the easy and fast calculation of
higher-order spectra. Reconstructing the lensing potential provides a measure
of $\Psi + \Phi$ which in combination with other probes that for example
masured $\Phi$, can yield constraints on general relativity
(Sect.~\ref{sec:mod_grav}).

Alternative techniques have been proposed as well, for example non-linear
methods \cite{2009MNRAS.395.1265P}, or the addition of extra-information, such
as weak gravitational flexion \cite{2010ApJ...723.1507P} or the observed galaxy
distribution as prior tracer of the total matter
\cite{2013arXiv1306.5324S,2013A&A...560A..33S}. Reconstructions in three
dimensions are briefly discussed in Sect.~\ref{sec:mass_recon}.

Results and cosmological applications of mass reconstructions are presented in
Sect.~\ref{sec:mass_maps_results}.


\section{Numerical simulations}
\label{sec:simuls}

\subsection{The necessity of simulations for weak lensing}
\label{sec:simuls_necessity}

Numerical simulations play a central role in the cosmological interpretation of
cosmic shear data. Realistic simulations of large volumes of the large-scale structure are
necessary for weak cosmological lensing for a variety of reasons, as follows.

\mylistemph{First}, the scales on which cosmic shear probes the LSS extend deep
into the highly non-linear regime. To make analytical predictions of the power
spectrum on those scales is very difficult. $N$-body simulations offer
ways to obtain the non-linear power spectrum (i) by
establishing fitting formulae based on the linear power
\citeaffixed{PD94,PD96,2003MNRAS.341.1311S,2012ApJ...761..152T}{e.g.}; (ii) by directly
providing the non-linear power using templates
\cite{CoyoteII,CoyoteIII,2013arXiv1304.7849H}; or (iii) by calibrating
semi-analytical, non-linear models such as the halo model
\cite{2002PhR...372....1C,2002ApJ...574..538J,2008ApJ...688..709T,2010ApJ...724..878T},
or hierarchical clustering models
\cite{1992A&A...255....1B,2001MNRAS.322..107M}. Simulations are even more
important for higher-order statistics, such as the bispectrum or peak counts
(Sect.~\ref{sec:peaks}), for which theoretical predictions are much harder to be
obtained compared to the power spectrum.

\mylistemph{Second}, the density field $\delta$, and consequently the weak-lensing
convergence $\kappa$, is highly non-Gaussian on those non-linear scales.
The distribution of $\kappa$ is not easily assessed. Information about this
distribution can however be estimated from a sufficiently large number of
independent numerical simulations. Even complex survey properties that affect
this distribution, like an inhomogeneous depth or a complicated mask geometry,
can be easily included in the simulations. From the distribution of lensing
observables, their covariance can be estimated, which is of great importance for
cosmological error analyses. This is discussed in more detail in
Sect.~\ref{sec:cov_estim}.

\mylistemph{Third}, baryonic effects have to be taken into account with
increasing necessity for current and future surveys. Hydro-dynamical
simulations with various details of baryonic physics have been run to quantify
the influence on the total matter power spectrum, and to implement ways to
take baryons into account for cosmological predictions
\cite{2006ApJ...640L.119J,2008ApJ...672...19R,2011MNRAS.417.2020S}.

\mylistemph{Fourth}, $N$-body simulations are an important tool for the
analysis of potential astrophysical and observational systematics. This is in
particular true for observational effects that are intertwined with
astrophysics and cosmology in a complex way. One example is the recently
discovered correlation between the weak-lensing galaxy selection function and
the background density field: Close galaxy pairs whose images are blended are
deselected by most lensing pipelines, since their shapes are not easily
measured.
This leads to an underrepresentation of
high density regions and therefore to a bias if not corrected
\cite{2011A&A...528A..51H}.
A similar bias arises from massive foreground galaxies blocking the
line of sight to background galaxies.
Another example is the assessment of systematics,
for which the cosmological shear signal has to be modeled as well: The
distribution of a systematics measurement can be altered by the latter: For
example a chance alignments between LSS and PSF pattern can create a non-zero
shape correlation between stars and galaxies (see Sect.~\ref{sec:error-model}).
To assess data quality using such a systematic test, this distribution has to
be accounted for, to compare with the simulated, systematic-free case
\cite{CFHTLenS-sys}. Other examples are contributions to lensing that depend on
complex non-linear or baryonic physics, such as intrinsic galaxy alignment or
source clustering.

\mylistemph{Fifth}, some mathematical approximations of weak lensing can be
tested with simulations, such as linearisation of the propagation equation
\citeaffixed{2009A&A...499...31H}{Born approximation and lens-lens coupling;},
or the reduced shear \cite{2006PhRvD..73b3009D}, see also
Sect.~\ref{sec:corrections}.

\subsection{Principles of ray-shooting and ray-tracing}
\label{sec:ray-tracing}

The simulation of gravitational lensing by large-scale structures can be
performed by propagating light rays through the particle distribution of an
$N$-body simulation. The techniques of calculating the light propagation depend
on the nature and accessibility of the $N$-body
simulation. In the following, we will present some of these methods.

\subsubsection{Projecting the density field}

In many cases, an $N$-body simulation is available as snapshots corresponding
to different redshifts, in the form of boxes in comoving 3D cartesian
coordinates. The density in each box is projected onto a 2D ``lens plane'',
yielding the convergence $\kappa$ and subsequently, the lensing potential
$\psi$ (\ref{eq:lensing_potential}), the shear $\gamma$, and the deflection
angle $\vec \alpha$ for a light bundle within a light cone are derived. This
provides a discretization of the light propagation equations
\citeaffixed{1994CQGra..11.2345S}{Sect.~\ref{sec:light_propagation}, see also}.
To avoid the repeated encounter of light rays with the same object at different
snapshot times, boxes are usually rotated and translated randomly. This reduces
but does not eliminate spurious correlations of structure across different
redshifts, the importance of which depends on the simulation size. 

Potential, shear, and the deflection angle can be calculated from the
convergence on a grid via FFT, making use of the periodic boundary conditions
of the box, or in real space via finite differences \cite{2009A&A...499...31H}.
Usually a smoothing step has to be involved here due to the discreteness of the
simulated density field represented by point masses. Subsequently, lensing
deflections are obtained from the smoothed density field. Alternatively, one
can adapt the resolution according to local density, allowing lensing
quantities to be obtained directly from the simulated particles
\cite{2007MNRAS.376..113A}. However, in any case smoothing of the density field
is necessary to reduce Poisson noise, and to avoid singularities from
deflection by point masses. Various smoothing schemes and density estimators
have been proposed
\cite{2000A&A...363L..29S,2004A&A...423..797B,2011MNRAS.414.2235K,2013arXiv1309.1161A}.

In the simplest case, the lensing contributions are added up along straight
lines of sight, corresponding to the Born approximation of light deflection
along the unperturbed path
\citeaffixed{1970ApJ...159..357R,1988ApJ...330....1S}{Sect.~\ref{sec:linear_lensing};}.

\subsubsection{Tracing the photons}

A further refinement of the ``ray-shooting'' method introduced in the last
section is ``ray-tracing'', where light rays are followed to the next plane
along the deflected direction calculated on the current lens plane
\cite{1986ApJ...310..568B,JSW00}. This multiple-plane approach takes into
account non-linear couplings between lens planes and generates a
non-symmetrical Jacobi matrix. With respect to the first-order cosmic shear
convergence power spectrum $P_\kappa$, those higher-order corrections are very
small on relevant scales: The E-mode power spectrum due to lens-lens couplings
is four orders of magnitude smaller than $P_\kappa$. B-mode power is created at
about the same level \cite{2009A&A...499...31H,2010A&A...523A..28K}. The
non-symmetrical, rotational contribution is about three orders below $P_\kappa$
\cite{JSW00}.

In the multiple-plane approach there is no one-to-one and onto mapping between the
light cone of emitted rays at high redshift and the observer's
field of view. To guarantee that each photon reaches the observer, 
light rays are traced backwards from the observer to the emitting redshift, instead
of forward propagation towards the observer.

For accurately simulating galaxy-galaxy lensing (Sect.~\ref{sec:ggl}),
ray-shooting is insufficient, since weak-lensing-induced changes in the
positions of background galaxies with typical deflection angles of order
several arcmin. To preserve the correlation between lens galaxies and
foreground matter structures, tracing of the deflected photon ray positions is
necessary \cite{2009A&A...499...31H}.

\subsubsection{Approximations}

Several subtleties and further implicit approximations have to be mentioned
here. \mylistemph{First}, a snapshot is a fixed point in time, neglecting any evolution or
redshift-dependence of light deflection for the duration of light crossing the
box. Thus, the box size should be not more than about 300 Mpc.
For larger simulations, boxes can be split and projected onto more than one
lens plane. This however causes problems when cutting through halos, and leads to loss of
large-scale power. Alternatively, the field of view is generated under a skewed angle with respect to the box. By 
choosing an appropriate angle, repeated structures can be largely reduced, and 
lens planes with periodic boundaries can be constructed \cite{2009A&A...499...31H}.

\mylistemph{Second}, the lens planes are parallel to each other, which means that the
projection of matter neglects the sky curvature, and the fact that the gradient of the
lensing potential is necessarily taken in the lens plane and not orthogonal to
the light ray. Both approximations remain accurate if the light cone is small,
on the order of a few degrees.

\mylistemph{Third}, most $N$-body codes only simulate Newtonian physics.
Relativistic corrections to lensing observables due to General Relativity are
however very small \cite{2014arXiv1403.4947T}. To explore deviations
from GR, simulations in modified gravity models have been introduced recently
\citep[e.g.~]{2012JCAP...01..051L}.

\subsubsection{Further methods}

To circumvent lens-plane projections altogether, the lensing signal can be
computed on the fly during the $N$-body run. For the ray-shooting technique,
where light trajectories are known before-hand, the lensing quantities can be
calculated by approximating the integral (\ref{eq:Jacobi}) with a discrete sum
evaluated at the time steps of the simulation \cite{WH00,2011MNRAS.415..881L}.
With relative scale factor ratios of typically $\Delta a / a \approx 0.01$ -
$0.03$, the time resolution is of the same order if slightly higher than usual
separations of lens planes with $\Delta a / a \approx 0.03$ - $0.05$. The
implementation of this method is straightforward for ray-shooting, but becomes
more difficult for propagation along the perturbed light ray
\cite{2011MNRAS.415..881L}. An alternative method to allow for ray-tracing is
to store the density field at each time step on a surface that moves at the
speed of light toward an observer in the centre of the box
\cite{2009A&A...497..335T}.


\subsubsection{Full-sky lensing simulations}

Upcoming and future large cosmic shear surveys require simulations covering a
substantial part of the full sky. Such simulations are also used for CMB
lensing. This requires taking steps beyond the small-angle fields of view and the
plane-parallel approximation. Spherical density shells generated on the fly are
transformed into lens spheres instead of Cartesian planes
\cite{2008MNRAS.391..435F,2008ApJ...682....1D,2009A&A...497..335T,2013MNRAS.435..115B}.
Lensing quantities are calculated using spherical geometry.

\subsection{Dark matter and hydro-dynamical simulations}
\label{sec:dm_hydro_sims}

Dark matter interacts gravitationally, which for weak fields is a linear
problem. Pure dark-matter $N$-body simulations are essentially solutions for
Newton's equations of motion. Fast, parallel computers and massive memory and
storage space allows for very accurate, high-resolution simulations spanning a
wide range of scales and dynamic range. For an overview over various methods
and codes, some of which publicly available, see \citet{2008SSRv..134..229D}. 

Baryonic interactions however are much more complicated and not well known
in detail. Already the simplistic approximation of the baryonic content as an
ideal fluid introduces a set of non-linear equations, which further couple
baryons and dark matter gravitationally. Additional physics is necessary
for realistic simulations, consisting for example in radiative cooling, star
formation, supernova feedback, magnetic fields, black hole and AGN feedback,
and cosmic rays. Many of those processes are only poorly known and understood.

\subsection{Baryonic effects on lensing observables}
\label{sec:baryonic_effects}

Baryons alter the profile of halos compared to pure colissionless dark matter.
The changes to the power spectrum happen on small scales. Baryons play only a
very minor role on large scales ($k \lsim 1 \, h \, \mbox{Mpc}^{-1}$ or $\ell
\lsim 800$), since most of the dissipational physics takes place within
virialized halos \cite{2008ApJ...672...19R}. On intermediate scales, the gas
distribution is more diffuse compared to dark matter, due to pressure which
suppressed the formation of structure in the range of $k \approx 1$ to $10 \, h
\, \mbox{Mpc}^{-1}$. On the other hand, baryonic cooling and dissipation leads
to the condensation of baryons into stars and galaxies, increasing the density
in the inner halo regions, leading to a stronger clustering at very small
scales of $k \gsim 10 \, h \, \mbox{Mpc}^{-1}$
\cite{2006ApJ...640L.119J,2008ApJ...672...19R,2011MNRAS.417.2020S}.

Future surveys require the knowledge of the total, dark + baryonic matter power
spectrum at the $1 - 2$ percent level, with the highest required sensitivity
being at scales between $k = 0.1$ and $10 \, h \, \mbox{Mpc}^{-1}$, corresponding
to a few to a few tens of arcmin on the sky
\cite{2005APh....23..369H,2011MNRAS.418..536E}. The convergence power spectrum
is significantly altered at angular scales corresponding to $\ell \le 1000$ to $\ell \le
3000$, depending on the assumed statistical uncertainty. This result has also
been found using semi-analytical models \cite{2004APh....22..211W,2004ApJ...616L..75Z}.

Some past and present cosmological results have been obtained by leaving out
small, very non-linear scales. Attempts to model baryonic effects on the total
power spectrum have been made in the framework of the halo model. Modifications
to halo properties like the density profile or the concentration parameter have
been calibrated with numerical simulations
\cite{2008ApJ...672...19R,2011MNRAS.417.2020S,2014JCAP...08..028F}. An
alternative ansatz of mitigating the uncertainty of baryonic effects is
self-calibration using additional information on the internal halo structure
\cite{2008PhRvD..77d3507Z}, or on the lensing bispectrum
\cite{2013MNRAS.434..148S}.


\section{Cosmology from cosmic shear}
\label{sec:cosmo_from_shear}

This section gives a brief overview of the techniques to obtain constraints on cosmological
parameters from cosmic shear.

\subsection{Covariance estimation}
\label{sec:cov_estim}

The covariance matrix of weak-lensing observables is an essential ingredient
for cosmological analyses of cosmic shear data. Shear correlations at different
scales are not independent but correlated with each other: The cosmic shear
field is non-Gaussian, in particular on small scales, and different Fourier
modes become correlated from the non-linear evolution of the density field.
This mode-coupling leads to an information loss compared to the Gaussian case
(unless higher-order statistics are included). If not taken into account
properly, error bars on cosmological parameters will be underestimated.

Additionally, even in the Gaussian case Fourier modes are spread on a range of
angular scales in real space, causing shear functions to be
correlated across scales. The correlation strength depends on the filter
function relating the power spectrum to the real-space observable
(Sects.~\ref{sec:real_space_2nd}, \ref{sec:other_2nd_order}). The broader the
filter, the stronger is the mixing of scales, and the higher is the
correlation.

For an observed data vector $\vec d = \{ d_i \}, i=1 \ldots m$,
the covariance matrix $\mat C$ is defined as
\begin{equation}
  C_{ij} = \langle \Delta d_i \Delta d_j \rangle =
     \langle d_i d_j \rangle - \langle d_i \rangle \langle d_j \rangle ,
  \label{covariance}
\end{equation}
where the brackets denote ensemble average.

In a typical cosmic shear setting, the data vector $\vec d$ consists of functions of
shear correlations (e.g.~the shear two-point correlation function at $m$ angular
scales $\theta_i$, or band-estimates of the convergence power spectrum
$P_\kappa$ at $m$ Fourier wave bands with centres $\ell_i$). Those functions are quadratic in
the observed galaxy ellipticity $\varepsilon$. The covariance then depends on
fourth-order moments of $\varepsilon$. From (\ref{eq:eps_eps_s_gamma}), one can
see that the covariance can be split into three terms: The shot noise, which is
proportional to $\langle | \varepsilon^{\rm s} |^2 \rangle^2 = \sigma_\varepsilon^4$,
and, in the absence of intrinsic galaxy alignment (Sect.~\ref{sec:ia}), only
contributes to the covariance diagonal; the cosmic variance term, which depends
on fourth moments of the shear; and a mixed term.

In particular the cosmic variance term is difficult to estimate since it
requires the knowledge of the non-Gaussian properties of the shear field. Note
that in the case of 3D lensing (Sect.~\ref{sec:3d_lensing}), the vector $\vec d$
consists of the observed galaxy shapes (in harmonic space), and the covariance
is quadratic in the shear, and does therefore consist of a signal covariance,
which is proportional to the convergence power spectrum $P_\kappa$, and a noise
contribution $\propto \sigma_\varepsilon$.

\subsubsection{The Gaussian approximation}

The covariance of the convergence power spectrum $P_\kappa$ at an individual mode $\ell$
in the Gaussian approximation is the simple
expression \cite{1992ApJ...388..272K,1998ApJ...498...26K,2008A&A...477...43J}
\begin{equation}
  \langle (\Delta P_\kappa)^2 \rangle(\ell)
  = \frac{1}{f_{\rm sky} (2\ell + 1)} \left( \frac{\sigma_\varepsilon^2}{2\bar n} + P_\kappa(\ell) \right)^2 .
  \label{eq:cov_P_kappa}
\end{equation}
Here, the survey observes a fraction of sky $f_{\rm sky}$, with a number
density of lensing galaxies $\bar n$. 
The quadratic expression expands into shot-noise (first term),
cosmic variance (second term), and a mixed term. In this Gaussian
approximation, the fourth-order connected term of $\kappa$ is zero, and the cosmic variance
consists of products of terms second-order in $\kappa$.

Analytical expressions for the Gaussian covariance of real-space second-order
estimators have been obtained in \cite{SvWKM02,KS04,2009MNRAS.397..608S}. The
power-spectrum covariance for shear tomography is easily obtained
\cite{2004MNRAS.348..897T}.

\subsubsection{Non-Gaussian contributions}

\Eref{eq:cov_P_kappa} can be extended to the case of a non-Gaussian
convergence field. For example, terms of order four in $\kappa$ can be
parametrized as integrals over the trispectrum $T_\kappa$
\cite{1999ApJ...527....1S,2004MNRAS.348..897T}.

Non-Gaussian evolution leads to a further coupling of small-scale modes with
long wavelength modes that are larger than the observed survey volume. These
super-survey modes were first introduced as \emph{beat coupling} in
\cite{2006MNRAS.371.1188H}, and later modeled in the halo model framework as
\emph{halo sample variance} 
\citeaffixed{2009ApJ...701..945S,2013MNRAS.429..344K}{HSV;}. Contrary to the other terms of
the covariance that scale inversely with the
survey area $f_{\rm sky}$, the super-survey covariance decreases faster. Therefore
it is important for small survey areas \cite{2009ApJ...701..945S}. A
rigorous treatment of the non-Gaussian covariance including super-survey modes
is presented in \citet{2013PhRvD..87l3504T}.


The non-Gaussian cosmic variance has been fitted to $N$-body and ray-tracing
simulations, providing fitting functions in terms of the Gaussian cosmic
variance \cite{2007MNRAS.375L...6S,2011ApJ...734...76S}. Using the halo model,
the trispectrum contributions to the covariance can be computed 
\cite{2009ApJ...701..945S,2013PhRvD..87l3504T}, including expressions for
the bispectrum covariance \cite{2013MNRAS.429..344K,2013PhRvD..87l3538S}. 

An alternative, non-analytic path to estimate the covariance matrix is
replacing the ensemble average in (\ref{covariance}) by spatial averaging. This
requires a large enough number of independent or quasi-independent lines of
sight $n$, to have a fair representation of the LSS. These lines of sight can
be either numerical simulations or the observed survey itself. For the
resulting matrix to be non-singular, the dimension of the data vector $m$ has
to be smaller than $n$. The corresponding estimator is unbiased; however, the
inverse of this estimate, being a non-linear operation, is not. It is the
inverse covariance that is needed for the likelihood function (see following
section). The bias can be calculated and removed in the Gaussian case
\cite{andersen03}. This debiasing also works reasonably well for non-Gaussian
fields, and overestimates error bars by less than $5\%$ if $m/n$ is smaller
than about $0.1$ \cite{HSS07}. For current surveys with a few redshift bins and
reasonably small number of angular scales, resulting in a total $m \lsim 100$,
a few hundred realisations are sufficient. Future surveys, with many redshift
bins and multiple galaxy population subsamples, have $m$ of order a few
thousand. To reach percent-level precision, the number of realisations has to
be at least a few times $10^4$ \cite{2013MNRAS.tmp.1312T,2013PhRvD..88f3537D}.

Estimating the covariance from numerical simulations offers the additional advantage
that systematic effects are relatively easy to include, for which analytical expressions
are difficult to obtain. By populating simulations with galaxies and modeling their
properties, effects such as photo-$z$ errors, shape measurement biases, or intrinsic alignment
can be included into the covariance matrix.

A further complication is that super-survey modes may not be fully captured by
sample variance from numerical simulations, since the latter often lacks
large-scale modes due to small box sizes or periodic boundaries. Including
these modes requires special care in the simulation set-up
\cite{2014PhRvD..89h3519L}.

\subsection{The likelihood function}
\label{sec:likelihood}

To compare weak-lensing observations to theoretical predictions, one invokes a
likelihood function $L$ as the probability of the observed data $\vec d$ given
a model $M$ with a set of parameters $\vec p$ of dimension $q$.

For simplicity, in most cases, the likelihood function is modeled as an $m$-dimensional multi-variate
Gaussian distribution,
\begin{equation}
\fl
L(\vec d | \vec p, M) 
=
(2 \pi)^{-m/2} |\mat C(\vec p, M)|^{-1/2}
\exp\left[ - \frac 1 2 \left( \vec d - \vec
    y(\vec p, M) \right)^{\rm t} \mat{C}^{-1}(\vec p, M) \left( \vec d - \vec
    y(\vec p, M) \right) \right].
  \label{eq:likelihood}
\end{equation}
The function $\vec y$ is the model prediction for the data $\vec d$, and depends on the model
$M$ and parameter vector $\vec p$.
This is only an approximation to the true likelihood function, which is unknown, since
shear correlations are non-linear functions of
the shear field, which itself is not Gaussian, in particular on small scales.

The true likelihood function can be estimated by sampling the distribution
using a suite of $N$-body simulations for various cosmological parameters.
Because of the high computation time, this has been done only for a restricted
region in parameter space \cite{2009A&A...504..689H,2009A&A...505..969P,2011ApJ...742...15T}.

In contrast, analytical approaches might be promising to determine the true
likelihood function. Transforming the data to obtain more Gaussian
distributions involves Gaussianizing the convergence
\cite{2011MNRAS.418..145J,2011ApJ...729L..11S,2012PhRvD..86b3515Y}, or
transforming the correlation function
\cite{2009A&A...504..705S,2013A&A...556A..70W}. The so-called copula can be
used to reconstruct the multi-variate probability distribution function (pdf)
from one-dimensional pdfs \cite{2011PhRvD..83b3501S}. Further, a lognormal
distribution \cite{2002ApJ...571..638T,2011A&A...536A..85H} might be a better
approximation to the convergence field.


The log-likelihood function can be approximated by a quadratic form, which is the
inverse parameter covariance at the maximum point, called the \emph{Fisher
matrix} \cite{KS69,TTH97}. This approximation is exact in the case of a
Gaussian likelihood function \emph{and} Gaussian distributed parameters, but
does not account for non-linear parameter degeneracies, nor non-Gaussian tails of
the likelihood. The diagonal of the Fisher matrix inverse represents minimum
parameter variances. This is very useful to quickly generate predictions of
parameter constraints without the need of a time-consuming exploration of the
parameter space (see following section). The Fisher matrix has become a
standard tool to assess the performance of planned surveys,
or to explore the feasibility of constraining new cosmological models, e.g.~\cite{DETF}.
Marginalising over an arbitrary number of
nuisance parameters, and modeling parameter biases \cite{2005APh....23..369H}
are easily incorporated. However, one has to keep in mind that the Fisher
matrix is often ill-conditioned, in particular in the presence of strong
parameter degeneracies, and its inversion requires a very high precision 
calculating of theoretical cosmological quantities \cite{WKWG12}.

In most cases, the parameter-dependence of the covariance in
(\ref{eq:likelihood}) is neglected, since the compuation of the covariance is
very time-consuming, e.g.~when derived from $N$-body simulations. When
estimated from the data themselves, the cosmology-dependence of the covariance
is missing altogether. This is a good approximation, as was shown in
\citet{2009A&A...502..721E} and confirmed in \citet{CFHTLenS-2pt-notomo}, in
particular when only a small region in parameter space is relevant, for example
in the presence of prior information from other cosmological data.

\subsection{Parameter estimation}
\label{sec:param_estim}

Theoretical models of cosmic shear observables can depend on a large number of
parameters. Apart from cosmological parameters, a number of additional,
nuisance parameters might be included to characterize systematics, calibration
steps, astrophysical contaminants such as intrinsic alignment, photometric
redshift uncertainties, etc. The number of such additional parameters can get
very large very quickly and reach of the order a few hundred or even thousands, for example if
nuisance parameters are added for each redshift bin \cite{2008arXiv0808.3400B}.

When inferring parameter constraints within the framework of a given
cosmological model, one usually wants to estimate the probability of the
parameter vector $\vec p$ given the data $\vec d$ and model $M$. In a Bayesian
framework, this is the \emph{posterior} probability $\pi$, which is given via
Bayes' theorem as
\begin{equation}
  \pi(\vec p | \vec d, M) = \frac{L(\vec d | \vec p, M) P(\vec p | M)}{E(\vec d | M)},
\end{equation}
which links the
posterior to the likelihood function (see previous section) via the
\emph{prior} $P$ and the \emph{evidence} $E$. In most cases, one wants to
calculate integrals over the posterior, for example to obtain the mean
parameter vector, its variance, or confidence regions.
Such integrals can be written in general as
\begin{equation}
I(h) = \int {\rm d}^q p \, h(\vec p) \pi(\vec p | \vec d, M),
\label{eq:I_h}
\end{equation}
where $h$ is a function of the parameter $\vec p$. To calculate the mean of the
$\alpha^{\rm th}$ parameter, $I(h) = \bar p_\alpha$, $h(\vec p) =
p_\alpha$. For the variance of $p_\alpha$, set $h(\vec p) = (p_\alpha - \bar
p_\alpha)^2$. For a confidence region ${\cal C}$ (e.g.~the 68\% region
around the maximum) $h$ is the characteristic function $1_{\cal C}$ of the
set ${\cal C}$, that is $h(\vec p) = 1$ if $\vec p$ is in $\cal C$, and 0 else. Note that
this does not uniquely define ${\cal C}$; there are indeed many different ways
to define confidence regions.

In high dimensions, such integrals are most efficiently obtained by
means of Monte-Carlo integration, in which random points are sampled from the
posterior density function. Many different methods exist and have been applied
in astrophysics and cosmology, such as Monte-Carlo Markov Chain \citep[MCMC; ]{cosmomc},
Population Monte Carlo \cite{WK09,KWR10}, or Multi-nested sampling
\cite{2008MNRAS.384..449F}. Monte-Carlo sampling allows for very fast
marginalization, for example over nuisance parameters, and projection onto
lower dimensions, e.g.~to produce 1D and 2D marginal posterior constraints.
MCMC provides a chain of $N$ points $\vec p_j$, which under certain conditions represent
a sample from the posterior distribution $\pi$. Using this Markov chain, integrals of the form (\ref{eq:I_h})
can be estimated as sums over the $N$ sample points $\vec p_j$, 
\begin{equation}
\hat I(h) = \frac 1 N \sum_{j=1}^{N} h(\vec p_j).
\label{eq:I_h_MC}
\end{equation}
Other Monte-Carlo sampling techniques might provide samples under a different distribution, and (\ref{eq:I_h_MC})
has to be modified accordingly.

Alternatively, in a frequentist framework, one can minimize the function
$\chi^2 = -2 \ln L$. This implicitely assumes flat priors on all parameters.

Cosmic shear using current data is sensitive to only a few cosmological
parameters, in particular $\Omega_{\rm m}$ and $\sigma_8$. Shear tomography is
beginning to obtain interesting results on other parameters such as $\Omega_K$,
or $w$. For parameters that are not well constrained by the data, for example
$\Omega_{\rm b}$ or $h$, the (marginal) posterior is basically given by the
prior density. Therefore, the prior should be chosen wide enough to not 
restrict other parameters, and result in overly optimistic constraints.


\section{Measuring weak lensing}
\label{sec:measuring_wl}

\subsection{Data analysis methods for weak lensing}
\label{sec:data_analysis}

The cosmological interpretation of cosmic shear measurements requires the observation
of large and deep sky areas in superb
image quality, together with sophisticated image analysis methods.
\mylistemph{First}, the cosmological distortion induced on an individual galaxy is typically much smaller than
the galaxy's intrinsic ellipticity. A significant detection of cosmic shear requires a very large number of
galaxies to high redshifts and low signal-to-noise ratios, necessitating
very wide and deep images at high resolution. The shapes of those faint galaxies have to be measured with
high accuracy (Sect.~\ref{sec:shapes}).
\mylistemph{Second}, galaxy images are
corrupted by the \emph{point-spread function} (PSF). The PSF is the combined
effect of the imaging system consisting of the atmosphere (for ground-based
surveys), telescope optics, and detector. The anisotropic part of the PSF creates
spurious correlations of galaxy shapes which, if uncorrected, are 
typically larger than the
cosmological shear correlations. To estimate the PSF, a very pure sample of stars,
uncontaminated by small galaxies, has to be selected (Sect.~\ref{sec:PSF}).
\mylistemph{Third}, any method of galaxy shape measurement and PSF correction has to be
calibrated, to ensure that measurement biases are small enough compared to the
statistical errors. This can be achieved by using large sets of image simulations that
include
properties of the survey and the lensing galaxy population as realistically as
possible (Sect.~\ref{sec:image_sims}).
\mylistemph{Fourth}, the interpretation of measured shape correlations depends crucially on the
redshift distribution of the lensed galaxy sample, see (\ref{eq:kappa}). The large number of
lensing galaxies prohibits obtaining this distribution using spectroscopy.
Instead, multiple optical band observations are necessary to estimate
photometric redshifts (Sect.~\ref{sec:photo-z}). 
%

\subsection{Galaxy shape measurement}
\label{sec:shapes}

One of the greatest challenges of cosmological weak lensing is shape
measurement. High-redshift lensing galaxies, carrying the bulk of the
cosmological signal, are faint objects with typical $i$- or $r$-band magnitudes
of around $24$, signal-to-noise ratios down to $S/N \sim 10$, and sizes of the
order arcsec, thus extending over just a handful of pixels. Those galaxies are
convolved with a spatially and temporally varying PSF of similar size. It is
not required that an individual galaxy shape be measured with high precision;
what is important instead is an unbiased measurement for a sample of galaxies.
How small the residual bias can be depends on the survey size and depth, which
drives the expected precision of cosmological parameters to be measured.
Current surveys measure the sample galaxy ellipticity to about $1\%$ accuracy,
but the next generation of surveys needs to improve this by an order of
magnitude
\cite{2006MNRAS.366..101H,2007MNRAS.381.1018A,2009MNRAS.399.2107K,2010MNRAS.404..926A,2014MNRAS.tmp..157C}.

There are two main families of methods to measure the ellipticity $\vec
\epsilon$ of a galaxy image with light distribution $I(\vec \theta)$.
The \mylistemph{first} family directly estimates the ellipticity from the
data, for example by measuring second moments of $I$, or by decomposing $I$
into basis functions, and extracting the ellipticity from the corresponding
coefficients (Sect.~\ref{sec:shapes_direct}). These direct techniques are
generally more sensitive to the pixel noise compared to the fitting methods,
but they also require fewer assumptions
about $I$. However, no method is model- or assumption-\emph{free} in principle.
The \mylistemph{second} approach is to assume a model for the surface brightness
$I$, including ellipticity parameters, and to fit the model to the observed
image (Sect.~\ref{sec:shapes_model}). One of the advantages of these
\emph{forward-fitting} methods is the straightforward treatment of the PSF: The
model is easily convolved with the PSF before comparing it to the data.

For all approaches, there are complications coming from the fact that
weak-lensing galaxy images usually consist of multiple exposures. This
observation strategy increases the image depth, and helps to fill gaps between
the CCDs of modern multi-chip mosaic wide-field cameras. Two options are
possible to perform the shape measurement: on single exposures images, or on a
stacked image. The stacked image is obtained after a complex co-addition
procedure, which consists in astrometrically aligning the individual images
first, and interpolating pixel values of the individual frames to a new, common
pixel grid. Although the stacking approach offers the advantage of higher
signal-to-noise ratio per image, it has been shown for some shape measurement
methods to limit their accuracy: The interpolation of the individual frames
necessitates small image transformations to match the astrometry between
exposures. This produces distortions in the PSF, and correlated pixel noise.
Further, PSFs from different exposures usually have different shapes. Averaging
over those PSFs is far from optimal, and can lead to a highly complex stacked
PSF \cite{CFHTLenS-shapes}.

To avoid these difficulties with stacked images, there are various
other ways to combine the individual exposures to yield a single galaxy shape
esimate: shapes can be measured on individual images and then be averaged
\cite{2002AJ....123..583B}; the (Bayesian) posterior probabilities of model fits on each
image can be multiplied \cite{2007MNRAS.382..315M}; a common
model can be fitted jointly to all exposures. See Sect.~\ref{sec:shapes_model}
for more details on model-fitting shape methods.

Galaxy shapes are usually measured in one optical band, which is observed in
the best-seeing nights. Combining different bands does 
improve the measurement, although the correlation of measured galaxy shapes
from different wavelengths has to be accounted for \cite{2008JCAP...01..003J}.

\subsubsection{Direct estimation methods}
\label{sec:shapes_direct}

One can further sub-classify the first class of direct measurements techniques
into perturbative and non-perturbative approaches. For the perturbative
approach we find most notably KSB \nocite{1995ApJ...449..460K} (Kaiser, Squires
\& Broadhurst 1995), which measures ellipticity using weighted second moments
of the galaxy light distribution, and approximates the convolution of the image
with the PSF by linear operations on the ellipticities. Various significant
improvements of KSB have been achieved
\cite{1997ApJ...475...20L,1998ApJ...504..636H,2000ApJ...537..555K,2002AJ....123..583B},
including an extension to higher perturbation order \cite{2009ApJ...699..143O},
originally developed for the measurement of flexion
\citeprefixed{2007ApJ...660..995O}{, see also Sect.~\ref{sec:flexion}}.
Alternatively, instead of correcting the ellipticities as in KSB, the
PSF-deconvolution can be done directly on the moments of the galaxy light
distribution \cite{2000ApJ...536...79R}. This is more rigorously explored using
a truncated hierarchy of higher-order moments in \citet{2011MNRAS.412.1552M}.

The non-perturbative approach consists in assuming that the noise-free lensing
galaxy profile can be decomposed into a set of orthogonal eigenfunctions,
dubbed \emph{shapelets} in the original paper \cite{2003MNRAS.338...35R}. If
the PSF can also be represented by the same set, the convolution of the galaxy
with the PSF is a simple sum over the eigenfunctions, reducing the shape
measurement and PSF correction process in principle to a simple matrix inversion
problem. Several families of basis functions have been proposed
\cite{2002AJ....123..583B,2003MNRAS.338...35R,2003MNRAS.338...48R,2005MNRAS.363..197M,2006A&A...456..827K,2009MNRAS.396.1211N,2011MNRAS.417.2465A}.
In practice one has to truncate the infinite eigenfunction expansion. This
leads to biases, since the coefficients representing the ellipticity and the
ones that have been truncated are typically correlated, due to noise and the
presence of the PSF. Therefore, the measured ellipticity is biased compared to
the true ellipticity from the hypothetical case of an infinite expansion and
perfect representation of the galaxy \cite{2010MNRAS.406.2793B}. Additionally,
limiting the number of degrees of freedom is necessary to avoid the
over-fitting of noise features. All shapelet methods seem to share the tendency
of not well representing the galaxy light profile \cite{2010A&A...510A..75M}.
To date, none of these techniques have been used on optical data for shape
measurement in the framework of cosmic shear. However, shapelets have been
employed using radio data (see Sect.~\ref{sec:radio}), galaxy-galaxy flexion
\cite{2011MNRAS.412.2665V}, and for weak lensing by clusters
\cite{2008MNRAS.385..695B,2010A&A...514A..88R}. Further, the orthogonal
eigenfunctions in the \citet{2002AJ....123..583B} variant have been employed to
characterise and interpolate the PSF across the field in
\citet{2003AJ....125.1014J}.

\subsubsection{Model-fitting methods}
\label{sec:shapes_model}

The first model-dependent, forward-fitting technique was proposed in \citet{1999A&A...352..355K}.
However, only in the last five years or so has model-fitting become a widely used and very successful
approach. It can be mathematically designed to work with low
signal-to-noise images, allowing for measuring shapes from
individual exposures instead of the stacked image.

A fully Bayesian forward-fitting method is \emph{lens}fit, which measures the
posterior distribution of ellipticity for galaxies on individual exposures, and
combines the results in a Bayesian way without information loss
\cite{2007MNRAS.382..315M,2008MNRAS.390..149K,CFHTLenS-shapes}. Further
notable model-fitting methods are \emph{gfit} \cite{2012arXiv1211.4847G},
\emph{im2shape} and \emph{im3shape} \cite{2013MNRAS.434.1604Z}, 
and \emph{StackFit}, the method used for the
latest \survey{DLS} (\survey{Deep Lens Survey}) cosmic shear results
\cite{2006ApJ...643..128W,2012arXiv1210.2732J}.


Challenges for model-fitting methods are to find a good model and
parametrization to match the wide observed range of galaxy light distributions. A
non-suitable profile will result in a \emph{model bias}
\cite{2010MNRAS.404..458V}. The number of parameters is equally important: A
model with too few free parameters is not flexible enough and can also give rise to model
bias \cite{2010MNRAS.406.2793B}. Note that this case of under-fitting is
similar to the truncation of the expansion in eigenfunctions for
non-perturbative methods discussed above. On the other hand, too many parameters risk to fit
the noise and can cause a bias from over-fitting.

For any Bayesian shear measurement method, a prior distribution of galaxy
properties is required on input. This prior may originate for example from
deep, high signal-to-noise images, either from a deeper sub-part of the survey,
from external observations, or from simulations. This prior distribution is
difficult to obtain, since one needs to sample a large and multi-dimensional
space of galaxy properties.

\subsubsection{Further approaches}

The auto-correlation of light over the whole image, without the necessity of
detecting galaxies, was proposed as an alternative, model-independent approach
for weak cosmological lensing \cite{1997A&A...317..303V}. This technique has
however never been tested on large surveys, and it is further lacking a formal
procedure for accurate shape measurement.

Recently, two Bayesian inference methods were proposed. One approach is
hierarchical multi-level Bayesian inference \citep[\emph{MBI};
]{2014arXiv1411.2608S}, that constructs a joint posterior of shear, galaxy
properties, and the PSF (along with other nuisance parameters) given the pixel
data. The posteriors of individual galaxies are combined to infer population
parameters of galaxies and the PSF, which are marginalised over to obtain the
shear. The other Bayesian method is an approach developed by
\citet{2014MNRAS.438.1880B}, which does not need to attribute ellipticities to
individual galaxies. It estimates an approximation of the posterior probability
of shear given observed properties of a galaxy population. The prior
distribution of those properties has to be inferred, for example from deep
images.

Further, ellipticities can be inferred by comparing observed shape parameters
to a learning sample of galaxies, using some classification scheme. In
principle, any method can be used to measure shape parameters, and since it is
applied to both observations and the learning set, potential shear biases cancel
in principle \cite{2012A&A...544A...8T}.

These last two methods are limited by the dimensionality, resolution, and
completeness of the training sample.

\subsubsection{Shape measurement biases}
\label{sec:shape_biases}

One can make the very general statement that the non-linear dependence of
ellipticity estimators on the light distribution in the presence of noise
creates a bias, the so-called \emph{noise bias}. This bias has been
investigated for methods based on moments \cite{2004MNRAS.353..529H} and model
fitting \cite{2012MNRAS.425.1951R,2014MNRAS.441.2528K}. Some work has been done
to explore methods that are linear in the pixel light to reduce the bias, for
example using unnormalised shape estimators
\cite{2011MNRAS.414.1047Z,2013arXiv1309.7844V,2014MNRAS.438.1880B}. However,
since the intrinsic flux variation of galaxies is very large, the price to pay
for a small bias is a very large variance.

A further source of shape measurement biases are \emph{ellipticity gradients},
which occur if the ellipticity is a function of the scale where it is measured,
for example for galaxies with a bulge that is more circular than its disk,
see \citet{2010MNRAS.404..458V}, \citet{2010MNRAS.406.2793B}, and
Sect.~\ref{sec:PSF_colour_effects}.

Shape biases can be characterised to first approximation
by a multiplicative component $\vec m$, and and additive term $\vec c$. These bias parameters are given
by the relation between observed and true ellipticity
\cite{2006MNRAS.366..101H,STEP1},
\begin{equation}
  \varepsilon_i^{\rm obs} = (1 + m_i) \varepsilon_i^{\rm true} + c_i; \quad i = 1, 2.
  \label{eq:ell_bias}
\end{equation}
The shear biases $\vec m$ and $\vec c$ are generally functions of galaxy properties and
redshift. Current shape measurement methods provide shear estimates with $m$ at
the $1$ to $10$ percent level, and $c$ between $10^{-3}$ and $10^{-2}$.
Typically, the measured shapes are corrected for those biases
using calibration image simulations (Sect.~\ref{sec:image_sims}).
Possible additional ways to mitigate shear biases  is their self-calibration using the weak-lensing
data themselves \cite{2006MNRAS.366..101H,2010ApJ...720.1090Z}.
Future surveys require the accuracy of calibrated shapes to be on the order of $0.1$\%
\cite{2006MNRAS.366..101H,2013MNRAS.429..661M,2013MNRAS.431.3103C}.




\subsection{PSF correction}
\label{sec:PSF}

The PSF is the response of the image system to a point source. 
Since stars are unresolved objects, they are
the local (noisy) representation of the PSF, and can be used to correct
galaxy images for the PSF.

The intrinsic size of the PSF from the optical system together with atmospheric
turbulence (\emph{seeing}) circularizes galaxy images, resulting in a decrease
of shape correlations. Ground-based sites with excellent conditions have seeing
disk of the order $0.5 - 0.7$ arcsec in the optical. A seeing of larger than an
arcsec or so dramatically reduces the depth and usability of cosmic shear data.
PSF anisotropies originate from the atmosphere, optical
aberrations, mirror deformations, tracking errors, CCD non-flatness and
misalignments on the
focal plane, and pixelation.
\cite{2008arXiv0810.0027J,2013PASJ...65..104H,2013MNRAS.428.2695C}.
The typical PSF anisotropy in ground-based images is up to 10\%. Therefore, for
a sub-percent measurement of a 1\% cosmic shear, the PSF induced on galaxy
images has to be corrected with a precision of one part in 100.

\subsubsection{The PSF model}
\label{PSF_model}

To estimate the PSF at the position of a galaxy, one has to select stars on the
image, measure their shape, and interpolate the resulting PSF to the position
of the galaxy. \mylistemph{First}, this requires a sample of suitable stars, i.e.~without
saturated pixels, not hit by cosmic rays, and uncontaminated by galaxies. A
common selection criteria is the identification of the \emph{stellar locus} in
a size-magnitude diagram. This is a region of bright and small objects that is
relatively well isolated from resolved galaxies and unresolved, dim objects such
as very faint galaxies and detection artefacts. Additionally, colour information
can be added to classify stars and galaxies.

\mylistemph{Second}, the shape of stars is measured and, in the simplest case, their
ellipticity parameters are extracted. To increase the precision of the PSF correction,
more parameters of the stellar shape have to be included. These can be
higher-order moments, or a pixellised postage stamp around each star, or
coefficients of a decomposition into a set of basis functions. 


\mylistemph{Third}, those parameters are interpolated onto the galaxy position. For
ground-based observations in the past, this has usually been done with a two-dimensional
polynomial or a rational function, and
the choice of the interpolating function can affect the measured cosmic
shear signal \cite{2004MNRAS.347.1337H}. A variety of interpolating schemes that
differ from simple fitting of a smooth function have been proposed
\cite{Kitching:2012fj,2012MNRAS.419.2356B,2013A&A...549A...1G}.
Alternatively, a physical model of the optics and its aberrations can be used
\cite{2008arXiv0810.0027J}. For mosaic multi-CCD cameras, discontinuities
between chips are common and have to be accounted for in the PSF model, for
example by performing fits on each chip individually \cite{CFHTLenS-shapes}.
By correlating the observed PSF to the PSF model residuals, cases of over-
and under-fitting can be diagnosed \cite{2010MNRAS.404..350R}.

For space-based observations the number of stars per field is very low, making
it difficult to fit an interpolating function over the currently available small
fields of view. However, due to the lack of the stochastic atmospheric
contribution to the PSF, and the high stability of the optical system, dense
stellar fields can be used to construct a model of the PSF, e.g.~using
a principal component analysis \citeaffixed{2004astro.ph.12234J,SHJKS09}{PCA; }. A PCA
model of the PSF on individual images is also a viable option for lensing
measurements on stacked images, where stacking can create PSF discontinuities
\cite{2012arXiv1210.2732J}.

\subsubsection{Colour effects}
\label{sec:PSF_colour_effects}

For a general image system, the PSF varies with wavelength $\lambda$. For
example for a diffraction-limited telescope, the PSF size is directly
proportional to $\lambda$. Further, atmospheric refraction is chromatic. Stars
have different spectral energy distributions (SED) than galaxies; in general
they are Milky Way disk stars and therefore bluer than high-redshift galaxies.
The PSF that is used to correct galaxy images is therefore not exactly the one
with which these images are convolved, introducing a bias. This bias
depends on the SED variation within the filter, which is the larger the broader
the filter is. In particular for the planned satellite mission \emph{Euclid} (Sect.~\ref{sec:upcoming_surveys})
for which 
the wide optical transmission spans the optical filter range $R+I+Z$, this
colour-dependence has to be taken into account, for example by adding
information from other wavelengths \cite{2010MNRAS.405..494C,2012PASP..124.1113A,2014arXiv1409.6273M}.

An additional complication arises because in general, the SED of a galaxy
varies spatially, and the projected 2D shape of a galaxy depends on the scale
where it is measured. For example, a bulge+disk galaxy often has a spherical,
red bulge, and a more elliptical, blue disk. The measured shape involves some
(radial) weight function of the galaxy's light distribution, the colour of which
in general does not correspond to the SED-weighted PSF. Thus, even if the PSF
for all wavelengths was known, the galaxy is corrected with the `wrong' PSF,
introducing a bias in the inferred shape. To account for this \emph{colour
gradient} and the associated bias, one can, as in the case of the
wavelength-dependent PSF, reduce the filter width, observe the sky in
additional filters, or use an external galaxy calibration set of sufficient
size \cite{2012MNRAS.421.1385V,2013MNRAS.432.2385S}

\subsection{Image simulations}
\label{sec:image_sims}

As we have discussed in Sect.~\ref{sec:shape_biases}, virtually all
shape measurement methods suffer from measurement biases and require a
large and representative sample of observed or simulated galaxies to calibrate
these biases. Such a calibration set is also needed for methods that are
constructed to provide (near-)unbiased estimates of shear. For example,
Bayesian methods, use it to estimate the prior distribution of galaxy
properties. An inappropriate prior will result in a biased shear estimate.

Such a calibration sample can be obtained from the
deep part of a survey with significantly longer integrated
exposure time than the main survey. The sky coverage of such very deep
observations is however very limited; in cases like the \survey{Canada-France
Hawaii Legacy Survey} (\survey{CFHTLS}, see Sect.~\ref{sec:early_era}) or
\survey{Euclid} the deep part covers only on the order of a percent of the main
survey.

An alternative calibration approach is to use image simulations, of which a
huge data volume can be created. These simulations need to have realistic survey
characteristics, for example concerning the noise. Additionally, the
distribution of galaxy properties need to be well covered, since shape biases
depend on galaxy size, signal-to-noise ratio, ellipticity, type, etc., and the
calibration needs to be done as a function of these properties.

Image simulations have been created as collaborative projects within the
weak-lensing community, such as the Shear TEsting Project (STEP) with the two
consecutive blind tests STEP1 \cite{STEP1} and STEP2 \cite{STEP2}. Public
challenges like the GRavitational lEnsing Accuracy Testing (GREAT) projects
have been launched to reach out to a larger community, in particular computer
science, to invite more ideas to tackle the problem of galaxy shape
measurement. This contains GREAT08
\cite{2009AnApS...3....6B,2010MNRAS.405.2044B}, GREAT10
\cite{2010arXiv1009.0779K,2012MNRAS.423.3163K,Kitching:2012fj}, and GREAT3
\cite{2013arXiv1308.4982M}, as well as two ``Kaggle'' crowdsourcing challenges
\cite{2012arXiv1204.4096K,2013arXiv1311.0704H}.

Those collaborative image simulation projects typically started under simple,
well-controlled conditions, for example, a constant PSF, constant shear over
the field, and analytical galaxy light distributions with high signal-to-noise.
They then progressed to more complex and more realistic images, for example
galaxy images based on observed HST deep fields. The purpose of those
simulations is to test estimates of shear with amplitudes of a few percent
to an accuracy at also the percent level. This is typically quantified in
terms of multiplicative and additive bias (\ref{eq:ell_bias}). The number
of simulated images is necessarily very large, producing hundreds of gigabytes
of data.

After the end of those projects, the true input parameters, and in some cases
the codes to generate the simulations \citep[e.g.~the galaxy image simulation
toolkit \texttt{GalSim}; ]{GalSim14}, were released to the public. This policy
has been proven to be of great value for the weak-lensing community, as those
simulations and results have been used extensively in subsequent work, to
better understand the performance of existing methods, and to scrutinize and
calibrate new shape measurement techniques.

\subsection{Photometric redshifts}
\label{sec:photo-z}

Weak lensing observables, being integrals along the line of sight weighted by
the source galaxy distribution $n(z)$ (\ref{eq:lens_efficiency}), require knowledge of
the latter if they are to be interpreted cosmologically. To first order, the mean redshift $\bar z$
has to be determined, but also the shape of $n(z)$ plays an important role.
The sensitivity of the cosmic
shear power spectrum (\ref{eq:p_kappa_limber}) to $\bar z$ is comparable to
its sensitivity with respect to cosmological parameters
\cite{1997ApJ...484..560J,1997A&A...322....1B}. For example,
\citet{2006MNRAS.366..101H} find a rough estimate of $P_\kappa(\ell \sim 1000)
\propto \Omega_{\rm de}^{-3.5} \sigma_8^{2.9} {\bar z}^{1.6} |w|^{0.31}$.
Clearly, for a desired accuracy on cosmological parameters, the mean redshift
of sources has to be known to at least that accuracy, and to a much higher
accuracy in the case of parameter on which the power spectrum has a weaker
dependence such as $w$.
For tomography, this is true for each individual bin, further exacerbating the
demands on photometric redshifts. For example, the centroids of each bin have to
be known to better than a per cent in order to limit the decrease in accuracy of
dark-energy parameters to $50\%$ \cite{2006MNRAS.366..101H}.

Spectroscopy of all the faint galaxies used for a typical weak-lensing survey is too costly,
and redshifts have to be estimated from broad-band photometry, using the technique of
\emph{photometric redshifts}, or photo-$z$s.
There are various methods to measure photometric redshifts. Template-based
approaches perform $\chi^2$-type fits of (redshifted) template SEDs to the flux
in the observed bands. Exemplary methods that have been used in a weak-lensing
context include \emph{LePhare} \cite{2006A&A...457..841I}, \emph{Bayesian
Photometric redshift estimation} \citeaffixed{2000ApJ...536..571B}{BPZ; }, and
\emph{HyperZ} \cite{2000A&A...363..476B}. In contrast, varous empirical
approaches exist that typically require a training set of galaxies with
spectroscopic redshifts, such as \emph{ANNz} \cite{2004PASP..116..345C}.
Other methods do not attempt to measure individual redshifts, but apply
statistical tools to the ensemble of observations, for example clustering in
colour space \cite{2007JCAP...03..013J,2008MNRAS.390..118L}.

Most photo-$z$ methods not only yield an estimate $\hat z$ of the redshift, but
provide more information about the distribution, for example error bars, or a
goodness-of-fit parameter. Ideally, they estimate the full probability
distribution function. Additional outputs may be the galaxy type, and the
probability of the object being a star.

The dispersion of current photometric redshifts is of the order $\sigma_z /
(1+z) = 0.03$ -- $0.06$ for typical multi-band optical surveys. The rate of
catastrophic outliers --- galaxies whose estimated redshift is off from the
true (spectroscopic) redshift by more than a couple of standard deviations ---
is between a few to a few tens of percent. All redshift estimates strongly
depend on galaxy type: Elliptical galaxies usually have much smaller
uncertainties than blue and irregular types. 

The currently-reached amplitude of dispersion $\sigma_z$ is sufficient for
future surveys. However, they require $\sigma_z$ to be known to sub-percent
accuracy \cite{2006ApJ...636...21M}. Catastrophic outliers can strongly bias
tomographic shear power spectra, and their rate has to be lower than a percent
\cite{2010MNRAS.401.1399B,2010ApJ...720.1351H}. 
This rate is applied to the galaxy sample after possible rejection of likely outliers
by the photometric redshift code.
The required precision of
photometric redshifts necessitates a very large spectroscopic calibration set.
For current surveys, this number is on the order of ten thousand, and a
magnitude larger for future surveys \cite{2006ApJ...636...21M}. The
spectroscopic survey has to be a representative sample of the lensed galaxy
population, covering all possible types and redshifts. In most cases however,
the spectroscopic surveys are too shallow to be complete down to the limiting
magnitude of weak-lensing galaxies, and suffer from a non-zero spectroscopic
redshift failure rate.

Additional methods to assess the quality of photometric redshifts, and to
recover the true redshift distribution, make use of the spatial clustering of
galaxies. From the amount of cross-correlation of samples between different
redshift bins one can deduce the amount of redshift outliers
\cite{BvWMK10,CFHTLenS-2pt-tomo}. Similarly, the cross-correlation of the
photometric with a spectroscopic sample can reveal the true redshift
distribution \cite{Newman08}, although this reconstruction is hampered by a
possible redshift-dependent bias of the photometric galaxy sample
\cite{2010ApJ...724.1305S}.

\subsection{Error modelling and residual systematics}
\label{sec:error-model}

Any weak lensing data analysis must be completed with a robust error modeling.
This step is necessary to quantify any residual systematics caused by an
imperfect PSF correction, since those residuals can mimic a cosmological
signal. The most commonly used approach is a null test of the correlation
between the stellar ellipticities $\varepsilon^\star$ (before PSF correction)
and the corrected galaxy shapes $\varepsilon$.  This star-galaxy ellipticity
correlation function is defined as
\begin{equation}
  \xi_{\rm sys}=\langle \varepsilon^\star \varepsilon\rangle .
  \label{eq:xi_sys}
\end{equation}
There are two sources of noise contributions, statistical noise and sample variance: 
(1) The statistical noise is easy to estimate for the shear because it is caused by
the uncorrelated intrinsic ellipticity dispersion $\sigma_\varepsilon^2$ of galaxies which can be
known very accurately either from the lensing data themselves, or from deep,
high-resolution space-based surveys.
(2) Sample variance can create local, random alignments between the PSF
ellipticity patterns and shear from the large-scale structure.
Even though a perfect PSF correction leads to
a zero expectation value of $\xi_{\rm sys}$, there is a non-zero scatter, which
exceeds the statistical-noise scatter of that estimator in the absence of cosmic shear.
\citet{CFHTLenS-sys} showed that the distribution of $\xi_{\rm sys}$ can be
measured including sample variance estimated from
$N$-body simulations. By comparing the observed scatter of $\xi_{\rm
sys}$ to the one from LSS simulations, one can statistically
estimate the amount of PSF residual systematics.

Other null tests for detecting potential systematics in weak-lensing data look at
correlations of the measured shear with quantities that should be independent of
the shear, such as PSF size and shape, position on the CCD or in the field, etc.

A further test that has been used frequently in the past is the null detection
of a B-mode dispersion, for example using the aperture-mass or the COSEBIs
(Sect.~\ref{sec:other_2nd_order}). Note however that a small non-zero B-mode is expected
from astrophysical sources (Sect.~\ref{sec:corrections}), and future surveys will have
large enough statistical power to detect this B-mode. Moreover, this test does not
catch systematic effects that only produce an E-mode pattern.

Additionally, it is important to test for potential redshift dependent
systematics \cite{2009MNRAS.397..608S}. For example, the shear biases $m$ and
$c$ (\ref{eq:ell_bias}) likely depend on galaxy properties such as size,
magnitude, or colour, which change with redshift.  A simultaneous test of
redshift-dependent shear biases and catastrophic photometric redshift outliers
is the increase of the galaxy-galaxy lensing amplitude for a fixed foreground
sample with increasing redshift of the background population.  This increase is
a purely geometrical effect and depends on the angular diameter distance
between lens and source. This redshift-scaling shear test can help to identify
the redshift range with the most reliable shear and photo-$z$ quality
\cite{CFHTLenS-sys}.


\section{Observational results and cosmological constraints}
\label{sec:obs_results}

\subsection{Basic results}
\label{sec:basic_results}

The following sections discuss the basic observational results from
second-order cosmic shear statistics.  The parameter combination that cosmic
shear (including non-linear scales) is most sensitive to is $\sigma_8
\Omega_{\rm m}^\alpha$, with $\alpha \approx 0.5$ - $0.7$. Fig.~\ref{fig:Sigma}
shows this combination measured in recent years. The corresponding data and
original results used for this plot are listed in Table \ref{tab:Sigma}.
Some data points in this figures are from third-order and 3D lensing. These
results are discussed in more detail in Sects.~\ref{sec:third_order} and \ref{sec:3D_lensing}.

\stoptwocol
\begin{figure}
  \begin{center}
    \resizebox{0.9\hsize}{!}{
      \includegraphics{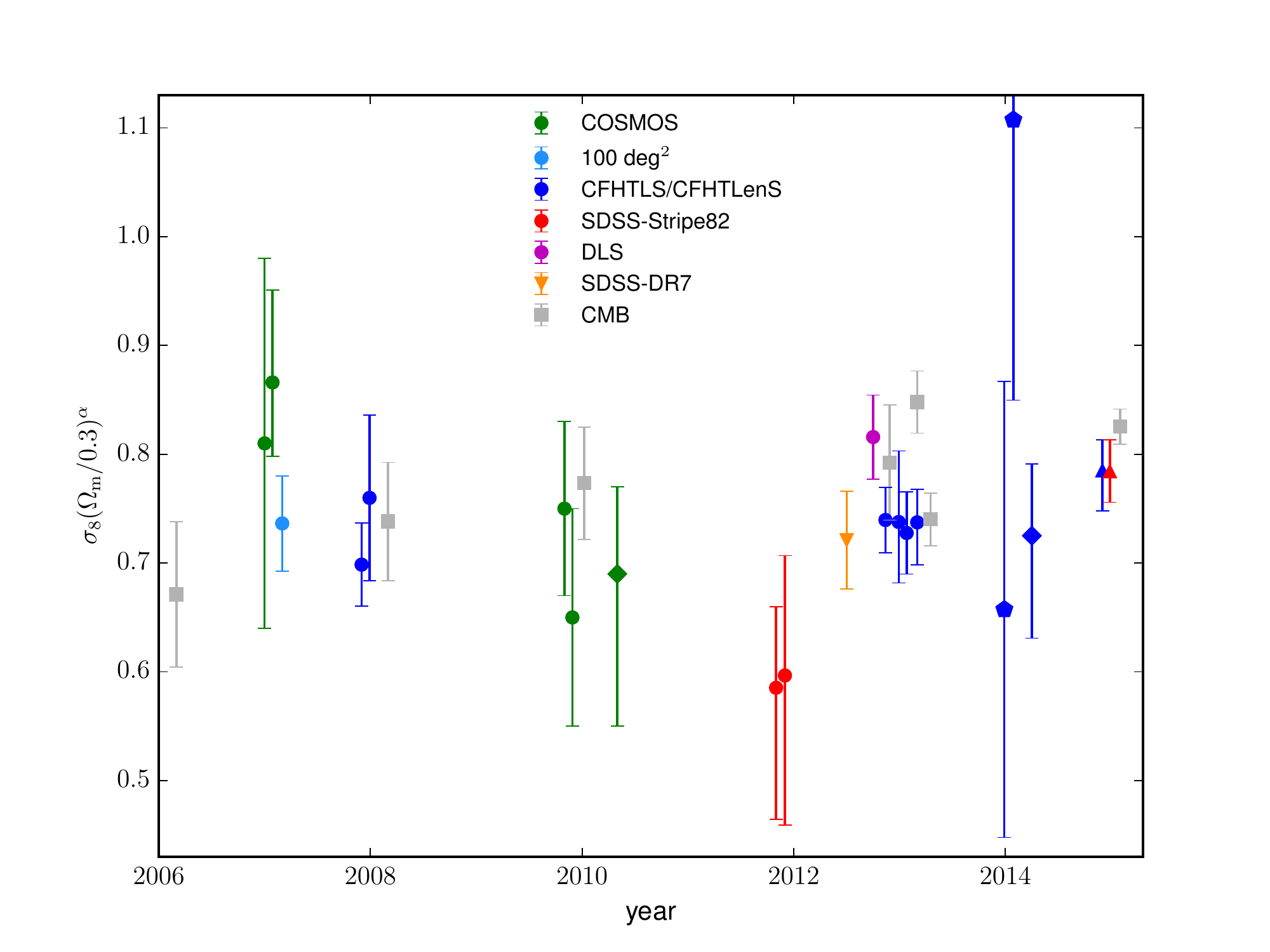}
    }
  \end{center}

    \caption{Mean and 68\% error bars for the parameter $\sigma_8 \left(\Omega_{\rm m}/0.3\right)^\alpha$,
    for various cosmic shear observations, plotted as function of their publication date
    (first arXiv submission). All parameter values are given in Table \ref{tab:Sigma}.
    Different surveys are distinguished by colour as indicated in the figure. Data points are shown
    for second-order statistics (circles), third-order (diamonds), 3D lensing (pentagons), galaxy-galaxy
    lensing (+ galaxy clustering; triangle), and CMB (squares).
    }

    \label{fig:Sigma}

  \end{figure}
\begtwocol

\stoptwocol
  \begin{table}

    \label{tab:Sigma}

    \caption{Parameter values of $\sigma_8 \Omega_{\rm m}^\alpha$
            used for Fig.~\ref{fig:Sigma}. For different surveys (first column)
            the second column is the value rescaled to $\Omega_{\rm m} = 0.3$ which is plotted in the figure,
            obtained from
            the original measurement (third column). The fourth column indicates
            the number of redshift bins, the column 5 is the reference.
            }

    \bigskip

    {\footnotesize
    \renewcommand{\arraystretch}{1.25}
\begin{tabular}{|l|l|r@{ = }l|l|l|}\hline\hline
\rule[-3mm]{0em}{8mm}Survey	 &$\sigma_8(\Omega_{\rm{m}}/0.3)^\alpha$	 &\multicolumn{2}{c|}{original\phantom{x}measurement}	 &$N_z$	 &Reference	\\\hline\hline
COSMOS	 &$0.81\pm{0.17}$	 &$\sigma_8(\Omega_{\rm{m}}/0.3)^{0.48}$ & $0.81\pm{0.17}$	 &1	 &\citet{2007ApJS..172..239M}	\\\hline
COSMOS	 &$0.87^{+0.09}_{-0.07}$	 &$\sigma_8(\Omega_{\rm{m}}/0.3)^{0.44}$ & $0.866^{+0.085}_{-0.068}$	 &3	 &\citet{2007ApJS..172..239M}	\\\hline
100\phantom{{x}}deg$^2$	 &$0.74\pm{0.04}$	 &$\sigma_8(\Omega_{\rm{m}}/0.24)^{0.59}$ & $0.84\pm{0.05}$	 &1	 &\citet{JonBen07}	\\\hline
CFHTLS	 &$0.70\pm{0.04}$	 &$\sigma_8(\Omega_{\rm{m}}/0.25)^{0.64}$ & $0.785\pm{0.043}$	 &1	 &\citet{FSHK08}	\\\hline
CFHTLS	 &$0.76\pm{0.08}$	 &$\sigma_8(\Omega_{\rm{m}}/0.25)^{0.53}$ & $0.837\pm{0.084}$	 &1	 &\citet{FSHK08},\phantom{x}large\phantom{{x}}scales	\\\hline
COSMOS	 &$0.75\pm{0.08}$	 &$\sigma_8(\Omega_{\rm{m}}/0.3)^{0.51}$ & $0.75\pm{0.08}$	 &$5\!+\!1^a$	 &\citet{SHJKS09}	\\\hline
COSMOS	 &$0.65\pm{0.10}$	 &$\sigma_8(\Omega_{\rm{m}}/0.3)^{0.62}$ & $0.65\pm{0.1}$	 &1	 &\citet{SHJKS09}	\\\hline
COSMOS	 &$0.69^{+0.08}_{-0.14}$	 &$\sigma_8(\Omega_{\rm{m}}/0.3)^{0.46}$ & $0.69^{+0.08}_{-0.14}$	 &1	 &\citet{2011MNRAS.410..143S},\phantom{x}$+$IA	\\\hline
SDSS-Stripe82	 &$0.59^{+0.07}_{-0.12}$	 &$\sigma_8(\Omega_{\rm{m}}/1.0)^{0.7}$ & $0.252^{+0.032}_{-0.052}$	 &1	 &\citet{2012ApJ...761...15L}	\\\hline
SDSS-Stripe82	 &$0.60^{+0.11}_{-0.14}$	 &$\sigma_8(\Omega_{\rm{m}}/0.264)^{0.67}$ & $0.65^{+0.12}_{-0.15}$	 &1	 &\citet{2011arXiv1112.3143H}	\\\hline
SDSS-DR7	 &$0.72\pm{0.05}$	 &$\sigma_8(\Omega_{\rm{m}}/0.25)^{0.57}$ & $0.8\pm{0.05}$	 &1	 &\citet{2013MNRAS.432.1544M}	\\\hline
DLS	 &$0.82\pm{0.04}$	 &$\sigma_8(\Omega_{\rm{m}}/0.265)^{0.5}$ & $0.868\pm{0.041}$	 &1	 &\citet{2012arXiv1210.2732J},\phantom{x}priv.~comm.	\\\hline
CFHTLenS	 &$0.74\pm{0.03}$	 &$\sigma_8(\Omega_{\rm{m}}/0.27)^{0.59}$ & $0.787\pm{0.032}$	 &1	 &\citet{CFHTLenS-2pt-notomo}	\\\hline
CFHTLenS	 &$0.74^{+0.07}_{-0.06}$	 &$\sigma_8(\Omega_{\rm{m}}/0.27)^{0.65}$ & $0.79^{+0.07}_{-0.06}$	 &1	 &\citet{CFHTLenS-2pt-notomo},\phantom{x}large\phantom{{x}}scales	\\\hline
CFHTLenS	 &$0.73\pm{0.04}$	 &$\sigma_8(\Omega_{\rm{m}}/0.27)^{0.55}$ & $0.771\pm{0.04}$	 &2	 &\citet{CFHTLenS-2pt-tomo}	\\\hline
CFHTLenS	 &$0.74^{+0.03}_{-0.04}$	 &$\sigma_8(\Omega_{\rm{m}}/0.27)^{0.46}$ & $0.774^{+0.032}_{-0.041}$	 &6	 &\citet{CFHTLenS-IA},\phantom{x}$+$IA	\\\hline
CFHTLenS	 &$0.66\pm{0.21}$	 &$\sigma_8(\Omega_{\rm{m}}/0.27)^{0.46}$ & $0.69\pm{0.22}$	 &$\infty^b$	 &\citet{CFHTLenS-3d}	\\\hline
CFHTLenS	 &$1.11\pm{0.26}$	 &$\sigma_8(\Omega_{\rm{m}}/0.27)^{0.44}$ & $1.16\pm{0.27}$	 &$\infty^b$	 &\citet{CFHTLenS-3d},\phantom{x}large\phantom{{x}}scales	\\\hline
CFHTLenS	 &$0.73^{+0.07}_{-0.09}$	 &$\sigma_8(\Omega_{\rm{m}}/0.27)^{0.57}$ & $0.77^{+0.07}_{-0.1}$	 &1	 &\citet{CFHTLenS-2+3pt},\phantom{x}$+$IA	\\\hline
CFHTLenS	 &$0.79^{+0.03}_{-0.04}$	 &$\sigma_8(\Omega_{\rm{m}}/0.27)^{0.64}$ & $0.84^{+0.03}_{-0.04}$	 &1	 &\citet{2014arXiv1412.0757L}	\\\hline
SDSS-Stripe82	 &$0.78\pm{0.03}$	 &$\sigma_8(\Omega_{\rm{m}}/0.27)^{0.42}$ & $0.82\pm{0.03}$	 &1	 &\citet{2014arXiv1412.3683L}	\\\hline
WMAP3$^c$	 &$0.67\pm{0.07}$	 &$\sigma_8(\Omega_{\rm{m}}/0.234)^{0.5}$ & $0.76\pm{0.05}$	 &-	 &\citet{2007ApJS..170..377S}	\\\hline
WMAP5$^c$	 &$0.74\pm{0.05}$	 &$\sigma_8(\Omega_{\rm{m}}/0.258)^{0.5}$ & $0.796\pm{0.036}$	 &-	 &\citet{WMAP5-Komatsu08}	\\\hline
WMAP7$^c$	 &$0.77\pm{0.05}$	 &$\sigma_8(\Omega_{\rm{m}}/0.273)^{0.5}$ & $0.811^{+0.03}_{-0.031}$	 &-	 &\citet{2010arXiv1001.4538K}	\\\hline
WMAP9	 &$0.79\pm{0.05}$	 &$\sigma_8(\Omega_{\rm{m}}/1.0)^{0.5}$ & $0.434\pm{0.029}$	 &-	 &\citet{2012arXiv1212.5226H}	\\\hline
Planck2013	 &$0.85\pm{0.03}$	 &$\sigma_8(\Omega_{\rm{m}}/0.27)^{0.46}$ & $0.89\pm{0.03}$	 &-	 &\citet{2013arXiv1303.5076P},\phantom{x}$C(\ell)$	\\\hline
Planck2013	 &$0.74\pm{0.02}$	 &$\sigma_8(\Omega_{\rm{m}}/0.27)^{0.3}$ & $0.764\pm{0.025}$	 &-	 &\citet{2013arXiv1303.5080P},\phantom{x}SZ	\\\hline
Planck2015	 &$0.83\pm{0.02}$	 &$\sigma_8(\Omega_{\rm{m}}/1.0)^{0.5}$ & $0.4521\pm{0.0088}$	 &-	 &\citet{2015arXiv150201589P},\phantom{x}$C(\ell)$	\\\hline
\end{tabular}

    \smallskip
     \renewcommand{\baselinestretch}{0.75}
    $^a$ 5 narrow photo-$z$ bins and one wide bin of faint galaxies.\\[0.1em]
    $^b$ 3D lensing, no $z$-binning.\\[0.1em]
    $^c$ The index $\alpha = 0.5$ is adopted
            from the WMAP9 measurement, published on \texttt{http://lambda.gsfc.nasa.gov}.
            The resulting values and errors are therefore only illustrative.
    \renewcommand{\baselinestretch}{1}
    }

  \end{table}
\begtwocol

\subsubsection{Early era, 2000 - 2006}
\label{sec:early_era}

The first detection of weak gravitational lensing by the large-scale structure
was reported in 2000 by four independent groups
\cite{2000MNRAS.318..625B,kaiser00,2000A&A...358...30V,2000Natur.405..143W}.
The observations were taken with different cameras and telescopes --- the
\instrument{Prime Focus Imaging Camera (PFIC)} on the
\instrument{William-Herschel Telescope (WHT)}, \instrument{UH8K} and
\instrument{CFH12K} on the \instrument{Canada-France Hawaii Telscope (CFHT)}, and the
\instrument{Big Throughput Camera (BTC)} on \instrument{Blanco} --- and covered sky areas
between $0.5$ and $1.5 \, \mbox{deg}^2$. These early analyses measured
correlations of galaxy ellipticities that were larger than the expected
residual systematics. Limits on $\Omega_{\rm m}$ and $\sigma_8$ could be
obtained.

Those exploratory results were very soon followed by other surveys from a wide
range of telescopes, for example \instrument{CFH12K/CFHT} with the
\survey{Red-sequence Cluster Survey (RCS)} and \survey{VIRMOS-DESCART}
\cite{2001A&A...374..757V,2002A&A...393..369V,2002ApJ...577..595H,2002ApJ...572...55H,vWMH05},
\instrument{FORS1 (FOcal Reducer and Spectrograph)/VLT}
\citeaffixed{2001A&A...368..766M}{Very Large Telescope; }, the \survey{75-deg$^2$ survey}
with \instrument{BTC/Blanco-CTIO}
\cite{2003AJ....125.1014J,JBBD06}, \instrument{PFIC/WHT}
\cite{2005MNRAS.359.1277M}, \instrument{ESI (Echelle Spectrograph and
Imager)/Keck II} \cite{2003MNRAS.344..673B}, \instrument{WFI} at
\instrument{MPG/ESO 2.2m} with the \survey{Garching-Bonn Deep Survey}
\citeaffixed{2007A&A...468..859H}{\survey{GaBoDS}; }, and
\instrument{Suprime-Cam/Subaru} \cite{2003ApJ...597...98H}.

Cosmic shear then was measured using \instrument{MegaCam/CFHT} on the
Canada-France Hawaii Legacy Survey (\survey{CFHTLS}). During five years this
large program observed $170$ square degrees in five optical bands. First
results from the first data release were published over 22 deg$^2$ of the wide
part \cite{CFHTLSwide} and the $3$ out of the $4$ deg$^2$ of the deep part
\cite{CFHTLSdeep}.

Apart from those ground-based observations, cosmic shear was successfully
detected with the \instrument{Hubble Space Telescope (HST)}, using parallel
archival data from the cameras \instrument{STIS}
\citeaffixed{2002A&A...385..743H,2005A&A...432..797M,2004ApJ...605...29R}{Space
Telescope Imaging Spectrograph; }, and \instrument{WFPC2}
\citeaffixed{2001ApJ...552L..85R,2002ApJ...572L.131R,2003ApJ...598L..71C}{Wide
Field Planetary Camera 2; }. Further \instrument{HST} cosmic shear results came
from the \survey{Galaxy Evolution From Morphology And SEDs (GEMS)} survey
\cite{2005MNRAS.361..160H,2007A&A...468..823S}.

Those early observations were done in only one optical filter, and the
photometric redshift distribution $n(z)$ of source galaxies
(Sect.~\ref{sec:photo-z}) could not be measured from the data themselves. To
estimate $n(z)$, external data were used by matching the survey depths. Those
data were typically very deep but covered a very small area, such as the
Hubble Deep Field (HDF) with $6.5$ arcmin$^2$, and the adopted redshift
distribution suffered from a large cosmic variance \cite{2006APh....26...91V}.
A notable exception were cosmic shear results from the \survey{COMBO-17} survey
observed with \instrument{WFI} at \instrument{MPG/ESO 2.2m}, %
\cite{2003MNRAS.341..100B}, for which accurate photometric redshifts were
available from 5 broad- and 12 narrow-band filters, with a precision of
$\sigma_z / (1+z) = 0.02$ at $R = 23$ \cite{2003A&A...401...73W}. From a
convergence band-power (2PCF) estimate using the method of
\citet{2001ApJ...554...67H}, the constraints $\sigma_8 (\Omega_{\rm m} /
0.3)^{0.49} = 0.72 \, (0.75) \pm 0.09$ were obtained
\cite{2003MNRAS.341..100B}. Intrinsic alignment contaminations to this result
were estimated to amount to $0.03$, one third of the statistical errors
\cite{2004MNRAS.347..895H}. Note that some of the five COMBO-17 fields were not
selected randomly, but included the very empty CDFS
(Chandra Deep Field South), and A901 with the $z=0.16$ super-cluster Abell
901/902.

By 2006, there was a small but systematic tension in the fluctuations amplitude
parameter $\sigma_8$ between different cosmic shear results.
Some of the above-mentioned surveys such as \survey{GaBoDS},
\survey{CFHTLS}, and \survey{COSMOS} yielded values of $\sigma_8$ of around
$0.85$, systematically higher than the CMB result from WMAP3 of
$\sigma_8 = 0.76 \pm 0.05$ \cite{2007ApJS..170..377S}. Other surveys such as
\survey{COMBO-17} and the 75-\mbox{deg}$^2$ survey were in agreement with
WMAP3. The scatter of $\sigma_8$ between cosmic shear results was larger than
the statistical errors, which were typically between $0.05$ and $0.15$ (for a
fixed $\Omega_{\rm m} = 0.3$), see \citet{2007A&A...468..859H} for a
compilation of pre-2007 results. These differences spurred a lot of activity
into finding possible lensing systematics. The comparison and calibration of
shape measurement methods had already started a few years earlier. The
STEP project (Sect.~\ref{sec:image_sims}), albeit using somewhat idealised simulations
showed that cosmic shear could be measured with current methods to better than
the statistical uncertainties \cite{STEP1}.

Many of those early surveys were clearly systematics-dominated, in some cases
showing significant PSF residuals and B-modes, or using incorrect shear calibrations, resulting
in biased measurements.
In addition, the cosmological interpretation of these measurements
requires the knowledge of the source galaxy redshift distribution. This
turned out to be one of the main causes of bias to cosmological
weak-lensing results. These problems were largely resolved in
subsequent years, as is discussed in the following section.

\subsubsection{Consolidating era, 2007 - 2012}
\label{sec:consol_era}

The main advancement during these years was the improved estimation of the
source galaxy redshift distribution. Large samples of multi-band observations
resulted in accurate photometric redshifts that were calibrated using deep
spectroscopic data from surveys that overlapped with the weak-lensing data.
Furthermore, more realistic error estimates were included in cosmic shear
analyses.

The third data release of the wide part of the Canada-France-Hawaii Telescope
Legacy Survey (\survey{CFHTLS}) with an observed area of $53$ deg$^2$ provided
2D cosmic shear results out to very large, linear scales,
\citeaffixed{FSHK08}{$7.7$ degrees, corresponding to $170$ Mpc at the mean lens
redshift of of $0.5$; }. This enabled cosmological constraints using large
scales only, thereby reducing uncertainties from non-linear and baryonic
physics on small scales. Using $\langle M_{\rm ap}^2 \rangle(\theta)$ on scales
$\theta > 85^\prime$ the authors obtained $\sigma_8 (\Omega_{\rm m} /
0.25)^{0.53} = 0.837 \pm 0.084$. Simultaneously, the third-release
\survey{CFHTLS} data was combined with the previously completed surveys
\survey{GaBoDS}, \survey{RCS}, and \survey{VIMOS-DESCART} into the \survey{100
$deg^2$ survey} \cite{JonBen07}, doubling the area that was available from
\survey{CFHTLS} at that time. Both these studies used the photometric redshifts
from the deep part of \survey{CFHTLS} \cite{2006A&A...457..841I}, taking into
account sampling variance. The deep fields have an area of $4$ square degrees,
an increase of nearly a factor $2500$ over the \survey{HDF}.

The space-based \survey{COSMOS}
\citeaffixed{2007ApJS..172....1S}{COSmological Evolution Survey; } was a
significant contribution to the field. This wide ($1.64$ deg$^2$) and very deep
\instrument{ACS (Advanced Camera for Surveys)/HST} survey provided a density of
source galaxies of about a factor of four larger compared to the deepest
ground-based surveys (e.g.~\survey{COMBO-17}). A large number of multi-band
follow-up observations from the ground allowed for accurate photometric
redshifts. A first analysis was presented in 2007
\cite{2007ApJS..172..219L,2007ApJS..172..239M}. Using 15-band photo-$z$'s
\cite{2007ApJS..172..117M}, the evolution of the shear signal between three
redshift bins was demonstrated, although with estimated relative calibration
errors of $5\%$. An independent re-analysis of the weak-lensing data
\cite{SHJKS09} used improved photo-$z$'s from twice the number of bands
\cite{2009ApJ...690.1236I}.  Due to the low number of high-$S/N$ stars in
\instrument{ACS} fields, and temporal instabilities of \instrument{HST}, the
PSF model was obtained by PCA of the PSF pattern from dense stellar fields.
This work presented a five-bin tomographic analysis, leading to constraints on
the deceleration parameter $q_0 = - \ddot a a / \dot a^2 = \Omega_{\rm m}/2 -
\Omega_\Lambda$, with a detection of acceleration ($q_0 < 0$) at 94.3\%
confidence, including additional priors on $h$ and $\Omega_{\rm b}$. The same
data were used in \citet{2011A&A...530A..68T} to verify the consistency with
GR.

By that time, ground-based surveys had become large enough to enable detailed
residual systematics tests. For \survey{CFHTLS}, \citet{KB09} as well as first
multi-colour observations \cite{FuPhD} revealed an anomalous shear amplitude
scaling with source redshift and a variance between \instrument{MegaCam}
pointings larger than expected.

\subsubsection{Survey era, 2013 - present}
\label{survey_era}

The survey era provided measurements that relied on independently cross-checked
photometric redshifts, and a robust estimate of residual systematics on
weak-lensing shear correlations (Sect.~\ref{sec:error-model}). In these
studies, the residual sytematics analysis were done completely independently
from the cosmological parameters analysis, in order not to bias the
cosmological results.

\stoptwocol
  \begin{figure}

     \centerline{Flat $\Lambda$CDM}
    \vspace*{-2.3em}

        \resizebox{1.0\hsize}{!}{	
          \includegraphics{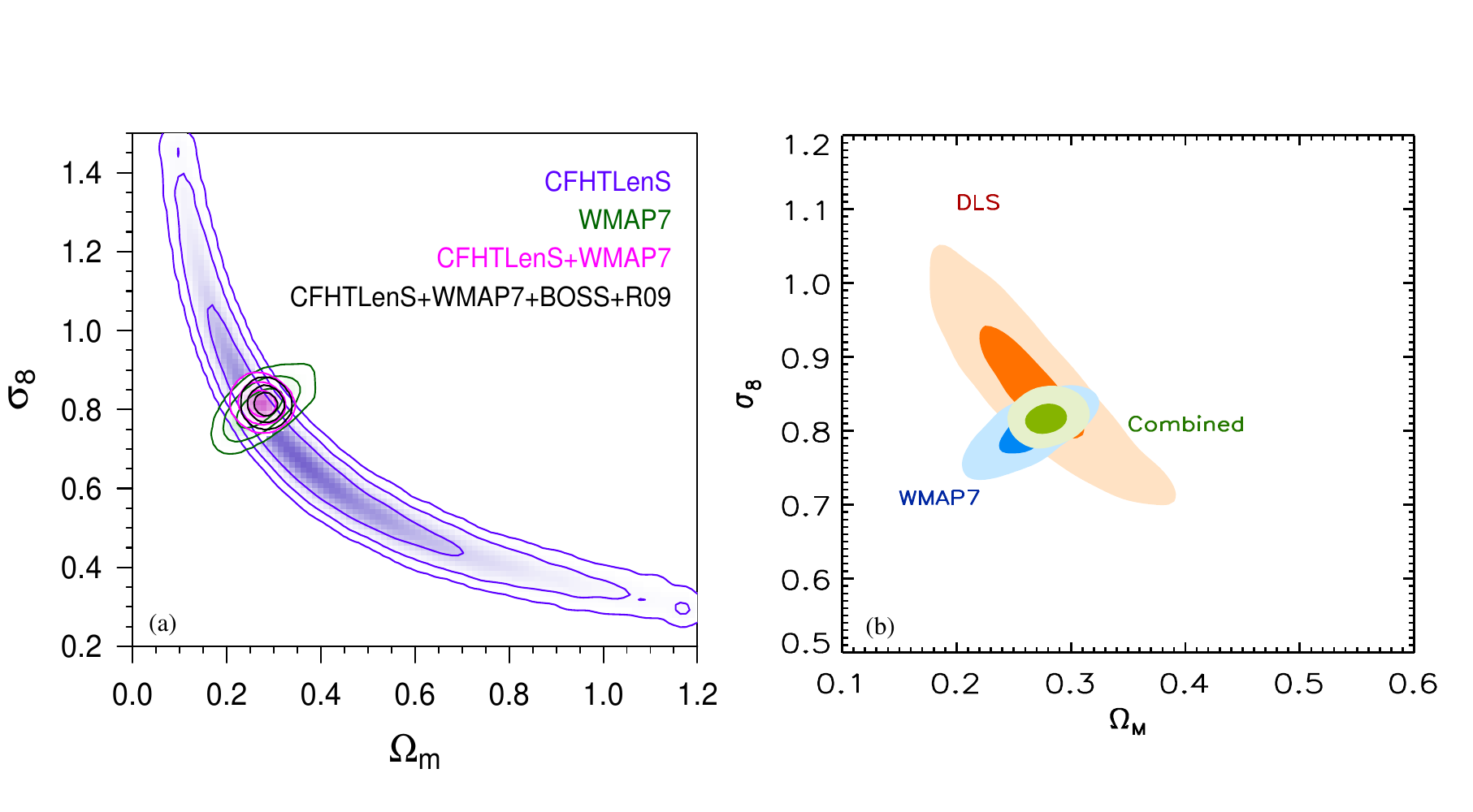}
        }

     \caption{The near-orthogonality of $\Omega_{\rm m}$ and $\sigma_8$ constraints from
     2D cosmic shear and CMB.
     (a) CFHTLenS, WMAP7, BAO from BOSS \cite{2012arXiv1203.6594A}, and a HST $H_0$ prior
     \cite[`R09']{2009ApJ...699..539R}. From \citet{CFHTLenS-2pt-notomo}.
     (b) DLS including tight priors on $\Omega_{\rm b}$ and
     $H_0$. From \citet{2012arXiv1210.2732J}. Figure used with permission
     from \citet{2012arXiv1210.2732J}, \emph{\apj}, \textbf{765}, 74. Copyright 2013 IOP.
     }

     \label{fig:Om_s8_flatLCDM}

  \end{figure}
\begtwocol

A milestone for cosmic shear represented the CFHT lensing survey
\citeaffixed{CFHTLenS-data}{\survey{CFHTLenS}; }. Photometric redshifts for
each source galaxy were obtained in \citet{CFHTLenS-photoz}, the robustness of
which was verified using spectroscopic redshifts, \survey{COSMOS} 30-band
photo-$z$s, and a cross-correlation analysis \cite{CFHTLenS-2pt-tomo}. Galaxy
shapes were measured on individual exposures with \emph{lens}fit and calibrated
using two independent suites of image simulations \cite{CFHTLenS-shapes}. An
excess correlation between star and galaxy shapes (\ref{eq:xi_sys}) was found
on $25\%$ of the observed fields, which in turn were discarded from the cosmological
analysis \cite{CFHTLenS-sys}. Two-dimensional cosmic shear correlation
functions from \survey{CFHTLenS} were presented in \citet{CFHTLenS-2pt-notomo}.
A two-bin tomographic analysis was performed by \citet{CFHTLenS-2pt-tomo}. The
same tomographic data were used to place constraints on modified gravity
\cite{CFHTLenS-mod-grav}. Further, a six-bin tomographic analysis was performed
where cosmological and intrinsic-alignment parameters were
constrained simultaneously \cite{CFHTLenS-IA}. Late-type galaxies were found to
not show any significant intrinsic alignment, while for early type galaxies IA
was detected at about $2\sigma$.

Cosmic shear results from the very deep, $20$ deg$^2$ wide \survey{DLS} (Deep
Lens Survey) were presented in \citet{2012arXiv1210.2732J}, observed with the
two $4 \, \mbox{m}$ telescopes Kitt Peak and Blanco. The PSF model was obtained
using PCA, and calibrated by minimizing correlations between PSF residuals and
PSF model \cite{2010MNRAS.404..350R}.

Cosmic shear results were also obtained by \survey{SDSS} (Sloan Digital Sky Survey).
\survey{SDSS} observations on $250$ square degrees ($168$ after
masking) of the \survey{Stripe-82} area are very shallow, with only $2$ lensing
galaxies per arc minute, and suffer from a large and strongly varying PSF
\citeaffixed{2014MNRAS.440.1296H}{between $0.8$ and $3.2$ arcsec; }. Despite
these challenges, cosmic shear results were obtained from stacked images,
presented in \citet{2012ApJ...761...15L} and \citet{2011arXiv1112.3143H}

\stoptwocol
  \begin{figure}

    \resizebox{1.0\hsize}{!}{
        \includegraphics{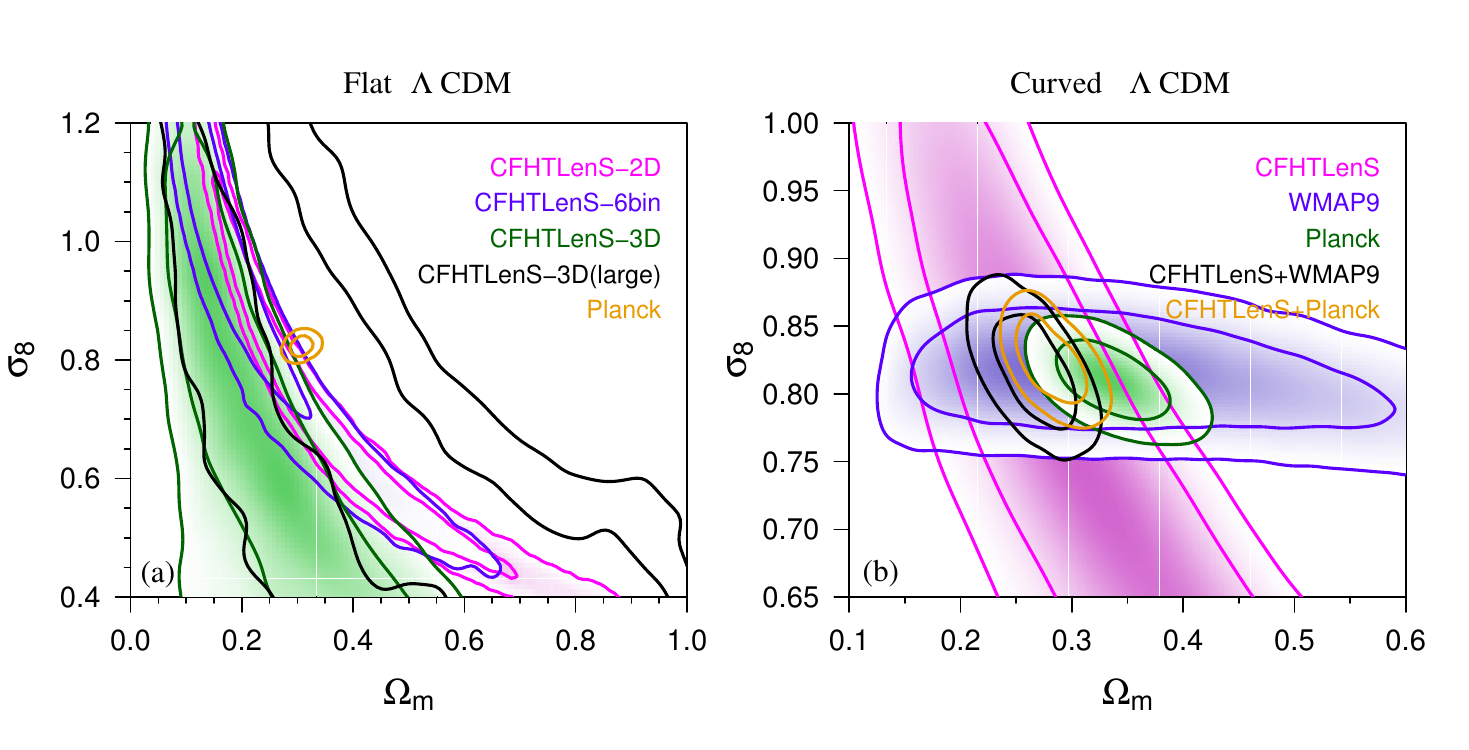}
      }

   \caption{Cosmic shear and CMB $68.3\%$ and $95.5\%$ confidence levels for $\Omega_{\rm m}$ and $\sigma_8$
  in a $\Lambda$CDM universe. (a) Assuming flatness.
  CFHTLenS 2D, 6-bin tomography, 3D, and 3D from large scales only are compared to Planck constraints.
  (b) With free curvature, showing CFHTLenS (joint second- and third-order), WMAP9, Planck,
  CFHTLenS $+$ WMAP9, and CFHTLenS $+$ Planck constraints,
	from \citet{CFHTLenS-2+3pt}. Figure used with permission from \citet{CFHTLenS-2+3pt}, \emph{\mnras},
  \textbf{441}, 2725. Copyright 2014 Oxford University Press.
  }

   \label{fig:Om_s8_Planck}

\end{figure}
\begtwocol

For a $\Lambda$CDM cosmology, cosmic shear constrains a combination of
$\Omega_{\rm m}$ and $\sigma_8$ that is perpendicular to the one obtained from
CMB \cite{Contaldi03}. Adding cosmic shear to \survey{WMAP (Wilkinson Microwave
Anisotropy Probe)} results in typical reduction of error bars on $\Omega_{\rm
m}$ and $\sigma_8$ of up to $50\%$, similar to other low-$z$ cosmological
probes such as Baryonic Acoustic Oscillations (BAO). For example, the
\survey{WMAP7} constraints of $\Omega_{\rm m} = 0.273 \pm 0.03$ and $\sigma_8 =
0.811 \pm 0.031$ \cite{2010arXiv1001.4538K} get tightened when adding
\survey{CFHTLenS}, resulting in $\Omega_{\rm m} = 0.274 \pm 0.013$ and
$\sigma_8 = 0.815 \pm 0.016$ \cite{CFHTLenS-2pt-notomo}. Similar constraints
were obtained with \survey{DLS} + \survey{WMAP7} + tight priors on $h$ and
$\Omega_{\rm b}$, with $\Omega_{\rm m} = 0.278 \pm 0.018$ and $\sigma_8 = 0.815
\pm 0.020$ \citeprefixed{2012arXiv1210.2732J}{; see
Fig.~\ref{fig:Om_s8_flatLCDM}}. Planck's cosmological findings from temperature
anisotropies (together with CMB lensing and WMAP polarization) correspond to a
higher matter density and normalization compared to most previous pobes, with
$\Omega_{\rm m} = 0.315 \pm 0.017$ and $\sigma_8 = 0.829 \pm 0.012$, or
$\sigma_8 (\Omega_{\rm m} / 0.27)^{0.46} = 0.89 \pm 0.03$
\cite{2013arXiv1303.5076P}. This is consistent with CFHTLenS at the $2\sigma$
level, see Fig.~\ref{fig:Om_s8_Planck}. Further, Planck's counts of
Sunyaev-Zel'dovich (SZ) clusters results in a lower normalization of $\sigma_8
(\Omega_{\rm m} / 0.27)^{0.3} = 0.78 \pm 0.01$ \cite{2013arXiv1303.5080P}.
Sect.~\ref{sec:follow_up_pub} discusses whether adding extra-parameters such as
massive neutrinos are needed to reconsile recent high- and low-$z$ data.

A model with variable curvature does not change the cosmic-shear constraints
on $\Omega_{\rm m}$ and $\sigma_8$ by a lot. Pre-\survey{Planck} CMB data alone
cannot constrain the curvature of the Universe, and adding other probes such as
measurements of $H_0$ or weak lensing are required. \survey{Planck} and
high-resolution ground-based millimetre-wavelength radio telescopes of similar
sensitivity and resolution such as \instrument{SPT (South Pole Telescope)} and
\instrument{ACT (Atacama Cosmology Telescope)} have detected weak-lensing of
the CMB by large-scale structures (\emph{CMB lensing}), which helps to break
the geometrical degeneracy. This results in tight constraints on $\Omega_{K}$
from CMB alone
\cite{2011PhRvL.107b1302S,2012ApJ...756..142V,2013arXiv1303.5076P}.
Fig.~\ref{fig:Om_s8_Planck} shows joint cosmic shear and CMB constraints for a
free-curvature model.

The dark-energy parameter of state $w_0$ has been measured with cosmic shear
already in 2006 \cite{CFHTLSwide,CFHTLSdeep}. However, since the effect of
dark energy on the supression of the growth of structure is relatively small,
2D weak lensing is not very sensitive to dark energy, and $68\%$ confidence
intervals on $w_0$ are typically of order unity, which furthermore is
degenerate with other parameters such as $\sigma_8$. However, weak lensing can
rule out some combinations of parameter values, and substantially reduce the
allowed region of parameter space when combined with other probes.
Fig.~\ref{fig:Om_w0} shows how CMB constraints from \survey{WMAP7} --- with an additional
prior on $H_0$ from \citep{2011ApJ...730..119R} --- are reduced by \survey{CFHTLenS} six-bin
tomography. The parameters $\Omega_{\rm m}$ and $w_0$ are measured to better
than $10\%$ accuracy, for both a flat and free-curvature $w$CDM model. The
improvement is similar to adding Baryonic Acoustic Oscillation (BAO)
data from the \survey{SDSS-III Baryon Oscillation Spectroscopic Survey}
\citeaffixed{2012arXiv1203.6594A}{\survey{BOSS}; } to CMB data.

\stoptwocol
\begin{figure}

   \begin{center}
      \resizebox{0.8\hsize}{!}{	
        \includegraphics{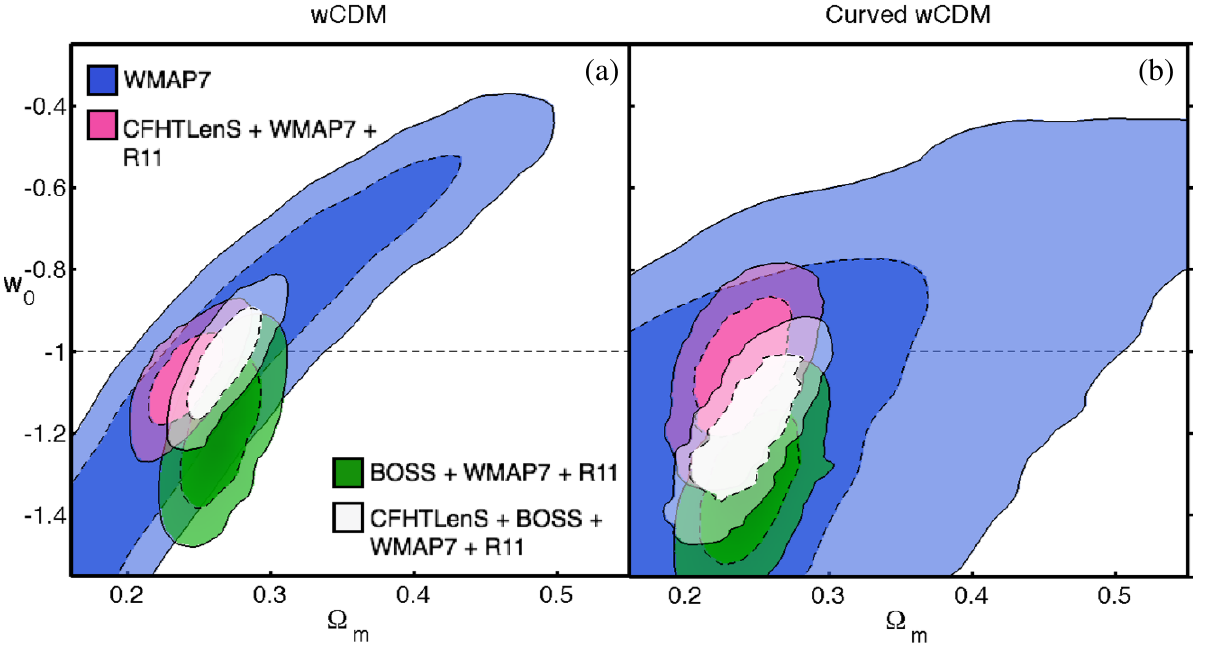}
      }
   \end{center}

   \caption{Combined constraints on $\Omega_{\rm m}$ and $w_0$ from cosmic shear, CMB, and BAO.
	The model is a $w$CDM universe with flat (free) curvature in 
	panel a (b). Cosmic shear is six-bin tomography from CFHTLenS. The CMB and BAO
	data are the same as in Fig.~\ref{fig:Om_s8_flatLCDM}. The HST $H_0$ prior is replaced with an 
	updated version \cite{2011ApJ...730..119R}. From \citet{CFHTLenS-IA}.
  Figure used with permission from \citet{CFHTLenS-IA}, \emph{\mnras}, \textbf{432}, 2249.
  Copyright 2013 Oxford University Press.
  }

   \label{fig:Om_w0}

\end{figure}
\begtwocol

Constraints on modified gravity using the parametrization in
(\ref{eq:Poisson_mod_Psi}, \ref{eq:Poisson_mod_Phi_plus_Psi}) showed
consistency with GR \cite{CFHTLenS-mod-grav}. A simple model was considered
where $\Sigma$ and $\mu$ did not vary spatially, and at early times tend
towards GR, so that deviations of GR are allowed at late times where the
accelerated expansion happens. The present-day values of those two parameters
were measured to be $\Sigma_0 = 0.00 \pm 0.14$, and $\mu_0 = 0.05 \pm 0.25$,
combining \survey{CFHTLenS} weak-lensing tomographic data
\cite{CFHTLenS-2pt-tomo}, redshift-space distortions from \survey{WiggleZ}
\cite{2012MNRAS.425..405B} and \survey{6dFGS} \cite{2012MNRAS.423.3430B}, \survey{WMAP7}
CMB anisotropies from small scales, $\ell \ge 100$ \cite{2011ApJS..192...16L},
and the \citet{2011ApJ...730..119R} $H_0$ prior (see Fig.~\ref{fig:modgrav}).

All measurements presented so far are based on real-space second-order shear correlations,
with the exception of \citet{2003MNRAS.341..100B} who directly estimated the shear power 
spectrum from the data. In the following sections, we present results from higher-order
shear and other, non-traditional lensing statistics.

\doifonecol{%
\begin{figure}

   \centerline{Flat $\Lambda$CDM}
  \vspace*{-0.0em}

   \begin{center}
      \resizebox{0.6\hsize}{!}{ 
        \includegraphics[bb=10 30 640 740, height=9em, width=10em]{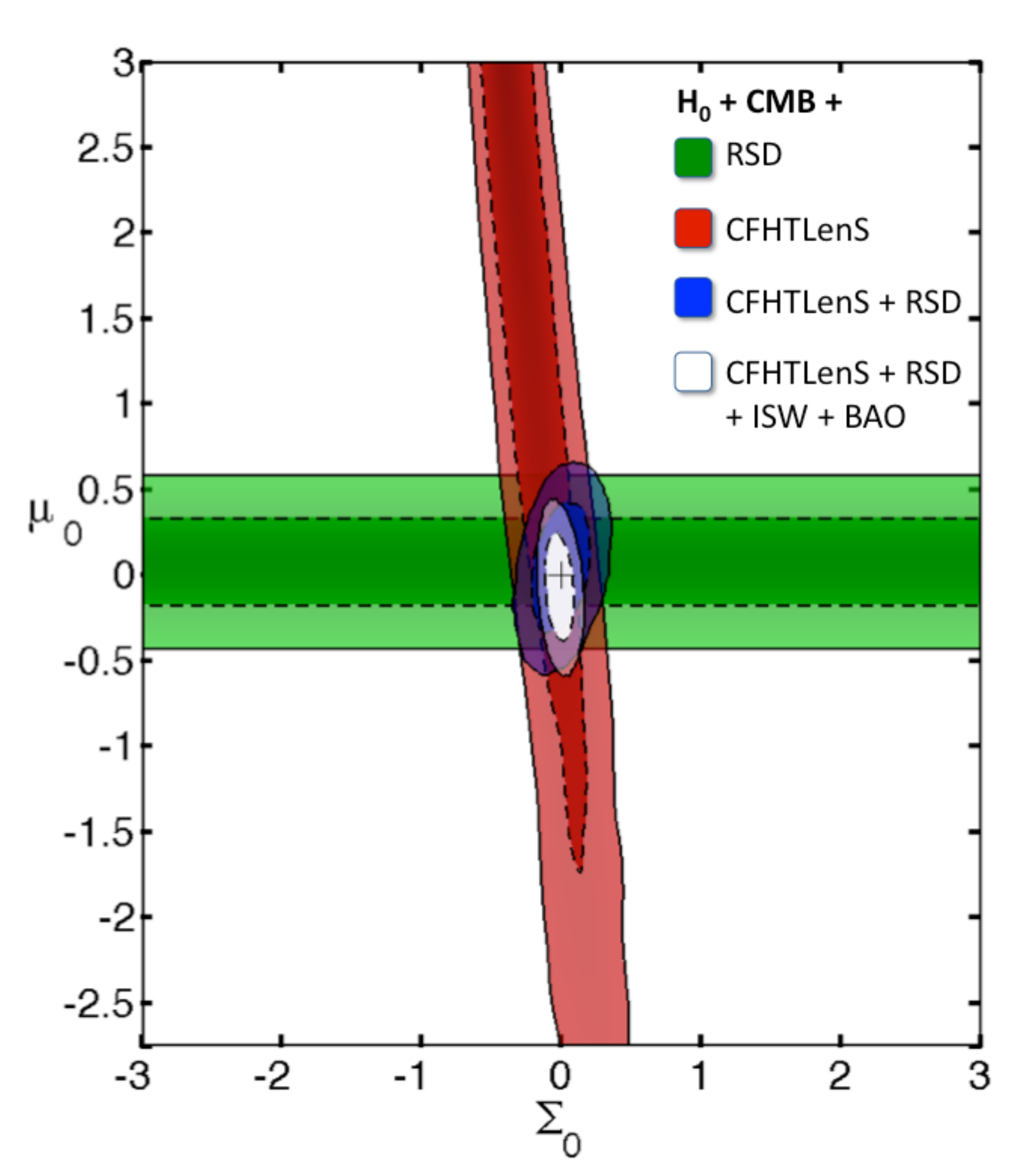}
      }
   \end{center}
}
\doiftwocol{%
\begin{figure}[H]

   \centerline{Flat $\Lambda$CDM}
  \vspace*{-0.0em}

   \begin{center}
      \resizebox{\hsize}{!}{ 
        \includegraphics[bb=10 0 640 740, height=9em, width=10em]{figures/S13-modgrav.png}
      }
   \end{center}
}

   \caption{Combined constraints on the present-day modified-gravity parameters
    $\Sigma_0$ and $\mu_0$, from redshift-space distortions (RSD), cosmic shear
    (CFHTLenS), and their combination, including the case of additional BAO
    \cite{2012arXiv1203.6594A} and large-scale WMAP7 (ISW) data. All data are
    combined with a $H_0$ prior and small-scale CMB data, see text. From
    \citet{CFHTLenS-mod-grav}. Figure used with permission from \citet{CFHTLenS-mod-grav},
    \emph{\mnras}, \textbf{429}, 2249. Copyright Oxford University Press.

}

   \label{fig:modgrav}

\end{figure}

\subsection{Third-order correlations}
\label{sec:third_order}

The motivation behind higher-order shear statistics has been argued for in
Sect.~\ref{sec:bispectrum}. Even though the measurement is challenging and the
overall signal-to-noise ratio is low, several significant detections of third-order
shear correlations have been made.

The first detection of a third-order cosmic shear correlation was obtained
with the \survey{VIRMOS-DESCART} survey \cite{BMvW02}, using a specific high
signal-to-noise projection of the 3PCF \cite{2003A&A...397..405B}. The
aperture-mass skewness $\langle M_{\rm ap}^3 \rangle$ was measured subsequently in
\survey{VIRMOS-DESCART} \cite{2003ApJ...592..664P} and the \survey{75 deg$^2$ survey}
\cite{JBJ04}.

The first space-based measurement of $\langle M_{\rm ap}^3 \rangle$ was
presented in \citet{2011MNRAS.410..143S}, based on the reanalysed HST \survey{COSMOS}
weak-lensing data \cite{SHJKS09}. Their cosmological constraints were consistent
with the \survey{WMAP7} best-fit cosmology.

Several higher-order measurements resulted from CFHTLenS.
\citet{CFHTLenS-kappa-maps} measured the skewness of reconstructed
convergence maps and found good agreement with WMAP7 predictions. After
validating the data for shear residual third-order correlations,
\citet{CFHTLenS-3pt} performed a cosmological analysis of the
aperture-mass skewness exploring a non-Gaussian likelihood.
\citet{CFHTLenS-2+3pt} combined the second- and third-order aperture-mass
combined with \survey{WMAP9} and \survey{Planck} to obtain cosmological
results, including models of intrinsic alignment and source-lens clustering as
astrophysical systematics.

\subsection{Follow-up publications}
\label{sec:follow_up_pub}

Past cosmic shear results have been subsequently used in a large number of
follow-up publications, often by people outside of the original collaboration.
In many cases, these works profited from public releases of the science
products of the cosmic shear survey in question. At a minimum, this includes
the shear correlation function, its covariance, and the source redshift
distribution. This is particularly important when new models are to be tested,
for which simply using the original cosmological constraints as priors, for
example the mean and error on $\Omega_{\rm m} \sigma_8^\alpha$ as Gaussian prior,
might not be valid. In the following, we point out some outstanding results.

Using the CFHTLS-T0003 data \cite{FSHK08}, which was available on request, models
of modified gravity (Sect.~\ref{sec:mod_grav}) were tested in
\citet{2009MNRAS.395..197T}. This work included potential systematics in the
weak-lensing data \cite{KB09}, and uncertainties on small scales to reduce the
uncertainty of non-linear model predictions and baryonic physics.
Further follow-up tests of modified gravity were performed using data from COSMOS
\cite{2007ApJS..172..239M} and CFHTLS-T0003 
\cite{2010PhRvD..81j3510Z,2010PhRvD..81l3508D,2010arXiv1011.2106S}. However,
those analyses did not take into account the anomalous redshift-scalings found
for both surveys. Weak-lensing tomographic data have been used to test models of
modified gravity, for example with the reanalysed COSMOS data \cite{SHJKS09} in
\citet{2011PhRvD..84b3012D}, and CFHTLenS data \cite{CFHTLenS-2pt-tomo} in
\citet{2013arXiv1310.4329D}.

The cosmological constraints from CMB temperature anisotropies measured by the
Planck satellite \cite{2013arXiv1303.5076P} were in slight tension with other
probes. In particular, Planck found a higher power-spectrum normalisation
$\sigma_8$. Several works proposed massive neutrinos to alleviate the tension
with low-$z$ probes such as weak lensing: Massive neutrinos are still
relativistic at recombination and do not significantly influence the CMB
anisotropies. They become however non-relativistic at late time, and dampen the
growth of structure, therefore reducing the low-$z$ clustering power.
Joint analyses including massive neutrinos from Planck and CFHTLenS
weak-lensing data \cite{CFHTLenS-2pt-notomo} were found to improve parameter
constraints with detections of non-zero neutrino masses
\cite{2014PhRvL.112e1303B,2014arXiv1403.4599B}, but the evidence still favors
a $\Lambda$CDM model without additional parameters for massive neutrinos
\cite{2014arXiv1404.5950L}.
Further, one has to note that for a reliable interpretation of small-scale
lensing correlations, baryonic suppression has to be accounted for, the effect
of which is somewhat degenerate with massive neutrinos \cite{2014arXiv1407.4301H}.
Earlier constraints on neutrino masses were obtained using
data from \citet{JonBen07} in \citet{2008arXiv0810.3572G}, and using
data from \citet{FSHK08} in \citet{TSUK09}.

Galaxy-galaxy lensing has been measured with CFHTLenS source galaxies and
\survey{BOSS} (Sloan Digital Sky Survey III Baryon Oscillation Spectroscopic
Survey) lens galaxies \cite{arXiv:1407.1856}. The addition of galaxy number
counts and galaxy clustering helped in independently constraining galaxy bias
and cosmological parameters.

\subsection{Convergence and mass maps}
\label{sec:mass_maps_results}

Mass maps from weak-lensing observations have been produced since the 1990s
for massive galaxy clusters, to study the total matter distribution, and the
relation between galaxies, hot intra-cluster gas, and dark matter
\citeptwo[e.g.~]{1994ApJ...427L..83B}{1996ApJ...461..572S}.
Reconstructing the projected mass (or related quantities such as the density or
the lensing potential) in large, blind fields, where the shear is an order of
magnitude smaller than for galaxy clusters, is more challenging.

\citet{2001ApJ...556..601W} constructed $\kappa$ maps on \instrument{UH8K/CFHT}
data, which were correlated with galaxy light to infer mass-to-light ratios on
large scales. Convergence maps from one of the \survey{DLS} fields
\cite{2006ApJ...643..128W} were confronted in \citet{2005ApJ...635L.125G} with a
velocity dispersion map from $10,000$ galaxy spectra obtained with
\instrument{Hectospec/MMT} (Magnum Mirror Telescope). A strong correlation
between the two was found, indicating that the lensing signal was produced by
groups and clusters.

\citet{2004MNRAS.353.1176T} and \citet{2012MNRAS.419..998S} presented a
reconstruction of the 3D potential
\cite{2001astro.ph.11605T,2003MNRAS.344.1307B} and 3D density
\cite{2009MNRAS.399...48S}, respectively, in the Abell 901/2 super-cluster
field of \survey{COMBO-17}, revealing new background structures behind A902.
Auto- and cross-correlation functions between the potential and galaxy density
and luminosity were found to be consistent with a halo and HOD (halo occupation
distribution) modeling.

The weak-lensing measurements in \survey{COSMOS} \cite{2007ApJS..172..219L}
were transformed into a convergence map using a non-linear, wavelet-based
method \cite{2006A&A...451.1139S}. Thanks to the wealth of multi-wavelength
observations over the same sky area, the mass maps could be compared to various
baryonic tracers such as stellar mass, optical and IR galaxy density, and hot gas
from X-rays. Cross-correlation factors between total mass and baryons out to
large scales between $0.3$ and $0.5$ were obtained. The redshift evolution of
structures was traced by splitting the background source galaxy population into
redshift bins. Finally, a full 3D reconstruction of the potential and mass was
obtained using again the method of \citet{2003MNRAS.344.1307B}.
Further 2D mass maps were obtained, with the addition of galaxy position
information in \survey{COSMOS} \cite{2012MNRAS.424..553A}.

\survey{CFHTLenS} convergence maps were reconstructed and analysed in
\citet{CFHTLenS-kappa-maps}. Moments of $\kappa$ up to order five were in
agreement with $N$-body simulations (see also Sect.~\ref{sec:third_order}). The
maps were cross-correlated with stellar-mass maps, and common over- and
also under-densities could well be identified.

\subsection{3D lensing}
\label{sec:3D_lensing}

\citet{2007MNRAS.376..771K} applied 3D lensing as a proof of concept to two out
of five fields of 0.26 deg$^2$ size each, from the \survey{COMBO-17} survey
\cite{2003A&A...401...73W}. Constraints on $\Omega_{\rm m}, \sigma_8$ and $w_0$
were obtained with error bars consistent with expectations. A more refined
analysis was performed on the \survey{CFHTLenS} survey \cite{CFHTLenS-3d}. This
included conservative cuts directly in $k$-space, to limit the uncertainty of
models of non-linear and baryonic physics on small scales. Cuts in $k$ of $1$
and $5 \, h$ Mpc$^{-1}$, respectively, were considered, and, for the latter, the
phenomenological halo model with baryonic effects from
\cite{2011MNRAS.417.2020S} was included. Constraints including small scales were in
better agreement with the large-scale result when baryonic damping was taken
into account. Moreover, early-type galaxies were excluded from the analysis to
minimise the contamination from intrinsic alignment.

Dark-energy models beyond a constant parameter $w_0$ were considered as well in
\citet{CFHTLenS-3d} but the resulting constraints did not represent an
improvement compared to Planck + BAO.

\subsection{Other weak-lensing techniques in a cosmological context}
\label{sec:other_wl_techniques}

\subsubsection{Shear-ratio geometry test}
\label{sec:shear-ratio}

The variation with redshift of weak gravitational shear produced by a given
foreground structure depends on distances between observer, lens, and
source galaxy. By taking the ratio of shears at different background redshifts,
the dependence on the properties of the foreground structures at fixed redshift, even in the
highly non-linear regime, cancels out. This \emph{shear ratio test} is
thus is a purely geometrical probe of cosmology \cite{PhysRevLett.91.141302,2004ApJ...600...17B,2007MNRAS.374.1377T}.

Since the shear ratio involves ratios of angular distances, which are very
slowly varying functions of redshift and cosmology, it is a relatively
insensitive probe. Even though shear enters in linear order in the background
shear - foreground position correlations, and, similar to galaxy-galaxy lensing
(Sect.~\ref{sec:ggl}) and peak counts (Sect.~\ref{sec:peaks}) many
PSF residuals may cancel from circular averaging, this method is all the more
sensitive to redshift-dependent systematics, e.g.~in the measured shapes.
Further, the quality of photometric redshifts must be excellent not to dilute
the shear ratio.

\citet{2007MNRAS.376..771K} obtained cosmological constraints using the shear
ratio behind three massive Abell clusters at $z=0.165$ in the A901/A902
super-cluster field, observed with \survey{COMBO-17}.
\citet{2012ApJ...749..127T} measured the shear-ratio from 129 \survey{COSMOS} groups 
for a source galaxy sample for which redshifts
were as accurate as $\sigma_z = 0.018 (1+z)$, at a very high mean redshift of
$\bar z = 0.95$.

\subsubsection{Galaxy-galaxy lensing}
\label{sec:ggl-res}

Cosmological results from GGL have been obtained by combining GGL and galaxy
clustering. The ratio of combinations of those observables measures the ratio
$b/r$ of the bias and correlation coefficient between galaxies and total matter
\cite{1998A&A...334....1V,1998ApJ...498...43S}. This has been measured early
on with \survey{RCS} \cite{2001ApJ...558L..11H}. Cosmic shear adds an additional
observable to allow for an independent measure of both $b$ and $r$,
with only little dependence on cosmology \cite{2003ApJ...594...33F}.
However, it also sets more stringent demands on image quality, and is more prone
to systematics. Using all three observables, $b$ and $r$ have been measured as
functions of scale and redshift in 
\survey{RCS}+\survey{VIMOS-DESCART} \cite{2002ApJ...577..604H},
\survey{GaBoDS} \cite{Sbias06},
and \survey{COSMOS} \cite{2012ApJ...750...37J}.

Third-order correlations between mass and galaxies
(\emph{galaxy-galaxy-galaxy lensing}), introduced in
\citet{2005A&A...432..783S}, has been measured in \citet{2013MNRAS.430.2476S}.
This technique allows to quantify the excess mass around pairs of galaxies, probing the
joint environment of correlated galaxies.

\citet{2010PhRvD..81f3531B} introduced an estimator of clustering and GGL that
removes small scales, retaining only information on mass coming from an
annulus. This leads to a measure that is not sensitive to the non-linearity and
stochasticity of bias. \citet{2013MNRAS.432.1544M} applied this method to
\survey{SDSS} data and, by marginalising over non-linear bias parameters, obtained
results $\sigma_8(\Omega_{\rm m}/0.25)^{0.57} = 0.80 \pm 0.05$.

\citet{2010Natur.464..256R} used the same estimator and added redshift space
distortions to 2D clustering and GGL. They formed a combined quantity that is
(near-)independent of cosmology and galaxy bias, and probes relations between
the Bardeen potentials (\ref{eq:metric_gen}). GR was tested and confirmed, and
interesting constraints on certain types of $f(R)$ and TeVeS modified gravity theories
were obtained.

\subsection{Intrinsic alignment}
\label{sec:ia_results}

To measure intrinsic alignment, one can use a galaxy sample for which $GG$
is a subdominant contribution compared to $GI$ or $II$ (Sect.~\ref{sec:ia}).
This can be a very shallow survey where $GI$ is the dominant signal, or a very
narrow redshift bin for which $II$ is large. Other techniques consist of
nulling out $GG$ \cite{2008AA...488..829J,2009A&A...507..105J}, or a joint
measurement of $GG + GI + II$ together with a joint model-fitting of the three
components \cite{2005A&A...441...47K}. Additional observations of galaxy-galaxy lensing and galaxy
clustering helps to separate the different components
\cite{2010A&A...523A...1J,2010ApJ...720.1090Z}.

$II$ has been measured early on using digitized photographic plates from the very
shallow $10,000$ deg$^2$
\survey{SuperCOSMOS}
survey \cite{brown02}, by noting an order-of-magnitude excess of shape correlations compared to
cosmic shear predictions at low redshift.
In a more model-independent way, $II$ has been detected from physically close
pairs in \survey{COMBO-17} with its very accurate redshifts
\cite{2004MNRAS.347..895H}.

A further method to estimate $GI$ is the measurement of the correlation between
intrinsic ellipticity $\varepsilon^{\rm s}$ and galaxy number density. This can
be achieved by correlating two galaxy samples with overlapping redshift ranges, one of
which has shape information, and which can be a sub-set of the full sample.
Knowing the galaxy bias, e.g.~from measuring the galaxy spatial correlation
function, one can relate this correlation to the intrinsic shear -- density correlation.
Using a model of intrinsic alignment to infer $\varepsilon^{\rm s}$ in terms of
the density \cite{2004PhRvD..70f3526H}, one then writes the observed
correlation in terms of the density power spectrum.

This correlation has been measured in \survey{SDSS}
\cite{2004MNRAS.353..529H,2006MNRAS.367..611M,2007MNRAS.381.1197H,2009ApJ...694L..83O,2011A&A...527A..26J,2012JCAP...05..041B}
and \survey{Wiggle-Z} \cite{2011MNRAS.410..844M}, both from spectroscopic and
photometric redshift samples. Significant signals have been obtained for
luminous red galaxy (LRG) samples. Faint red galaxies and blue galaxies show a $GI$
correlation consistent with zero. This is consistent with the theoretical
expectation that mainly bright, central cluster galaxies are aligned with the
cluster potential. IA on large, super-cluster scales is produced by the
alignment of clusters with each other and with the surrounding dark matter.

A $2\sigma$ detection of intrinsic alignment from early-type galaxies was
obtained by jointly fitting cosmology and the \citet{2004MNRAS.353..529H}
linear IA model to \survey{CFHTLenS} cosmic shear tomographic data
\cite{CFHTLenS-IA}. Fig.~\ref{fig:II-GI} shows a systematic lower amplitude of
shear correlation for cross-redshift correlations compared to the cosmic-shear
prediction, as expected from a negative $GI$ contribution. No detection was
found for the late-type sample.

\doifonecol{%
\begin{figure}

  \centerline{\resizebox{0.6\hsize}{!}{
    \includegraphics{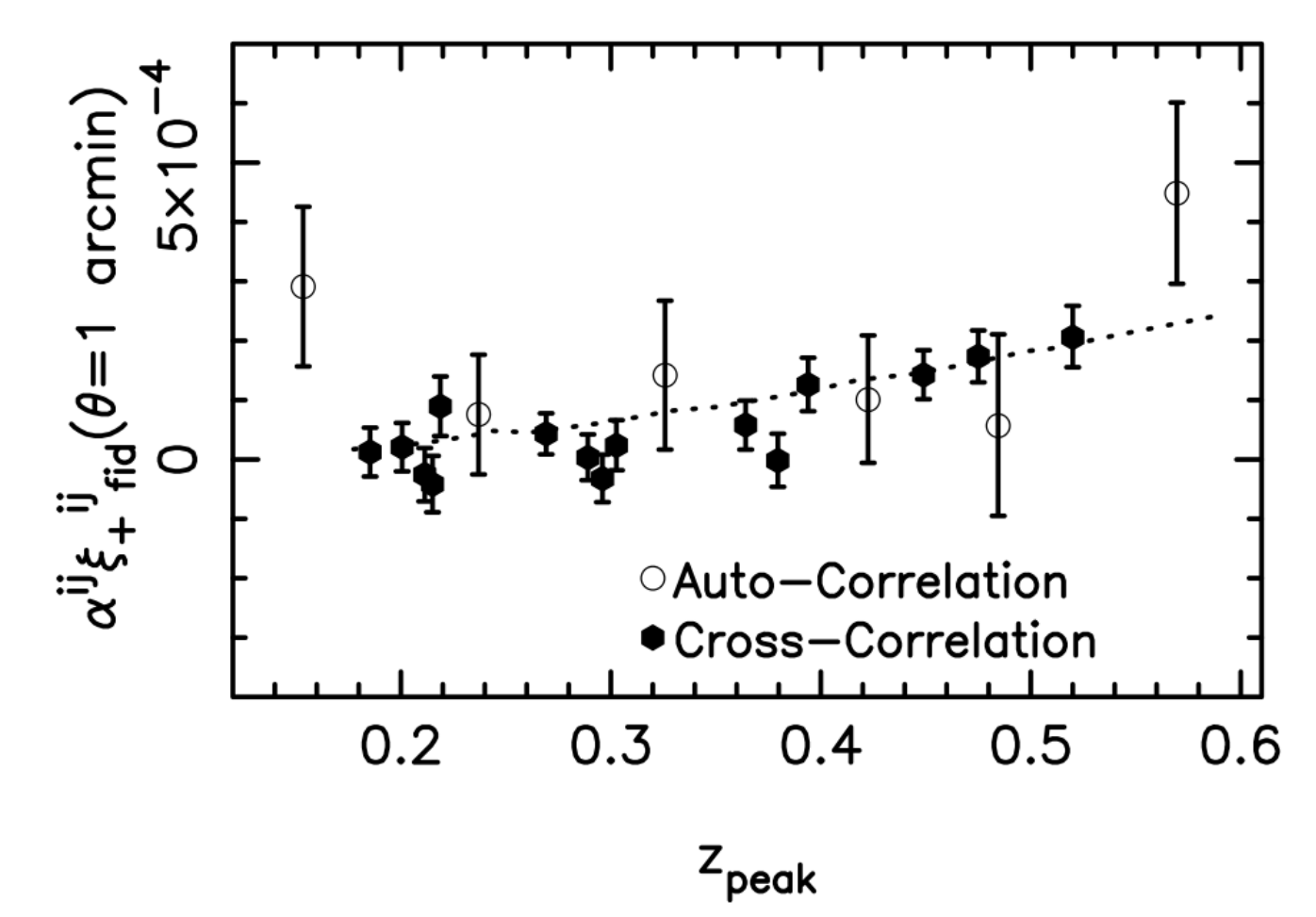}
  }}
}
\doiftwocol{%
\begin{figure}[H]

  \centerline{\resizebox{\hsize}{!}{
    \includegraphics[bb=20 0 385 280, height=10em, width=10em]{figures/fig12.pdf}
  }}
}

  \caption{Amplitudes of tomographic measures of $\xi_+$ at $\theta = 1$ arcmin
for redshift bins $(ij)$, against peak lensing efficiency redshift $z_{\rm
peak}$ for early-type galaxies, from \survey{CFHTLenS}. The free parameters
$\alpha^{ij}$ multiplied with a WMAP7 fiducial $GG$ model $\xi_{+ \rm fid}$
were fitted to $\xi_+$ and $\xi_-$, simultaneous for all redshift bins and
angular scales. At low $z$, the auto-correlations ($i = j$, open circles) lie
above the fiducial dashed line ($a^{ij} = 1$), as expected for a $II>0$
contribution. The cross-correlations ($i \ne j$, filled circles) lie
systematically below the prediction, indicating a $GI<0$ contamination. From
\citet{CFHTLenS-IA}.
Figure used with permission from \citet{CFHTLenS-IA}, \emph{\mnras}, \textbf{432}, 2433.
Copyright Oxford University Press.
}

  \label{fig:II-GI}

\end{figure}

Satellite galaxies are expected to be radially aligned towards the host cluster centre. This follows from
tidal-torque interactions with the cluster potential, and has been observed in the past for a small cluster samples,
e.g.~\citet{2005ApJ...627L..21P}.
A recent measurement using a much larger sample from \survey{SDSS} \cite{2014arXiv1411.1755S} detects radial alignment on small
scales. However, other measurements of cluster satellite galaxies 
are consistent with no alignment,
using spectroscopic galaxies in massive clusters \cite{2014arXiv1406.5196S}, and photometric
samples in groups and clusters \cite{2014arXiv1407.4813C}.


\section{Future cosmic shear expectations and forecasts}
\label{sec:future_cs}

\subsection{Upcoming and future surveys}
\label{sec:upcoming_surveys}

New instruments, either cameras, telescopes, or both, are being designed and
built specifically for the purpose of weak-lensing observations. Their design
is driven by the goal to provide superb image quality with very small, uniform,
and well-understood image distortions. The pixel scale is chosen to
sufficiently sample the PSF. In view of the enormous costs of new experiments,
in particular space missions, the instruments have to be thoroughly and
carefully designed to guarantee the desired scientific outcome, for example,
the measurement of dark-energy properties with a given accuracy.

An instrument designed for weak lensing allows at the same time pursuing other
scientific goals, for example the search for exoplanets via micro-lensing
\cite{2010arXiv1010.0002G}, or supernovae surveys, with the caveat that
detecting transient objects requires a specific survey strategy that is most
likely not optimal for weak lensing. Further cosmological probes
and techniques can be explored using photometric redshifts from a weak-lensing
survey such as galaxy clustering, galaxy clusters, and strong lensing.
Last but not least, a lensing survey offers a huge legacy value, providing
high-resolution multi-band images of galaxies at high redshift.

\subsubsection{Ground-based surveys}

An order-of-magnitue increase in area compared to current surveys is being
undertaken since 2012 by the Kilo Degree Survey
\citeaffixed{2013ExA....35...25D}{\survey{KiDS; }}. $1,500$ square degrees are
being mapped with the recently commissioned wide-field camera
\instrument{OmegaCAM} on \instrument{VST} (VLT Survey Telescope). Four optical
bands are complemented by five deep IR bands observed at \instrument{VISTA}
(Visible and Infrared Survey Telescope for Astronomy) within the survey
\survey{VIKING} (VISTA Kilo-degree Infrared Galaxy survey), providing excellent
photometric redshifts. The VST mirror with its $2.6 \, \mbox{m}$
diameter is on the small side to carry out a large survey. This is however
compensated by the small PSF and excellent observing conditions at the Cerro
Paranal site. With a projected limiting magnitude of $i =
24.2$\footnote[1]{if not stated otherwise, all limiting magnitudes quoted in
this section correspond to a $5\sigma$ extended source, measured in the AB
system}, \survey{KiDS} is shallower than \survey{CFHTLS-WIDE} ($i = 24.5$).
Note however that weak-lensing galaxy shapes are measured in the $r$ band,
which is observed longer and in better seeing conditions, leading to $r =
25.2$.

The \survey{Dark Energy Survey} \citeaffixed{2005astro.ph.10346T}{\survey{DES};
}, started in 2013 will observe $5,000$ square degrees in the South, using the
newly constructed $3 \,\mbox{deg}^2$ field-of-view \instrument{DECam} at the $4
\, \mbox{m}$ \instrument{Blanco} telescope. The seeing at Cerro Tololo is
larger compared Cerro Paranal, and the planned depth of \survey{DES} with $i =
24.5$ ($10\sigma$ extended source) is similar to \survey{CFHTLenS}. The survey
area overlaps with many observations in other wavelengths, e.g.~with the
\instrument{South Pole Telescope} (SPT), and the (shallow) infrared
\survey{Vista Hemisphere Survey} (VHS).

A smaller area but significantly deeper limiting magnitude is provided
the recently built \instrument{HyperSuprimeCam} (\instrument{HSC}) on the $8.2
\, \mbox{m}$ \instrument{Subaru} telescope. Around $1,500$ square degrees with
excellent image quality in multiple optical bands will be used for weak
lensing, with a very high planned depth down to $i = 26$
\cite{2012SPIE.8446E..0ZM}\footnote{see also the (unpublished) HSC white paper
\texttt{\url{www.astro.princeton.edu/~strauss/hsc_main.pdf}}}.

These current and near-future surveys will be followed by the next generation
of experiments that will cover most of the extragalactic sky of $15,000$ square
degrees and more. From the ground, the $8.4 \, \mbox{m}$ Large Synoptic Survey Telescope
\citeaffixed{2009arXiv0912.0201L}{\instrument{LSST}; } will provide extremely
deep images down to $r = 28$. Since individual
exposures are very short, on the order of $15$ seconds to discover transient
objects, they will have to be stacked or otherwise combined to do weak-lensing
measurements.

\subsubsection{Space-based surveys}

Going to space offers the two major advantages: Escaping atmospheric turbulence
leads to a stable and small PSF, and infrared observations provide photo-$z$s to
significantly higher redshifts than from the ground.

About $15,000$ deg$^2$ will be observed from space with the ESA satellite
mission \instrument{Euclid} \cite{2011arXiv1110.3193L}. The two main science
drivers for Euclid are cosmic shear and galaxy clustering, which will be
observed using three instruments, an optical imager, a near-infrared imager,
and a near-infrared slitless spectrograph. The optical imager on board Euclid
is designed to have a very stable PSF, both spatially as well as in the time
domain. To collect enough galaxy light from billions of high-redshift galaxies
($30 \, \mbox{arcmin}^{-2}$), the transmission curve is very broad,
corresponding to the combined $R + I + z$ filters, with a required depth of
$R+I+z = 24.5$. This poses new challenges to overcome, in particular galaxy colour gradients
and PSF calibrations (Sect.~\ref{sec:PSF_colour_effects}). Further obstacles
unique to space-based observations will have to be tackled
\cite{2013MNRAS.431.3103C}: For example, the very small PSF will be
undersampled by the pixels of size $0.1$ arcsec. From these undersampled
stellar images, a reliable, high-resolution PSF model has to be reconstructed.
Furthermore, the detector degrades with time due to the bombardment with cosmic
rays, and the shapes of objects get distorted by the so-called \emph{charge
transfer inefficiency} (CTI). Corrections as function of time, position on
chip, and brightness of the objects have to be applied
\cite{2010MNRAS.401..371M,2014MNRAS.439..887M}.

A space mission proposed by NASA is \instrument{WFIRST--AFTA} \citeaffixed{2013arXiv1305.5422S}{Wide-Field
Infrared Survey Telescope -- Astrophysics Focused Telescope Asset; }.
WFIRST--AFTA uses a $2.4$ m mirror with near-infrared imaging and
spectroscopy capabilities. Around $2,400$ square degrees will be imaged for
weak lensing in the near infrared, with $50$ galaxies per square
arcmin, corresponding to $J = 25.7$.
The smaller area but higher depth compared to \instrument{Euclid} will
probably result in a similar expected performance of \instrument{WFIRST--AFTA}
with respect to constraining cosmological parameters. Apart from weak lensing,
\instrument{WFIRST--AFTA} will measure galaxy clustering and SNe as
cosmological probes, and exoplanets will be hunted for with
microlensing and coronagraphy, using a coronagraph as additional instrument.
 
\subsubsection{Further ideas}

Further non-space-based proposals have been put forward to reduce the
atmospheric influence but not going into orbit. These include the
balloon-borne experiment High Altitude Lensing Observatory
\citeaffixed{2012APh....38...31R}{\instrument{HALO}; }, which can fly above
99.9\% of the atmosphere, but which in turn has to solve the problem of limited
flight duration and pointing stability. A further possibility is an
optical/infra-red telescope at the South Pole. This site offers nearly
space-based observing conditions, once the boundary layer turbulence with a
height of $30$ -- $40$ m above the ground is surpassed, with seeing as small
as $0.3$ arcsec at optical wavelengths \cite{2009PASA...26..397L}. The main
obstacles to a survey telescope in Antarctica are the missing infrastructure
and difficult access for a ground-based facility.

\subsection{Radio lensing}
\label{sec:radio}

Weak gravitational lensing from radio wavelengths was measured for the
first time in \citet{2004ApJ...617..794C}. Shapes of galaxies were obtained
using the shapelet method applied to interferometric radio data
\cite{2002ApJ...570..447C}. Compared to the optical, the number density of
current radio surveys is much smaller, typically by two to three orders of
magnitude, leading to a very low detection significance in particular for small
survey areas \cite{2010MNRAS.401.2572P}. Moreover, the redshift distribution of
radio galaxies is not known accurately, making the interpretation of the
measurement very challenging. However, radio weak lensing offers several
advantages: PSF effects are much smaller, since there is virtually no
atmospheric stochastic blurring. Well-known and stable beams of radio
interferometers facilitate accurate PSF models. Furthermore, radio galaxies are
typically at much higher redshifts than galaxies from wide-field optical
surveys.

Future radio surveys such as
the \instrument{Low-Frequency Array for Radio Astronomy (LOFAR)}
and the \instrument{Square Kilometre Array (SKA)}
will reach a sufficient sensitivity to resolve radio emission of ordinary galaxies and
therefore provide a large number density ($\ge 100$ arcmin$^{-2}$ for SKA), at
a resolution corresponding to optical observations from space. Furthermore, source
redshifts will be available, although not at high redshifts, from HI 21 cm line
observations \citep[e.g.~]{2004NewAR..48.1063B}.

Observations in polarized light can help to reduce the noise due to the intrinsic
ellipticity of galaxies: Gravitational lensing does not change the position
angle of the polarization emission of a galaxy. Since there exists a relatively
strong correlation between polarization and galaxy morphology, for example from
magnetic fields that are aligned with the disk, the measurement of polarized
emission is an estimator of the unlensed galaxy orientation
\cite{2011MNRAS.410.2057B}. Moreover, a prior on the unlensed orientation,
together with an unpolarized shear estimator, helps to correct for intrinsic
alignments.

\subsection{3D mass reconstruction}
\label{sec:mass_recon}

Weak-lensing reconstructions of the potential and mass distribution in three
dimensions are possible albeit very challenging. Even though the 3D information
is present in the shear field in the presence of photometric redshifts, the
lensing kernel is very broad, which makes it difficult to estimate
the redshift of the lensing structures. 3D potential or mass reconstructions
from shear data are ill-posed inverse problems, and to solve them, a
regularization scheme or prior has to be introduced. A commonly used
regularization for linear methods is Wiener filtering
\cite{2002PhRvD..66f3506H,2003MNRAS.344.1307B,2009MNRAS.399...48S}. This
increases the signal-to-noise ratio of the reconstruction, but also introduces
systematic biases by shifting and elongating the structures along the redshift
direction. These biases are general limitations of linear methods, and are also
found using other regularisation schemes such as singular value decomposition
\citeaffixed{2011ApJ...727..118V}{SVD;}.  Non-linear methods are able to 
significantly reduce those biases
\cite{2012A&A...539A..85L,2013arXiv1308.1353L}.

A 3D mass map allows for detailed comparison between dark matter and
baryonic tracers, including redshift evolution, and could serve as a direct
cosmological probe of the halo mass function.

\subsection{Magnification}
\label{sec:magnification}

Gravitational lensing conserves surface brightness, a consequence from
Liouville's theorem which holds in any passive optical system. Since the
apparent size of resolved background objects change, their flux changes as
well. These two effects are manifestations of \emph{gravitational
magnification}, and can be used as weak-lensing observables in addition to the
deformation (shear) of galaxy shapes.

Two competing effects are at play: Lensing not only changes the object sizes,
but stretches the region of sky behind a lens, thereby on the one hand
reducing the number density of source galaxies. On the other hand,
for a flux-limited galaxy sample which is the case for most optical surveys,
the flux of background objects are pushed above the limit, thereby increasing
the local number density. Whether the first, purely geometrical dilution effect
wins over the second, astrophysical effect depends on the intrinsic slope
$\alpha$\footnote{not to be confused with the deflection angle
(\ref{eq:alpha_hat})} of the cumulative flux distribution. The larger the ratio
of faint to bright objects in the sample, the steeper the distribution, and the
stronger is the number density increase.

The magnification of an object is defined as the ratio of the lensed to the unlensed flux.
An object's flux can be obtained by integrating over the 2D brightness
distribution. The magnification $\mu$
is therefore the determinant of the inverse Jacobian $\mat A^{-1}$ (\ref{eq:jacobi}),
\begin{equation}
   \mu =  \det {\mat A}^{-1} = \left[ (1-\kappa)^2 - |\gamma|^2 \right]^{-1} \approx 1 + 2 \kappa;
   \label{eq:magnification}
\end{equation}
the latter approximation holds in the limit of weak lensing.

Magnification comes for free from any weak-lensing survey, since fluxes have to
be measured with high accuracy anyway, to determine photometric
redshifts. The requirements on image quality are similar for measuring shapes, sizes,
or fluxes.

The signal-to-noise ratio of flux magnification for an ensemble of galaxies is
lower than the one for shear \cite{BS01,2010MNRAS.401.2093V}, since the former
signal scales with $|\alpha - 1|$, which typically is not much larger than unity,
even for very steep flux distributions. Size magnification is expected to have
slightly smaller signal-to-noise ratio than shear: A parameter $r$ describing
the size of a galaxy, for example its half-light radius, scales as $r \sim 1 +
\kappa$, therefore an estimator of the convergence is $\log r/r_0 \approx \kappa$,
where $r_0$ is the unlensed size.
Its intrinsic distribution has a width of $\sigma_{\log r} \approx 0.3 - 0.5$
\cite{2003MNRAS.343..978S}, slightly higher than the intrinsic ellipticity
dispersion $\sigma_\varepsilon \approx 0.3 - 0.4$. 

Flux measurement however can be made on galaxies that are too small or faint
for accurate shape measurements, and therefore reach higher redshifts.
Magnification suffers from different systematics which makes the two techniques
complementary.

There are several difficulties of the measurement and interpretation of
magnification compared to cosmic shear:

\begin{itemize}

\item Contrary to ellipticity, the expectation value of the unlensed magnitude
or size does not vanish. This results in a challenge to interpret \emph{cosmic
magnification} which, analogous to cosmic shear is the correlation of galaxy
magnitudes in blind fields.

\item The analogue to intrinsic alignment are intrinsic magnitude and size
correlations, which are caused by the clustering of galaxies. Intrinsic
magnitude correlations are about a factor of 10 larger than IA, whereas size
correlations are probably smaller \citeprefixed{2013MNRAS.433L...6H}{ and
references therein}.

\item Non-lensing contributions to number density fluctuations are dust
absorption, and crowded fields for example near cluster centres. The first
contamination can be corrected for in principle by measuring a colour-dependent
magnification signal \cite{2010MNRAS.405.1025M}.

\end{itemize}

Several of the above difficulties can be mitigated by correlating background to
foreground tracers of dark matter, analogous to galaxy-galaxy lensing for
shear. One either correlates the background magnitude (or size) around low-$z$
galaxies, or computes the angular number count cross-correlation function. For
faint background samples, the corrsponding shallow slope $s$ causes a negative
number count correlation which is a very distinct sign for gravitational
magnification. In the absence of gravitational lensing, this correlation
vanishes.

Due to photometric redshift errors, catastrophic outliers scatter objects
between the foreground and background sample, which can produce a physical,
intrinsic size and magnitude correlation that is an order of magnitude larger
than the gravitational magnification. To prevent this contamination, background
galaxy samples with very confident and well-controlled photometric redshifts
have been employed in prior robust detections of magnification, such as Lyman
Break Galaxies
\citeaffixed{2009A&A...507..683H,2012ApJ...754..143F,2012MNRAS.426.2489M}{LBG;
}. Other work found a magnification signal from foreground X-ray groups in
COSMOS from a large background sample, however, it is unclear how strongly
contaminated the measurement was from physical clustering due to photo-$z$
errors \cite{2012ApJ...744L..22S}.

\subsection{Cosmic flexion}
\label{sec:flexion}

Flexion denotes the second-order distortion terms in the
expansion of the lens equation (\ref{eq:lens}). Whereas shear parametrizes
first-order, linear distortions (\ref{eq:jacobi}), flexion is given as third
derivatives of the lensing potential. It has two (complex) components that can
be written as the derivatives of the shear field. While shear gives rise to
elliptical images with two-fold symmetry, the two flexion components produces
images with one- and three-fold symmetry, respectively. The combined effect of
shear and flexion is a skewed, arc-like image
\cite{2002ApJ...564...65G,2006ApJ...645...17I,2005ApJ...619..741G,2007ApJ...660..995O,2008A&A...485..363S,2009MNRAS.396.2167B}.

Flexion measures local changes of the shear field and is significant only on
small, non-linear scales. On larger scales, cosmic flexion has a lower S/N
ratio than cosmic shear \cite{2006MNRAS.365..414B}. For that reason, the
usefulness of cosmic flexion is doubtful, since it requires models of the LSS
on very small, non-linear scales. Moreover, contrary to shear, flexion is more
sensitive to photon pixel noise than to the intrinsic galaxy shape dispersion.
This leads to a strong increase of the measurement uncertainty towards low S/N
where photon Poisson noise becomes high. It also adds to the highly
non-Gaussian distribution of flexion components, which are difficult to
parametrize \cite{2013MNRAS.435..822R}.

\subsection{Peak statistics}
\label{sec:peaks}

In weak-lensing data one can identify projected over-densities by isolating
regions of high convergence, or enhanced tangential shear alignments. The
statistics of such weak-lensing \emph{peaks} are a potentially powerful probe
of cosmology, since peaks are sensitive to the number of halos and therefore
probe the halo mass function, which strongly depends on cosmological
parameters,
\citeaffixed{1986MNRAS.222..323K,1989ApJ...347..563P,1989ApJ...341L..71E}{e.g.}.
A \emph{shear-selected} sample of peaks is a tracer of the total mass in halos,
and does not require scaling relations between mass and luminous tracers, such
as optical richness, SZ or X-ray observables.

The relation between peaks and halos is complicated because of projection and
noise. Several small halos in projection or filaments along the line of sight
can produce the same lensing alignment as one larger halo. Noise in the form of
intrinsic galaxy ellipticities produces false detections, and alters the
significance of real peaks \cite{2007A&A...462..875S}. Because the number of
halos strongly decreases with mass, noise typically results in an up-scatter of
peak counts towards higher significance, which has to be carefully modeled.

Numerical simulations have shown a large potential of peak counts to constrain
cosmological parameters \cite{2010PhRvD..81d3519K,2012MNRAS.423.1711M}. Shear
peaks single out the high-density regions of the LSS, and therefore probe
the non-Gaussianity of the LSS. Despite peak counts being a non-linear probe of weak
lensing, they require the measurement only to first order in the observed shear, an
similar to galaxy-galaxy lensing (Sect.~\ref{sec:ggl}) this technique potentially suffers
from less systematics than higher-order shear correlations. Peak counts are
complementary to second-order statistics, and both probes combined are able to
lift parameter degeneracies
\cite{2010MNRAS.402.1049D,2009A&A...505..969P,2011PhRvD..84d3529Y,2012MNRAS.423..983P}.
In addition to peak counts,
the two-point correlation function of lensing peaks carries cosmological
information \cite{2013MNRAS.432.1338M}.

Theoretical predictions for peak counts are difficult to obtain, in particular
at high signal-to-noise. Recent approaches are based on Gaussian random fields
\cite{2010ApJ...719.1408F,2010A&A...519A..23M}, but more work is needed to
interpret weak-lensing peak count data in a cosmological framework. A
new, flexible model of peak counts is based on samples of halos drawn from the
mass function, which can be generated very quickly without the need to run
time-consuming $N$-body simulations \cite{peaks1}.

The first measurement of weak-lensing peaks including their counts and angular
correlation function was obtained in \citet{2013arXiv1311.1319S}. The observed
peak counts in the \survey{CFHT/MegaCam Stripe-82} survey (\survey{CS82}) were
found to be in good agreement with the theoretical prediction from
\citet{2010ApJ...719.1408F}. 
Two groups have obtained cosmological constraints from peak counts, from
\survey{CFHTLenS} (Liu, Petri et al.~\citeyear*{2014arXiv1412.0757L})
and \survey{Stripe-82} (Liu, Pan et al.~\citeyear*{2014arXiv1412.3683L}) data.

\subsection{Outlook}
\label{sec:outlook}

In 2000, cosmic shear was first measured over a few square degrees
of observed sky, from some ten thousand galaxies. Fifteen years later,
surveys have increased these numbers by a factor of $100$, imaging a few million
galaxies over ${\cal O}(100)$ square degrees. Many challenges were met to analyse these
data, taking years of work. This resulted in constraints on
cosmological parameters that are competitive compared to other
cosmological probes.

In another ten years, upcoming and future experiments will cover a substantial
fraction of the entire sky, measuring billions of galaxies. This signifies
yet another data volume increase of a factor of $100$,
not to mention the data quality improvement due to
instruments dedicated to weak lensing. The formidable challenge here is reducing 
systematic errors to an acceptable level when analysing these large data sets.
New, unprecedented difficulties
have to be overcome, for example CTI for Euclid, and
blended galaxy images for LSST. To fully exploit those surveys, large
follow-up programs are needed to obtain the
necessary large samples of photometric and spectroscopic redshifts.
In addition, to interpret the results of those surveys, the accuracy of
theoretical predictions of the non-linear power spectrum including baryonic
physics need to be significantly improved.

If all these challenges can be overcome, weak cosmological lensing has the
great potential to advance our understanding of fundamental physics. It can
explore the origin of the recent accelerated expansion of the Universe, and
distinguish between dark energy models and theories of modified gravity. Cosmic
shear can measure initial conditions of the primordial Universe, constrain the
mass of neutrinos, and measure properties of dark matter. Not only that, the
study of intrinsic galaxy alignments has provided insights into the formation
and evolution of high-redshift galaxies in their dark-matter environment,
proving that cosmic shear does not only probe cosmology, but influences and
enriches other areas of astrophysics. Thus, over the last fifteen years, weak
cosmological lensing has established itself as a major tool in understanding
our Universe, and with upcoming large surveys, it will continue to be of great
value for astrophysics and cosmology.
\bigskip

\ack

I am very grateful to Catherine Heymans, Chieh-An Lin, Peter Schneider, and
Melissa Thomas for their detailed comments that helped to improve the
manuscript.
I would further like to thank Jean Coupon, James Jee, Fergus Simpson, and
Massimo Viola for helpful discussions, and Ludo van Waerbeke for valuable
contributions to early stages of this article.
The comments of the two anonymous referees helped to improve the review, and
I am very thankful for their diligence.
Two figures in this review were generated using data that were kindly
provided by Liping Fu and Tom Kitching.
Several figures were taken from work published
elsewhere, and I am thankful to the authors and publishers for authorizing the
reproduction of the original figures.

\section*{References}
\addcontentsline{toc}{section}{References}

\bibliographystyle{jphysicsB}

\bibliography{astro}

\stoptwocol

\end{document}